\documentclass[useAMS,usenatbib,fleqn]{mnras}
\usepackage{graphicx, amsmath, caption, subcaption} 
\usepackage{newtxtext,newtxmath, color, xcolor}
\usepackage[pagewise]{lineno}
\usepackage[normalem]{ulem}
\usepackage[flushleft]{threeparttable}
\usepackage[nameinlink,capitalise]{cleveref}
\usepackage{orcidlink}
\def\visbnut{\mathcal{V}\left( \mathbfit{b}, \nu, t \right)}
\def\visbnutm{\mathcal{V}\left( \mathbfit{b}, \nu_m, t_m \right)}
\def\r{\hat{\mathbfit{r}}}
\def\rt{\hat{\mathbfit{r}}(t)}

\def\rcm{\hat{\mathbfit{r}}_c(t_m)}
\def\Irnut{\mathcal{I}\left( \r, \nu, t \right)}
\def\Brnut{\mathcal{B}\left( \r, \nu, t \right)}
\def\Brpnut{\mathcal{B}\left( \r_p, \nu, t \right)}
\def\Irnutm{\mathcal{I}\left( \r, \nu_m, t_m \right)}
\def\Brnutm{\mathcal{B}\left( \r, \nu_m, t_m \right)}
\def\bt{\mathbfit{b}(t)}
\def\utm{\mathbfit{u}(t_m)}
\def\btm{\mathbfit{b}(t_m)}
\def\dudtm{\frac{d\mathbfit{u}}{dt}\vert_{t=t_m}}
\def\dbdtm{\frac{d\mathbfit{b}}{dt}\vert_{t=t_m}}
\def\dbdt{\frac{d\mathbfit{b}}{dt}}
\def\nubc{\frac{\nu}{c}}
\def\numbc{\frac{\nu_m}{c}}
\def\dt{\delta t}
\def\dnu{\delta \nu}
\def\sinc{{\rm sinc}}
\newcommand{\HI}{{\rm H\hspace{0.5mm}}{\scriptsize {\rm I}}}
\def\oc{{\mathbf{\omega}}_{\oplus}}
\def\eg{{\it e.g.}\,}
\def\ie{{\it i.e.}\,}
\def\sigx{\sigma_{\nu, x}}
\def\sigy{\sigma_{\nu, y}}
\def\Dx{\Delta x}
\def\Dy{\Delta y}

\title{The MeerKLASS On-the-Fly continuum survey: pipeline design and validation }
\author[S. Chatterjee et al.]{Suman Chatterjee$^1$\thanks{sumanchttrj05@gmail.com}\,\orcidlink{0000-0001-8852-5888},
Mario G. Santos$^{1,2}$,
Kristof Rozgonyi$^3$,
Keith Grainge$^4$,
Sarvesh Mangla$^3$,
\newauthor
Joseph J. Mohr$^3$,
Sourabh Paul$^4$,
Yvette Perrott$^{5}$,
Oleg M. Smirnov$^{6,2,7}$,
Cyril Tasse$^{8,9}$,
Laura Wolz$^4$
\\
\\
$^1$Department of Physics and Astronomy, University of the Western Cape, Robert Sobukwe Road, Cape Town 7535, South Africa\\
$^2$South African Radio Astronomy Observatory, Cape Town, 7925, South Africa\\
$^3$University Observatory, LMU Faculty of Physics, Scheinerstr. 1, 81679, Munich, Germany\\
$^4$Jodrell Bank Centre for Astrophysics, Department of Physics \& Astronomy, The University of Manchester, Manchester M13 9PL, UK\\
$^{5}$School of Chemical and Physical Sciences, Victoria University of Wellington, Wellington 6012, New Zealand\\
$^6$Centre for Radio Astronomy Techniques and Technologies (RATT), Department of Physics and Electronics, Rhodes University, Makhanda, 6140, South Africa\\
$^7$Institute for Radioastronomy, National Institute of Astrophysics (INAF IRA), Via Gobetti 101, 40129 Bologna, Italy\\
$^8$GEPI \& ORN, Observatoire de Paris, Université PSL, CNRS, 5 Place Jules Janssen, 92190 Meudon, France\\
$^9$Department of Physics \& Electronics, Rhodes University, PO Box 94, Grahamstown, 6140, South Africa\\
}

\date{Accepted XXX. Received YYY; in original form ZZZ}

\pubyear{\the\year{}}

\begin{document}
\label{firstpage}
\pagerange{\pageref{firstpage}--\pageref{lastpage}}
\maketitle
\begin{abstract}

The MeerKAT Large Area Synoptic Survey (MeerKLASS) is designed to map large areas of the Southern sky for cosmology using the single-dish \HI~intensity mapping (IM) technique, while simultaneously delivering a wide, high angular-resolution interferometric survey. We present the design and first results of the MeerKLASS On-the-Fly (OTF) continuum data, which exploits the visibilities recorded during fast, constant-elevation scans. This observing mode enables fast commensal imaging over several hundred of square degrees on a nightly basis.

We describe the OTF survey strategy and pipeline, focusing on handling challenges introduced by the current MeerKAT fixed-delay correlation observing mode, which causes decorrelation (smearing). We implement a correction scheme based on time-dependent phase rotation, direction-dependent PSF modeling, and wide-band faceted deconvolution with \texttt{DDFacet}. Using UHF-band and pilot L-band data, we demonstrate the recovery of high-quality 2-second snapshot images and deep mosaics over hundreds of square degrees. After smearing correction we are able to achieve a resolution of $23.3\arcsec$ and $14\arcsec$ with an rms sensitivity of $35 \mu {\rm Jy\,beam}^{-1}$ and $ 33 \mu {\rm Jy\,beam}^{-1}$ in the UHF and L-band respectively.

The full survey will cover $10,000 \, {\rm deg}^{2}$ at 544-1088 MHz, and after the delay tracking fix implemented we expect to reach $\sim 25 \mu {\rm Jy\,beam}^{-1}$ at $14\arcsec$ resolution. The continuum OTF data products will support diverse science goals, including galaxy and AGN evolution, diffuse cluster emission, large-scale structure and cosmology, rotation-measure synthesis, and transient searches. MeerKLASS-OTF thus establishes an efficient path to wide-area commensal surveys with MeerKAT and provides a key technical precursor for SKA-Mid.

\end{abstract}


\section{Introduction}
Neutral hydrogen (\HI) 21-cm intensity mapping (IM) in the post-reionisation era, will allow us to place novel cosmological constraints on dark energy and modified gravity. The technique uses \HI~ as a tracer for galaxies, which in turn are a tracer for large scale structure. Observations of the redshifted 21-cm line allow us to build up a 3D picture of the Universe across cosmic time. Rather than attempting the very time consuming approach of identifying individual galaxies through their 21-cm emission, IM seeks to measure the bulk \HI~on cosmologically significant scales, therefore requiring access to large angular scales \citep{Bharadwaj2001b, Battye2004, Chang2008, Wyithe2008}. 

MeerKAT \citep{Jonas2009, Jonas2016}, the precursor instrument for the upcoming SKA-Mid\footnote{\href{https://www.skao.int/en}{https://www.skao.int/en}} \citep{Braun2019}, is using the single-dish data from each antenna of the array to perform a state of the art \HI~IM survey. This approach can provide us with access to the large cosmological scales inaccessible from the interferometric data. The ongoing MeerKAT Large Area Synoptic Survey (MeerKLASS; \citealt{Santos2017}) plans to observe a total of $\sim 10,000$ deg$^2$ in the Southern sky outside the galactic plane over the next few years. The focus is the UHF band ($544-1088$ MHz) corresponding to the \HI~redshift range: $0.40 < z < 1.45$. Earlier MeerKAT observations in the L-band have already shown promising results, including a detection of the \HI~cosmological power spectrum in cross-correlation with galaxy surveys at $z\sim 0.4$ \citep{Wang2021, Cunnington2022, Cunnington2025, Cunnington2025b}. 

Although this survey is using the auto-correlation data, all the interferometric cross-correlations are being saved at the same time. Using all these data, would allow us to perform a high angular resolution interferometric survey together with the single-dish \HI~IM survey. Such commensality would dramatically increase the legacy science provided by MeerKLASS without using more observing time. However, due to stability requirements for single dish data, MeerKLASS has adopted a fast scanning strategy, where the dishes move back and forth at constant elevation, instead of the standard tracking observations. This imposes challenges to the interferometric data processing, requiring the adoption of the so called On-the-Fly (OTF) imaging technique. This is similar for instance to what the VLASS survey is doing (\citealt{Lacy2020}). Although challenging, this also means that MeerKLASS has the capability to observe wide areas of the sky quickly (about 300 to 600 deg$^2$ per night). In this paper we describe the pipeline required to perform a commensal radio continuum survey using the interferometric visibilities recorded during MeerKLASS observations. The goal is to cover the 10,000 deg$^2$ with 9 sub-bands in the range $544-1088$ MHz, with a target resolution of $14\arcsec$ and 25 $\mu{\rm Jy\,beam}^{-1}$ rms in continuum.

Low frequency radio continuum surveys are powerful resources to understand galaxy formation and evolution \citep{Heckman2014}. Radio surveys below a few GHz, at relatively high flux densities are dominated by sources with strong synchrotron emission from Active Galactic Nuclei (AGN). Due to strong radio emission (radio-loud) and long wavelengths compared to the dust grains around AGN, radio surveys are able to probe these AGN across a substantial duration of cosmic time. Unlike AGN, star forming galaxies (SFGs) in shallow surveys are primarily from low redshifts ($z<0.1$) \citep{Mauch2007}. However, deeper surveys ($S_{\rm 1.4 GHz} < 1{\rm mJy}$) can potentially detect SFGs from higher redshifts  ($z<0.8-0.9$) \citep{Kondapally2022, Whittam2023}. Studying these variety of objects is expected to advance our understanding of the Universe.

A number of dedicated large continuum surveys of the southern sky have been completed or are ongoing using other telescopes, each offering different resolutions and sensitivities (see Table~\ref{tab:survey_SH}). At very low frequencies (72-231 MHz), the GaLactic and Extragalactic ALL-sky MWA survey (GLEAM; \citealt{Wayth2015} and GLEAM-X; \citealt{Hurley-Walker2022}) covers the declinations ($\delta$) below $+30^{\circ}$. A combination of Sydney Molonglo sky survey (SUMSS; \citealt{Bock1999, Mauch2003}), the Evolutionary Map of the Universe (EMU; \citealt{Norris2011, Hopkins_2025}) and RACS-Low (\citealt{McConnell2020, Hale2021}) covers the southern sky at or below 1GHz. Between 1 and 2 GHz we have RACS-Mid (\citealt{Duchesne2023}) and RACS-High(\citealt{duchesne2025}) covering a significant fraction of the southern hemisphere, with NVSS (\citealt{Condon1998}) covering north of declination $-40^\circ$. VLASS (\citealt{Lacy2020}) is another Northern survey, covering the sky at frequencies $2-4$ GHz, again down to the limit of $-40^\circ$ declination.

Up to now, there was no wide survey with the MeerKAT telescope. This array consists of $64 \times 13.5$-m dishes with offset Gregorian optics, providing an unblocked aperture. It is equipped with three receiver bands: UHF ($544-1088$ MHz), L  ($856-1712$ MHz), and S ($1750-3500$ MHz, split into sub-bands). Three-quarters of the collecting area is within a dense, 1 km diameter core region, and the remaining dishes are situated around the core, providing a maximum baseline of 8 km. The large number of baselines, wide field of view (1 deg at L band), and low ($\sim 20$ K) system temperature all conspire to make MeerKAT an exceptionally capable synthesis imaging telescope with large survey speeds.

Notable examples of continuum surveys from previously approved MeerKAT ``Large Survey Projects'' are the MeerKAT International GHz Tiered Extragalactic Exploration Survey (MIGHTEE) using the L-band ($~1.2-1.3$ GHz) that covers $~20 \, {\rm deg}^2$ across four fields focusing on deep imaging \citep{Heywood2021, Hale2024} and the MeerKAT Absorption Line Survey (MALS; \citealt{Gupta2016, Deka2024}) that covers 1000 deg$^2$ at 1.4 GHz over 500 non-contiguous shallow pointings targeting specific point sources.  The continuum MeerKLASS survey will be in an unique sweet spot in the current ensemble of wide surveys, extending to lower frequencies with high resolution and sensitivity while covering most of the Southern sky outside the galactic plane. Its data products will provide a wealth of legacy science, enhanced by synergies with other surveys both in the radio and at other wavelengths. This setup developed with MeerKAT will also be crucial for any future commensal wide survey with the upcoming SKA-Mid.
 
The paper is divided into several sections. In \cref{sec:meerklass} we introduce the MeerKLASS On-the-Fly survey. \cref{sec:smearing} describes the mathematical formulation for issues and corrections for our observations. \cref{sec:otf_data} and ~\cref{sec:img} describes the data processing and imaging pipeline that we have developed for the MeerKLASS. Results with image products are discussed in \cref{sec:result}. Finally we discuss the upcoming MeerKLASS data release plans in \cref{sec:DR} and summarise the paper in \cref{sec:summary}.

\section{The MeerKLASS OTF continuum survey}\label{sec:meerklass}

\subsection{MeerKLASS specs}
\begin{figure}
    \centering
    \includegraphics[width=0.85\linewidth]{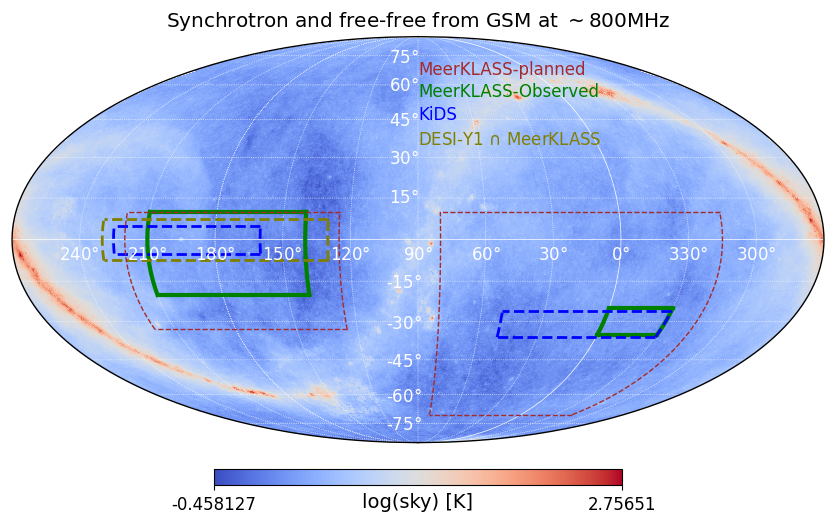}
    \caption{The approximated expected footprint for the full MeerKLASS 2023-2028 campaign (dashed dark red lines). The green lines mark the region already observed.
    The Galaxy at 800 MHz is shown for reference, together with available galaxy survey footprint from DESI (orange) and KiDS (blue). The location of the L-band 2021 observations is also shown (green lines).}
    \label{fig:sur_plan}
\end{figure}

The ongoing MeerKLASS-UHF survey, one of MeerKAT's approved XLP projects, aims to cover about 10,000 deg$^2$ of the Southern sky using up to 2,500 hours. The target sky area has significant overlap with several optical/NIR,  wide, galaxy surveys and is expected to provide invaluable legacy data sets (see \cref{fig:sur_plan}). The survey is performed both in the auto-correlation (single dish mode) and the cross-correlation (interferometric mode). With the single dish observations, the survey aims to perform \HI~IM up to redshift of $z = 1.44$. Although the majority of the MeerKLASS observation is planned to be done in MeerKAT UHF-band, we have also performed several pilot observations in MeerKAT L-band such as $200\,{\rm deg}^2$ around WiggleZ 11-h field \citep{Wang2021, Cunnington2022} (15 hours) and $236\,{\rm deg}^2$ around KiDS-South field \citep{Cunnington2025} (85 hours). In the UHF band, observations are ongoing, and we have already observed approximately $2,500\,{\rm deg}^2$ over 650 hours, covering a sky area that has strong overlap with DESI-Y1 observations \citep{DESI_Y1-2025} (mostly around the equator but away from the galactic centre). 
The future plan is to continue observing for next few observing cycles (see \cref{fig:sur_plan}) as part of the approved MeerKAT ``eXtra Large Projects'' program (XLPs). 

\begin{figure}
    \centering
    \includegraphics[width=1.0\linewidth]{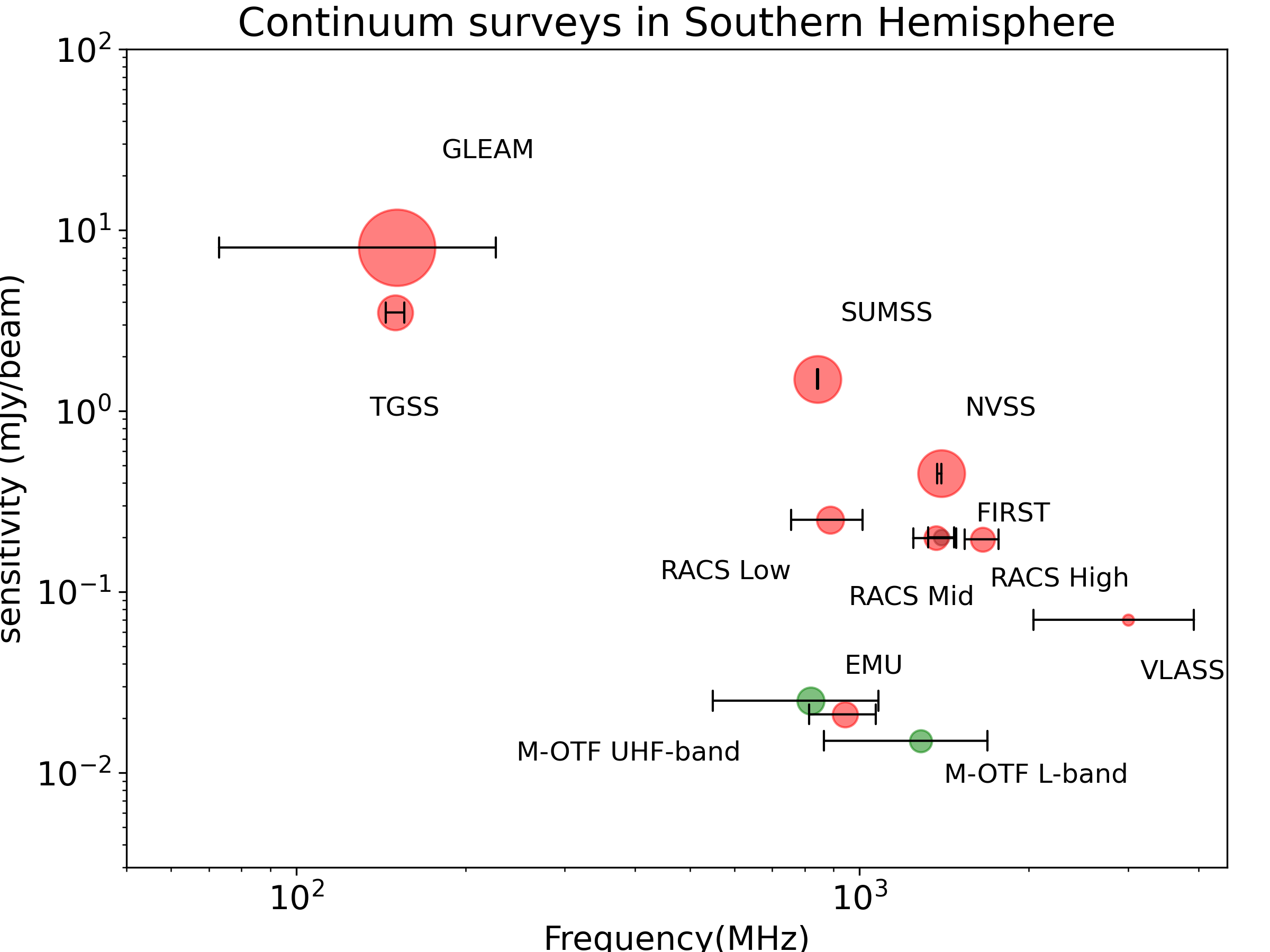}
    \caption{A summary of recent large area low-frequency surveys in the Southern hemisphere including their sensitivity, frequency, and resolution (see \cref{tab:survey_SH}). The size of the markers is proportional to the square root of the survey resolution. The horizontal lines show the frequency coverage for surveys. The green circles show the expected sensitivity and resolution of MeerKLASS OTF surveys without smearing incorporated.}
    \label{fig:comp_sensitivity}
\end{figure}

MeerKLASS observations are optimised for the single-dish mode, however, visibilities are recorded for all scans and using the OTF mapping technique, continuum images can be produced from all data. For the rest of this paper we will refer to the MeerKLASS OTF data products as ``M-OTF". Each M-OTF observation is made with the full instantaneous bandwidth of the telescope divided into 4096 contiguous frequency channels and with a sampling  of $\dt =$2-second. All four polarization of the visibilities (XX, XY, YX, YY) are recorded to allow images to be made in stokes parameters I, Q, U and V. The total M-OTF survey is expected to conservatively achieve a flux sensitivity of $25\mu{\rm Jy}$ in stokes I and an angular resolution of $14 \arcsec$ (robust weighted 0). \cref{tab:survey_SH} shows the expected sensitivity of the survey in the Southern sky and \cref{fig:comp_sensitivity} describes the same. The target resolution and sensitivity are estimated considering optimal declination and correct instrumental setup. We shall come back to this in a later section (\cref{sec:img}) to discuss current issues, mitigation and the impact on the survey goals. 
\subsection{Scanning strategy: Constant Elevation Scans} \label{sec:HRD}

\begin{figure}
    \centering
    \includegraphics[width=.9\linewidth]{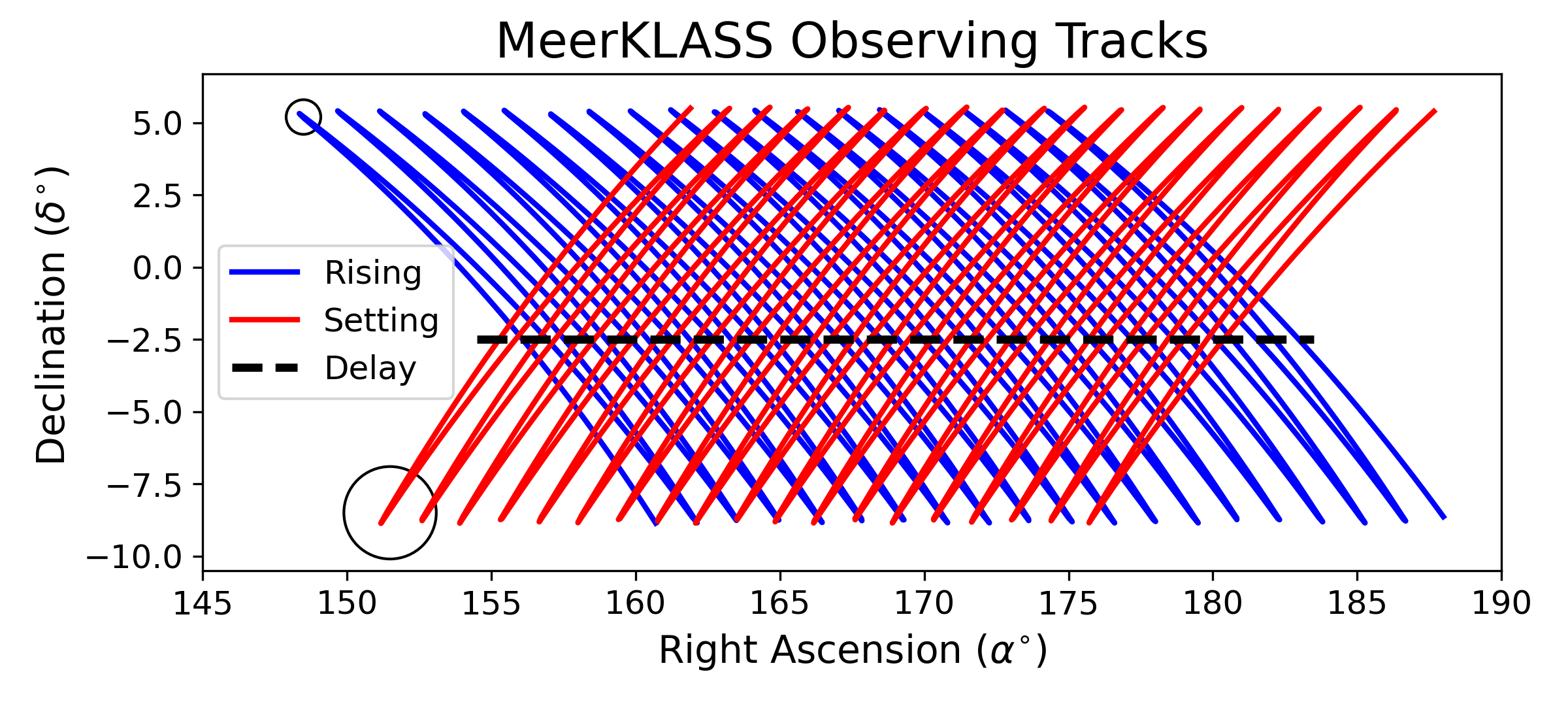}
    \caption{An example of MeerKLASS constant elevation observing strategy where ``rising" and ``setting" epochs are shown in blue and red respectively. L and UHF-band primary beam FWHM at the nominal frequencies are shown using circles. Each epoch lasts approximately 2 hours. The dashed horizontal line show the track for the correlator delay centre.}
    \label{fig:scan}
\end{figure}

A constant elevation fast scanning strategy is adopted to optimise the coverage of the relevant large angular scales and the stability time-scale of the instrument. MeerKAT dishes are moved back and forth to scan in the azimuth direction at a constant elevation (el) with a slew of $\sim 10\ {\rm deg}$ in sky (\cref{fig:scan}). This also minimizes fluctuations in the ground spill and air-mass \citep{Wang2021}. 
The telescope speed is set to  $\Theta = 7/\cos({\rm el}) \,{\rm arcmin \, s}^{-1}$ so that the projected scan speed on the sky is fixed at $7 \,{\rm arcmin \, s}^{-1}$. The data dump rate for the observation is set to 2-second, giving a $10 \,{\rm arcmin}$ scan 
per time sample which ensures that the telescope pointing does not shift significantly compared to the full-width-half-maximum (FWHM) of the primary beam (at L-band $\sim 1\ {\rm deg}$, at ${\rm UHF {\text -} band} \sim 2\ {\rm deg}$) during a single time dump. Due to smaller FWHM of the primary beam in L-band the scanning speed is set to $5 \,{\rm arcmin \, s}^{-1}$. Noise diodes attached to the receivers are fired for 0.585s every 19.5s during the observation for calibrating the single dish observations. Each block is observed for about 2h (an epoch). Sky rotation during this constant elevation scan provides the length in RA of the block. The same area is observed with a rising sky and setting sky as these scans cross each other and provide us with good sky coverage in the region of overlap (\cref{fig:scan}). This strategy is adopted to accommodate the challenging requirements for the total power \HI~IM observations. Considering the interferometric observation, the delay centre is currently set to a fixed azimuth-elevation (az-el) at the beginning of the constant elevation scan and kept fixed throughout the observation. This will create smearing on currently observed data as discussed later.

\begin{figure}
    \centering
    \includegraphics[width=0.95\linewidth]{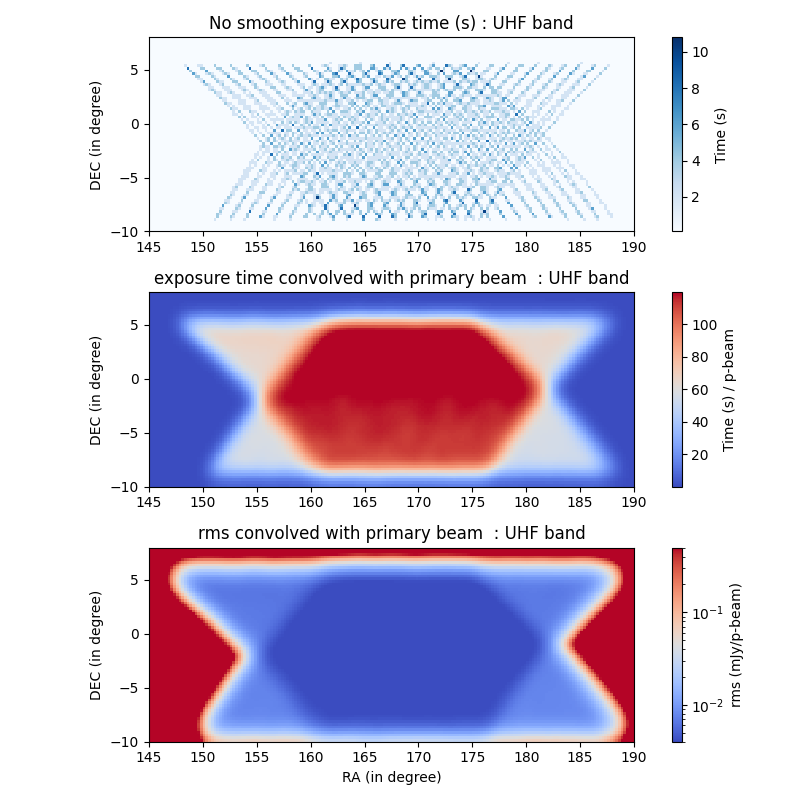}
    \caption{Top panel shows the density of the 2-second pointing when a rising and a setting scans are combined for an observation box. The middle panels shows the effective exposure time convolved with the primary beam assuming every 2-second snapshot contributes to the imaging. The bottom panel show the expected sensitivity in UHF-band, a optimistic scenario, where all the 2-second snapshots are used to construct image from the visibilities.}
    \label{fig:sen2}
\end{figure}

 \cref{fig:sen2} shows a prediction of our pipeline when combining (mosaicking) the 2-second pointings for the combination of one rising and one setting scan from one observation box. The top panel shows the expected observation duration as a function of RA-DEC. The middle panel shows an optimistic scenario where we combine each 2-second snapshot measurement from a box and compute the primary beam weighted exposure time at different RA-DEC. To estimate this we assumed a Gaussian primary beam with ${\rm FWHM} = 2^{\circ}$, which is independent of frequency. We perform a convolution of the primary beam and the scanning pattern shown in the top panel to compute this. The beam weighted exposure time gives us an optimistic estimate of the expected integration time at different RA-DEC while combining all the different 2-second measurements from an observation box. The bottom panel shows the expected average r.m.s. achievable in continuum for this scenario. We have only considered the MeerKAT system temperature \footnote{\href{https://skaafrica.atlassian.net/wiki/spaces/ESDKB/pages/277315585/MeerKAT+specifications}{MeerKAT UHF-band specifications.}} as a source of noise and combined the measurement from  all the frequency channels to estimate the rms. The middle and the bottom panel show that the combination of a rising and a setting scan form an hexagonal region in the middle where the average time spent is larger when compared with the edges of the observation box. Further we assume that the rms ($\sigma_N$) change with integration time by an inversely proportional relation ($\sigma_N \propto 1/\sqrt{\dt}$) and, as a consequence, we expect the M-OTF continuum images to be most sensitive in this hexagonal region. However as we continue with the survey the edges from different observation boxes are expected to overlap and provide us with relatively uniform rms throughout the survey area. We note however that the current observing setup introduces an unique set of challenges for continuum imaging that are different from traditional tracking observation. We discuss this in details at a later section.

\begin{table*}
\caption{A summary of recent large area low-frequency surveys (see also Figure \ref{fig:comp_sensitivity}). We have attempted to provide a fair comparison of sensitivities and resolutions but we note that both the sensitivity and resolution achieved varies within a given survey.}
 \centering
 \label{tab:survey_SH}
\begin{tabular}{lcccccc}
\hline 
Survey & Resolution & Noise & Frequency  & Declination & Area \\
                            & ($\arcsec$) & ($\mu$Jy/beam) & (MHz) &  & (${\rm degree}^{2}$)\\ \hline
TGSS ADR (\citealt{Intema2016}) & 25  & 3500            & 140--156 & $\delta>-53^\circ$ & 36,900\\
GLEAM (\citealt{Wayth2015}) & 150  & 5000 & 72--231 & $\delta<+25^\circ$ & 24,831\\
RACS-Low (\citealt{Hale2021}) & 18 & 260           & 743.5--1031.5 & $\delta<+30^\circ$ & 30,480\\
RACS-Mid (\citealt{Duchesne2023}) & 11.2 & 140           & 1151.5--1439.5 & $\delta<+30^\circ$ & 36,200\\
RACS-high (\citealt{duchesne2025})  & 11.8 & 195           & 1511.5--1799.5 & $\delta<+48^\circ$ & 37,000 \\
EMU (\citealt{Norris2011, Hopkins_2025}) & 15 & 20-30           & 800--1088 & $\delta<+30^\circ$ & 20,626\\
SUMSS (\citealt{Mauch2003}) & 45 & 1500           & 843 & $\delta<-30^\circ$ & 8,000\\
NVSS (\citealt{Condon1998}) & 45 & 450           & 1400 & $\delta>-40^\circ$ &33,000\\
VLASS (\citealt{Lacy2020}) & 2.5 & 70           & 2000 -- 4000 & $\delta>-40^\circ$ &33,885\\
MeerKLASS (UHF) & $14[23.3]^a$  & $25[35]^c$ & 580--1015 & $\delta<+10^\circ$ &10,000\\
MeerKLASS (L) & $11[14]^b$  & $20[33]^c$ & 856--1712 & $\delta<+10^\circ$ & 500\\ \hline

 \end{tabular}

\begin{tablenotes}[flushleft]
	\item \footnotesize{a: The target resolution is $14''$ with the delay tracking fix implemented, whereas `[]' show the average beam after the smearing correction in the current observations, where ${\rm beam} = \sqrt{b_{\rm maj}b_{\rm min}} = \sqrt{32'' \times 17''}$.}
	\item \footnotesize{b: Same as  mentioned in `a', where ${\rm beam}  = \sqrt{25'' \times 8''}$.}
    \item \footnotesize{c: r.m.s. achieved in data release 1.}
\end{tablenotes}
\end{table*}

\subsection{The MeerKLASS OTF survey: scientific goals}
The MeerKLASS's\footnote{\href{https://meerklass.org/}{https://meerklass.org/}}  target sky area overlaps with several optical/NIR wide galaxy surveys, providing an invaluable legacy dataset. The M-OTF survey primary data products include full polarisation continuum images (stokes I, Q, U, V), 9 sub-band images, spectral index maps and snapshot images. The survey also aim to provide a standard source and Gaussian component catalogueues as the derived data products. Here we outline the scientific plans for the M-OTF continuum survey.

\textbf{Galaxy evolution and Cosmology:}
The continuum survey aims to detect about 8 million galaxies (mostly star-forming) across a wide area spanning a unique frequency range down to 580 MHz. This number can be smaller depending on how much data is observed without the correlator delay fix which currently leads to smearing as described below. The survey can be used to explore the formation and evolution of galaxies, clusters of galaxies, and large-scale structure. Further we aim to study the star forming galaxy population to $z\sim 1$ and AGNs to $z\sim 6$ using stacking techniques. Due to its wide area and sensitivity at low frequencies, this survey can also be used to search for rare (massive) objects and even high-redshift AGNs from the epoch of reionization.

By combining the continuum sample from MeerKLASS with EMU \citep{Norris2021} and GLEAM-X \citep{Hurley-Walker2022}, we will have broad frequency coverage from 100–1000 MHz, important for spectral analysis. Together with morphological information, this will allow a better splitting of the galaxy sample into tracers with different biases. With this sample, and after cross-matching part of it with available photometric data, we can perform accurate clustering statistical analysis in 2D in (wide) redshift bins that can further constrain dark energy evolution and non-Gaussianity\citep{SKA2020}. The clustering on large angular scales will also be used to constrain the cosmological dipole \citep{SKA2020}. Because of the large-area coverage of the survey we plan to perform a cross-correlation between M-OTF radio galaxies with Cosmic Microwave Background  lensing maps to understand the redshift evolution of the galaxy bias \citep{Alonso2021}.

\textbf{Clusters:}
The M-OTF data will also have high legacy value for studies of galaxy clusters both directly and through stacking, in particular due to the MeerKAT configuration having a large number of short baselines making it sensitivity to diffuse emission.  The survey footprint overlaps with large Sunyaev-Zel'dovich (SZ) effect galaxy cluster surveys conducted with the South Pole Telescope, Atacama Cosmology Telescope (ACT), and future Simons Observatory, which are able to detect massive galaxy clusters at any redshift. For example, the ACT DR5 catalogue \citep{Hilton2021, ACT2025} contains $>$ 2800 clusters out to $z < 2$ within the MeerKLASS footprint. In addition, optical surveys such as DES\footnote{\href{https://www.darkenergysurvey.org/}{https://www.darkenergysurvey.org/}} and the future LSST/VRO\footnote{\href{https://www.lsst.org/}{https://www.lsst.org/}}  and X-ray surveys by eROSITA\footnote{\href{https://erosita.mpe.mpg.de/}{https://erosita.mpe.mpg.de/}}, provide samples of tens of thousands of clusters that probe lower mass systems over our proposed MeerKAT survey region. Example applications of the M-OTF data to cluster studies include: (i) using the radio continuum luminosity as a dust-unbiased tracer of star formation in and around galaxy clusters; (ii) investigating the impact of radio source contamination on SZ cluster surveys; (iii) probing diffuse radio emission in clusters, such as radio halos and relics, which are seen in colliding clusters.

\textbf{Transients with MeerKLASS:}
The fast scanning strategy of MeerKLASS and its 2 seconds time resolution also provides exciting opportunities for transient searches on the image plane. A given point in the sky will be observed with different cadences, from seconds to minutes, to several hours, weeks, and months, as each 300 deg$^2$ sky patch (or box) is observed up to $\sim 26$ times. Each point in the survey field is observed for $\sim 10$ min in total (300, 2-second snapshots). Each patch is usually observed within a 5 month period but it could be split over more years depending on time allocation constraints.

Astrophysical transients are the signposts of some of the most extreme physics in the Universe since the big bang. Serendipitous, commensal, discovery of radio transients in MeerKAT fields has been successfully achieved on both long timescales for synchrotron transients \citep[e.g.][]{Andersson2023} and short timescales for coherent transients \citep[e.g.][]{Caleb2022}. Furthermore, citizen science discovery of transients \citep[e.g.][]{Andersson2023} provides training sets for future Machine Learning (ML) searches for image plane variables. We will apply these ML techniques to the search for variable and transient sources in all our M-OTF imaging, aiming to do so with a low latency.

For the detection of short-duration transients, we aim to deploy the new ``\texttt{PARROT}'' fast imaging pipeline, based on \cite{Smirnov2024}. This is already being ramped up for mining archival MeerKAT data for transients and variables, and has already yielded detections. \texttt{PARROT} requires a set of calibrated visibilities and a deep continuum sky model, which is a natural by-product of the self-calibration and imaging with pipeline. These visibilities are then imaged at 2-second cadence, per every pointing, and transient searches are performed at various timescales of interest. The output of the pipeline is a catalogueue of detections, associated light-curves, and cube/image cutouts. This pipeline in the imaging domain is opening a new window in the search for short-duration transients and has the potential to find many previously undetected objects.

\section{The MeerKAT fixed-delay correlation observing mode}\label{sec:smearing}

\subsection{Theoretical formulation}
\begin{figure*}
    \centering
    \includegraphics[width=1.1\linewidth,trim={3.0cm 0.0cm 1.0cm 0},clip]{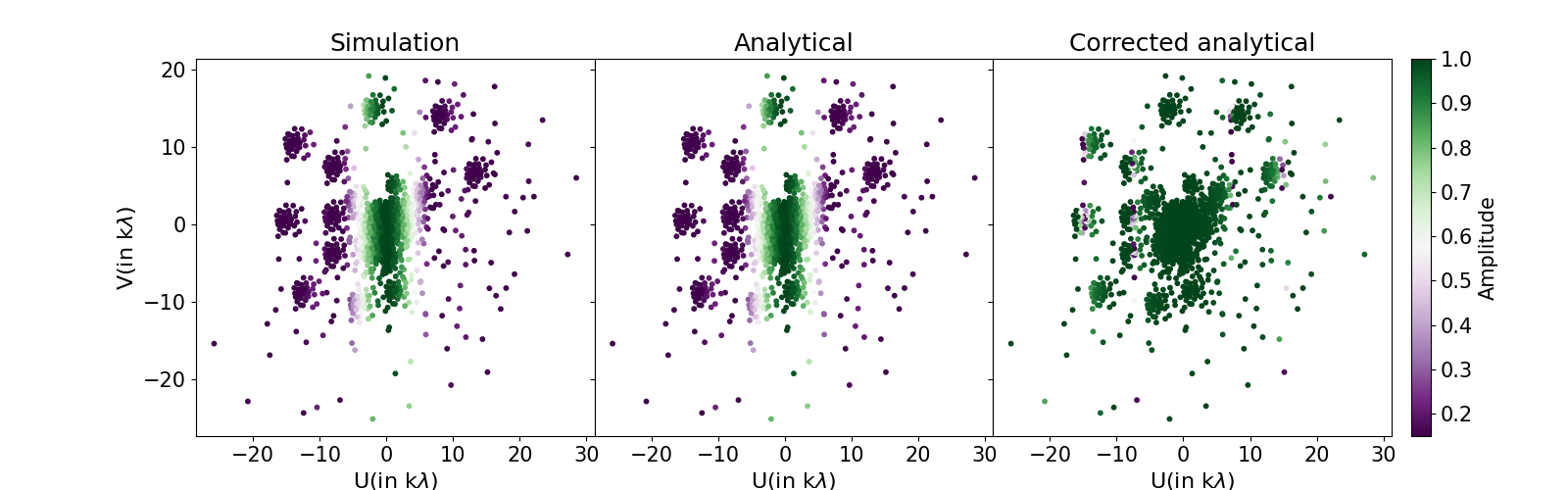}
    \caption{This plot shows the effect of smearing in a snapshot observation of the M-OTF. The results for simulated and analytical smearing computation are shown in left and the middle panels. The right panel shows the ratio of the analytical and the simulated prediction for the snapshot.}
    \label{fig:smbl1}
\end{figure*}

\begin{figure*}

  \begin{subfigure}{.49\textwidth}
    \includegraphics[width=0.95\linewidth,trim={1.05cm 0.85cm 3.5cm 1.0cm},clip]{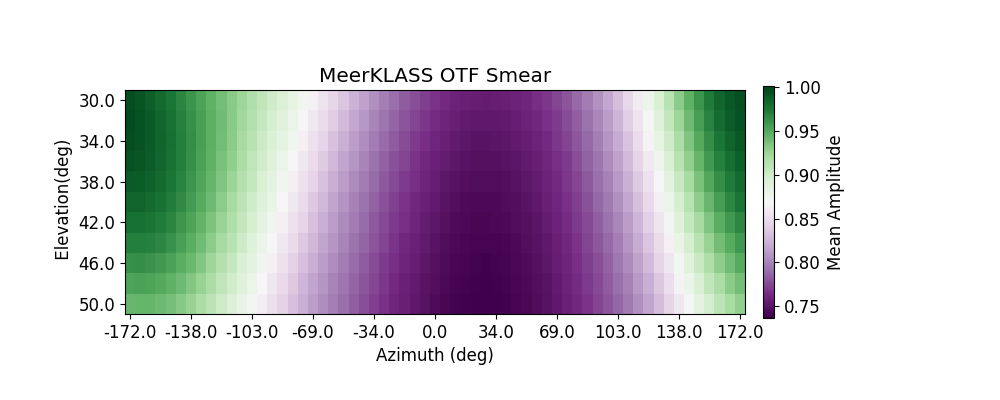}
  \end{subfigure}
  \begin{subfigure}{.49\textwidth}
    \includegraphics[width=0.95\linewidth,trim={1.05cm 0.85cm 3.5cm 1.0cm},clip]{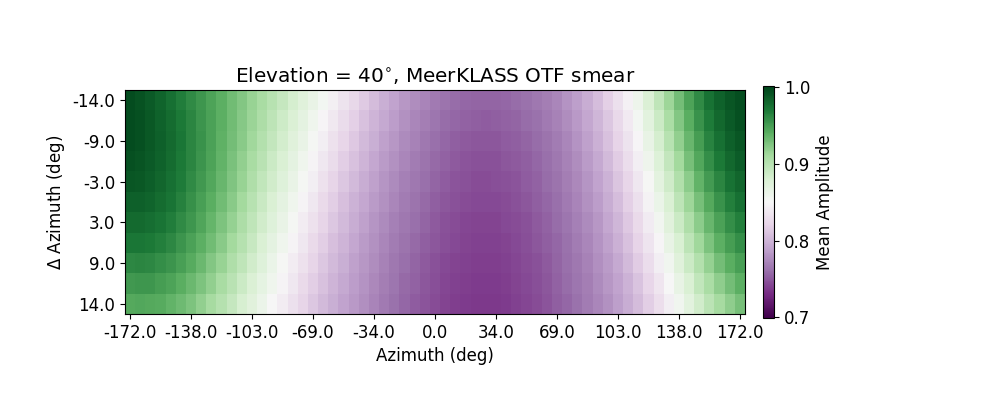}
     \end{subfigure}
    \caption{ The left panel shows the impact of the M-OTF smearing on a point source average amplitude as a function of az-el. The right panel shows the same amplitude as a function of steering in azimuth at a fixed elevation of $40^{\circ}$. }
    \label{fig:sm4}
\end{figure*}

The primary technical drawback of the current MeerKLASS observation setup is the inability to track the sky within the integration time interval ($\dt$). In the present survey mode the delay remains fixed at an az-el point during the time dump. During this $\dt$ interval, the sky rotates but the per-baseline correlations do not track this, leading to baseline dependent reduction of the visibility amplitudes, which is manifested as a smearing effect in the M-OTF images. As the antenna voltages have already been correlated and averaged, this effect is irreversible and cannot be fixed by correcting the delays. We note that a new fix is being implemented that will allow to track the sky with relative accuracy, dramatically reducing this smearing. However, data taken so far has been observed without this fix. We next discuss in detail the impact of the current mode on our data.

The mathematical details of the smearing effect are described in \cref{appen0}. The dominant effect can be seen by considering a single point source at the centre of the pointing. For a standard tracking observation with delay correction, the visibilities should just measure the point source flux ($S_{\rm p}$). For the M-OTF case, ignoring an overall phase, we get instead 
\begin{equation}
    \visbnutm = S_{\rm p}\, \sinc(\delta \Psi) \, \sinc(\delta \Phi),
\end{equation}
with,
\begin{align}
    &\delta \Psi  = \pi \dt \numbc \dbdtm \cdot \r_p  \label{eq:dpsi} \\
    &\delta \Phi = \pi \frac{\dnu}{c} \btm \cdot [\r_p - \rcm] \label{eq:dphi} , 
\end{align}
where $\btm$ is the baseline vector at time $t_m$, $\r_p$ the source position and $\rcm$ the time dependent delay position. The $\sinc$ functions will reduce the amplitude of the source flux. Naively we could try to correct this effect by just dividing the measured visibilities by these $\sinc$ functions. But in practice a proper deconvolution will be required as described later (see ~\cref{sec:img}). 

Obviously this effect can be small for very short time and frequency intervals, but even for a 2-second integration interval, the effect is evident from visual inspection of the snapshot images. The main effect comes from the time smearing. If we take the phase centre towards the point source, this can be further written in known quantities:
\begin{equation}
    \delta \Psi  = -\pi \dt u \oc \cos\delta_0 \label{eq:dpsi2},
\end{equation}
where $u$ is taken from the standard baseline $(u,v,w)$ coordinates, $\delta_0$ is the declination of the phase centre/point source and $\oc = 15 \arcsec/{\rm s}$  is the angular velocity of the Earth. In the map plane this phenomenon can be understood as follows. The telescope integrates signals over 2 seconds of time while the correlator delay centre is fixed at a certain azimuth and elevation and the sky is rotating round the South Pole. In the map plane this is equivalent to a convolution with a top-hat function of width $2 \times 15 \times\cos\delta_0$ arcseconds of angle in RA. This appears as a source smearing. Using the convolution theorem, the convolution in the uv-plane is a multiplication by a $\sinc$ function dependent on the u-coordinate as above.

\subsection{Comparison to simulations}
To investigate the impact of the smearing and assess the accuracy of the theoretical formulation we simulate M-OTF observations for a single source of 1~Jy at the phase-centre without adding system noise. The simulations were performed around a fixed arbitrary time $T_{0}$. To capture the smearing effect we generate visibilities with an artificially high time resolution of $\Tilde{\delta t} = \dt/100$ and frequency channel width of $\Tilde{\dnu} = \dnu/100$ for a bandwidth of 1 MHz. An unitary primary beam is assumed. We then vector sum these visibilities to the equivalent channel and time resolution to that of the actual observation. As explained, the key difference between a tracking and an M-OTF observation is that the delays are not corrected for Earth's rotation within the integration time and as a consequence the aforementioned source moves away from the delay centre (which ideally should be the same as the phase centre). The point source is at the phase-centre of the observation at the time $T_0$ and incorporating the Earth's rotation, visibilities simulated between $T_0 - \dt/2$ to $T_0 + \dt/2$ are summed to obtain the final visibility. Since the delay applied within the time resolution $\dt$ remains constant, we expect the summed visibility amplitude at large baselines to decrease from unity due to phase errors. This is in contrast to a tracking observation where we expect phases of all the component visibilities to be the same and so the amplitudes of the visibilities to be unity at all the observed baselines. 

The results for one 2-second snapshot observation at RA=$170^{\circ}$ and DEC=$-2^{\circ}$ are shown in \cref{fig:smbl1}.  The left panel shows the expected visibility amplitude as a function of baseline distribution estimated with the simulations, the middle panel show the theoretical prediction (see Eq.~\ref{eq:dpsi} and Eq.~\ref{eq:dphi}) for the same and the right panel show the ratio of the two predictions. Considering the left panel we can see that the smearing effect is pronounced along the $U$ direction whereas it is constant the $V$ direction. This is primarily due to the fact that the East-West baselines are affected significantly due to the fixed delay applied to the visibilities, whereas the North-South baselines are not affected significantly. In other words, the $U=0$ visibilities have zero fringe rate and so do not change during the integration period, as a consequence no time dependent delay compensation is required (see Eq. \ref{eq:dpsi2}).

The smearing pattern found in the middle panel is very similar to what we see in the left panel. To estimate the accuracy of the analytical modelling we show the  ratio of the predictions in the right panel. We find that the analytical prediction agrees with the simulations with-in $5\%$ at the majority of the baselines.  However there are vertical stripe-like features at long baselines where the deviations can go up to $\sim 20\%$. These stripe-like features coincide with the zero crossing of the $\sinc{\rm - functions}$ 
described in the \cref{eq:vis_def5b}. As the value of the $\sinc{\rm - functions}$ is very close to zero at these baselines, the ratio diverges for tiny differences in the location of these minima. It is worth noting that  MeerKAT is a telescope with core-heavy antenna configuration and the smearing on short baselines is smaller than that on long baselines. One can flag the long baselines to minimise the smearing, however this will degrade the resolution and increase the noise in the image. We discuss this further in \cref{sec:img}.

\subsection{Sky dependence}
The MeerKLASS survey is expected to cover a significant fraction of the Southern sky. From \cref{eq:dpsi2}, we can see that the M-OTF smearing depends on the declination of the observations so the effect will be variable over the survey area. As an example the smearing should be minimal if we observe close to the South Pole whereas the effect is expected to maximize at the Equator. To assess the impact of the smearing on the overall survey, we repeat a similar simulation described earlier for a range of azimuth and elevations. By design M-OTF observations are restricted between  the elevation of $30^{\circ} {\rm to }\, 60^{\circ}$, whereas the central azimuth can vary between the range of $- 180 ^{\circ}$ to  $180 ^{\circ}$. Throughout the simulation we use the same fixed arbitrary time for observation around $T_0$. Different combination of azimuth and elevations are expected to point towards a combination of unique RA and DEC on the celestial sphere. This mapping depends on the telescope location and time of observation (fixed to $T_0$). However, changing $T_0$ will not change the results we present here, since the smearing does not depend on the particular RA.

The left panel of the \cref{fig:sm4} shows the average impact of smearing as a function of observing azimuth and elevation using simulations. In this case, the constant delay is at each of the az, el coordinates (but not tracking the sky). The quantity plotted in the colour axis is the amplitude of the point source averaged over all the visibility measurements during 2 seconds, and we expect this is equivalent to a natural-like weighting scheme employed during imaging. The simulation show that as expected the impact of the M-OTF time smearing is maximum close to equator (towards azimuth zero and higher elevations) and reduces as we move away in declination (towards higher positive or negative azimuths). In the worst case scenario we see that on  average we are losing approximately 26 percent of the amplitude of the point source at the equator. Whereas the amplitude is close to 1 at the azimuth $\pm 180^{\circ}$ which points around declination of $-80^{\circ}$. As mentioned earlier, we expect the smearing effect to be negligible while observing towards the poles. Currently the declination range of our planned survey vary between $+10^{\circ}$ to $-60^{\circ}$ and we expect these observations to be affected by this smearing effect.

During M-OTF observations, the antennas are steered $\pm 7^{\circ}$ rapidly at a constant elevation around the central azimuth and the constant delay is set at that central azimuth instead of following the az-el coordinates. This introduces further smearing, although smaller, from the frequency smearing. The impact of this is shown in the right panel of the \cref{fig:sm4}. Here the simulation is performed for a fixed elevation of $40^{\circ}$, the x-axis shows the central azimuth where the constant delay is set, and the y-axis the variation around that.  We see a similar trend dominating the smearing connected to how the azimuth relates to elevation, but the smearing goes up to 30 percent from the extra frequency smearing due to the angular separation towards the central azimuth. 
 These simulations give us an idea about the degree of impact on the images due to the smearing. The qualitative impact can be understood as follows. Since the shortest spacing fluxes are largely unaffected by the smearing, the total fluxes of point sources on the map are expected to remain the same if a suitable weighting scheme is chosen, while the reduction in the long spacing fluxes results in the peak fluxes being reduced. This in-turn will impact the correct identification of the point like and extended sources. We introduce a fix to the peak versus integrated flux problem which we discuss in \cref{sec:img} together with its limitations.


\section{OTF data processing} \label{sec:otf_data}
\begin{figure*}
    \includegraphics[width=.9\linewidth]{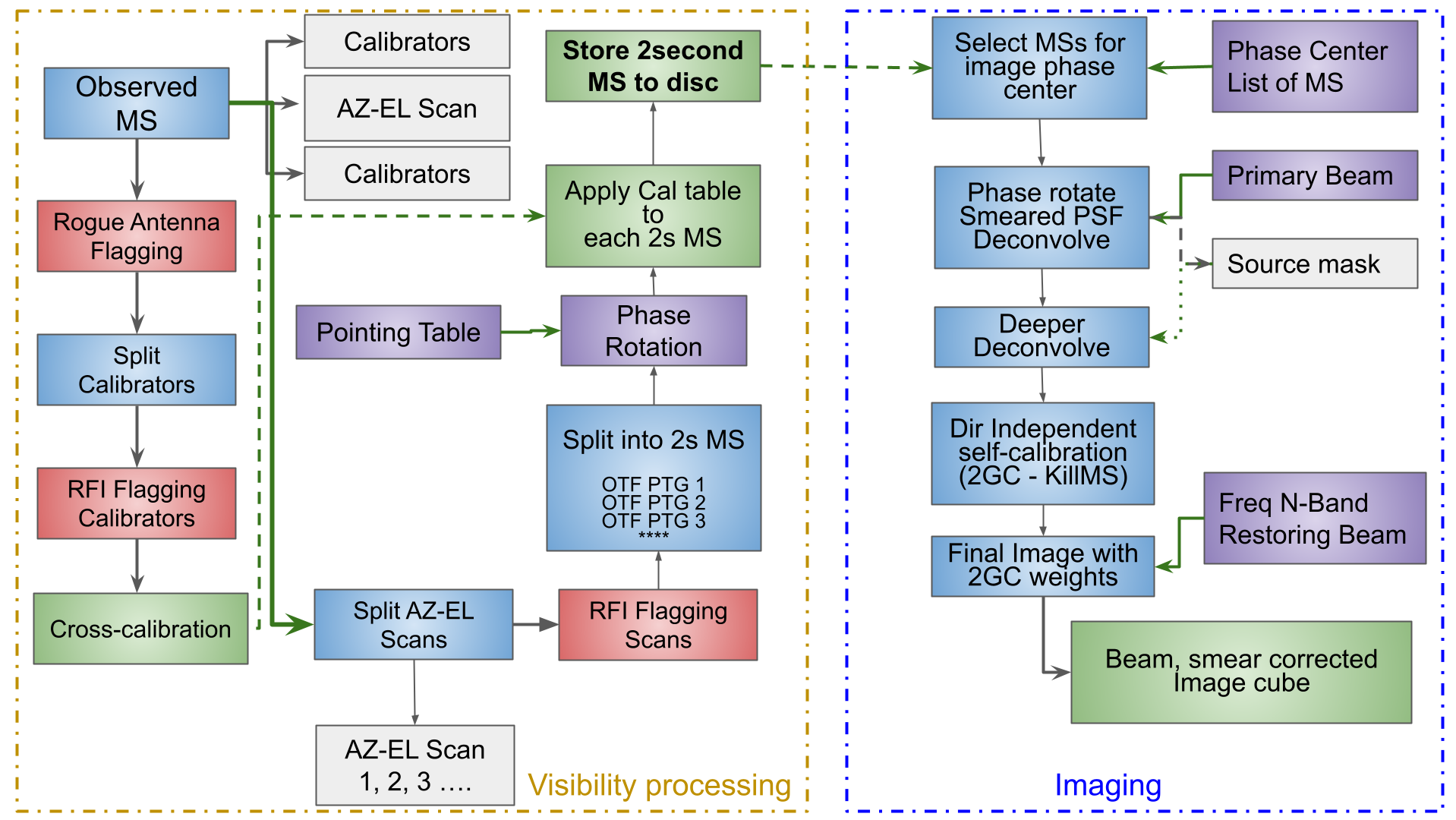}
    
    \caption{This shows the flowchart of MeerKLASS OTF end-to-end pipeline.}
    \label{fig:flow}
\end{figure*}
Processing of the M-OTF visibilities is different compared to the traditional tracking observation. In this section we describe the process of flagging and calibration in our pipeline for the observed visibilities. 
\subsection{Flagging }
\subsubsection{Rogue antenna flagging}
The MeerKLASS scanning strategy involves driving the antennas back and forth rapidly in azimuth while the elevation remains fixed. However we have found that during these fast scanning some of the antennas can fall behind or just stop moving for some time. It is important that we identify these misbehaving or `rogue' antennas and flag them from the visibilities. To identify these rogue antennas we compare the pointing of each antenna as a function of time with the median of the antenna pointing distribution. If an antenna deviates $\delta \theta \geq 0.1^{\circ} $ from the median pointing of the antennas at any time stamp, it is flagged. 
\begin{figure}
\centering

\begin{subfigure}{.49\textwidth}
\hspace{-0.35cm}
\includegraphics[width=0.99\linewidth, trim={0.cm 1cm 0 0.1cm},clip]{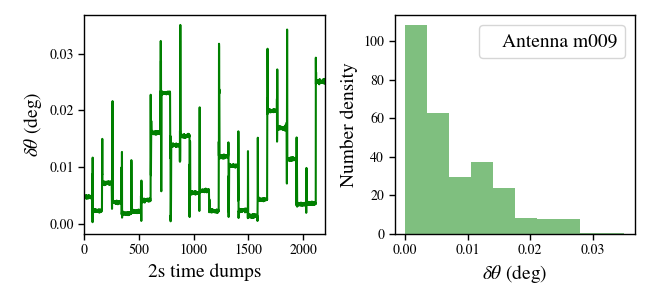}
\end{subfigure}

\begin{subfigure}{.49\textwidth}
\includegraphics[width=0.95\linewidth, trim={0 1cm 0 0.1cm},clip]{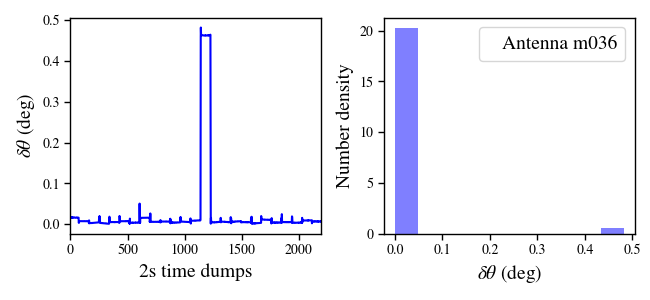}
\end{subfigure}

\begin{subfigure}{.49\textwidth}
\includegraphics[width=0.95\linewidth, trim={0 0.0cm 0 0.1cm},clip]{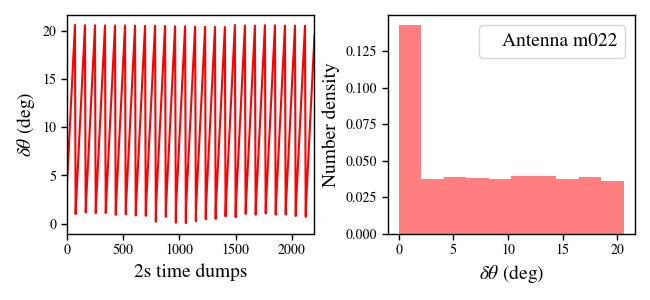}
\end{subfigure}
\caption{Here we show some examples of ``rogue" antennas. For reference, in the top row we show a well behaving antenna, where the left panel shows the deviation ($\delta \theta$) of the particular antenna from the median pointing of all the antennas as a function of the observation time and the right panel shows a histogram of the $\delta \theta$. Middle row shows the deviation $\delta \theta$ for an antenna that is unable to follow other dishes for small amount of time. The bottom row shows an antenna that remain ``rogue" through out the observation \ie could not keep-up with other antennas.}
\label{fig:rant}
\end{figure}
\cref{fig:rant} shows an example of such antennas that are identified as good and rogue. Left and right columns show the deviation ($\delta \theta$) of a particular antenna from the median antenna pointings as a function of time dumps and a histogram of $\delta \theta$ respectively. The top panels show an antenna that remains in sync throughout the observation and we identify it as a good antenna. In the middle panels we show an antenna that fails to follow the median of the antenna pointing for a part of the scan and we flag this particular antenna at these identified time stamps. The bottom panels show an antenna that has stopped driving completely during the observation and so this antenna is completely flagged from the particular observation epoch. Approximately $5\%$ of the antennas get flagged due to this issue and we lose about $10\%$ of the visibilities. This is the first step in our pipeline (see \cref{fig:flow}). 

\subsubsection{RFI flagging}
Radio Frequency Interference (RFI) is emitted from various sources, terrestrial or orbital, and create a nuisance for ground-based radio observations.  Since MeerKAT has a large effective collecting area of approximately 9160 square meters, it has high sensitivity and as the sensitivity of the instruments increases, so does the sensitivity to unwanted signals \citep{Harper2018, Engelbrecht2024}. Primary RFI contributors to the MeerKLASS observations are Digital TV (UHF), GSM (Mobile phones) (UHF + L-band), Aircraft transponders, GPS, GLONASS, Galileo and Inmarsat\footnote{\href{https://skaafrica.atlassian.net/wiki/x/AQAzEg}{RFI statistics for MeerKAT }.}.  MeerKAT has RFI mitigation systems to prevent RFI before and during the observation \citep{Jonas2016}. However, as even the best RFI mitigation methods cannot completely prevent all RFI \citep{Baan2010}, we must employ methods to reduce the effect of RFI after observation.

\begin{figure*}

  \begin{subfigure}{.5\textwidth}
  \centering
    \includegraphics[width=.95\linewidth,trim={0.0cm 0.0cm 0.0cm 0},clip]{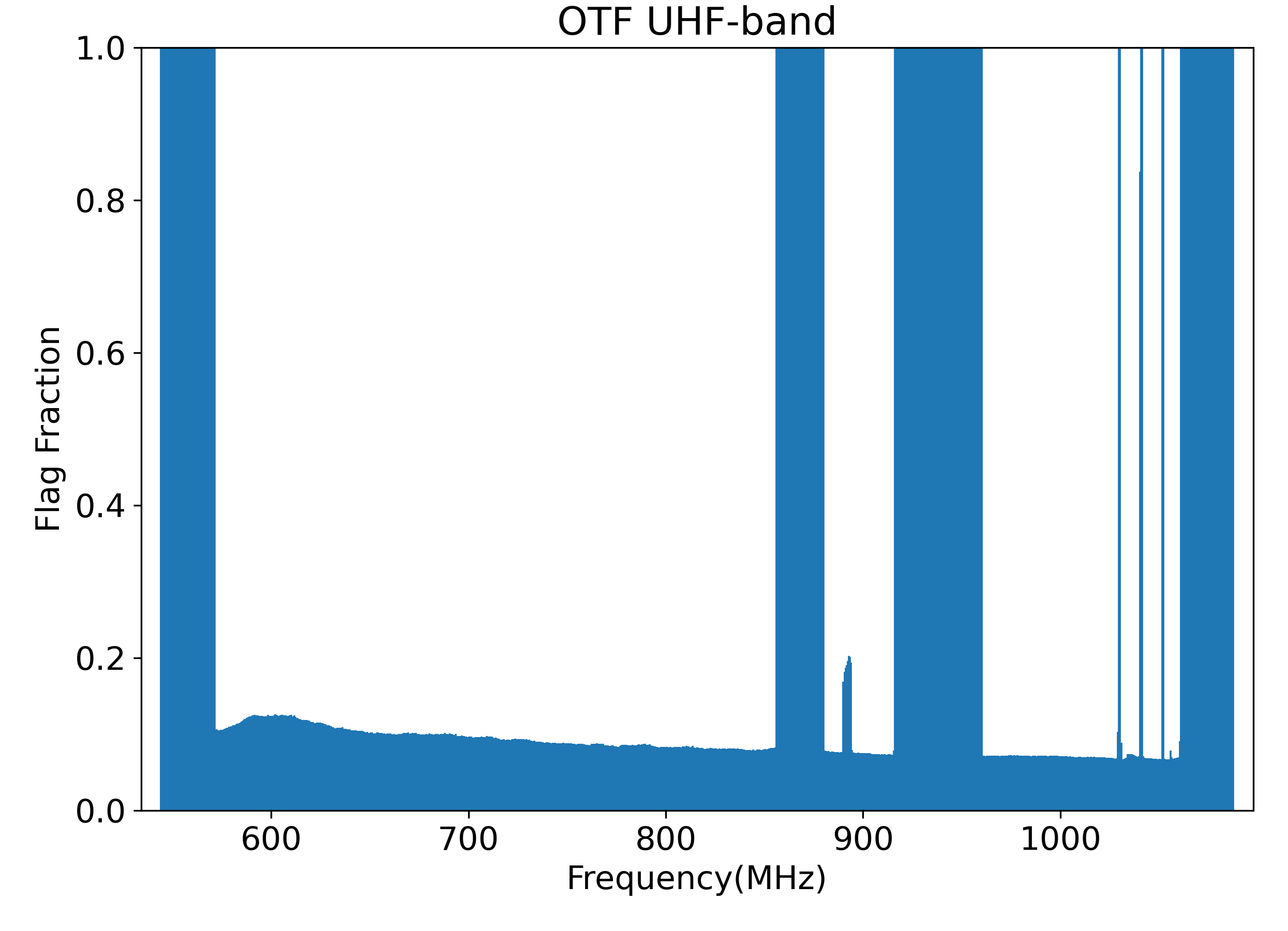}
  \end{subfigure}%
  \begin{subfigure}{.5\textwidth}
  \centering
    \includegraphics[width=.95\linewidth,trim={0.0cm 0.0cm 0.0cm 0},clip]{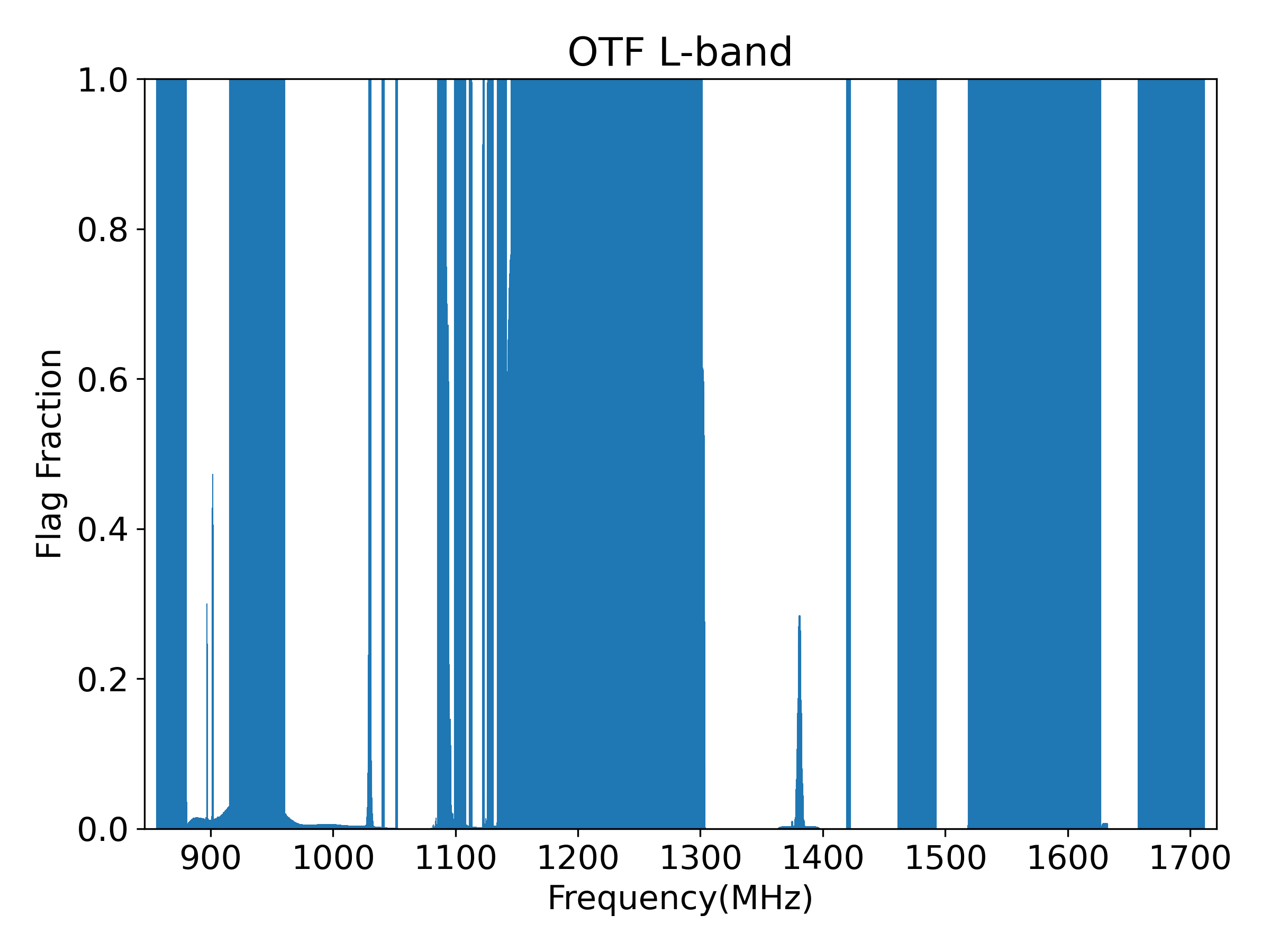}
  \end{subfigure}
  \caption{Here we show the flagging fraction of the OTF visibilities in UHF (Block ID: 1680626188, Observation date: 2023-04-04) and L-bands (Block ID: 1630519596; Observation date: 2021-09-01 ) for a typical observation block. Given that the observations have been performed }
  \label{fig:ff}
\end{figure*}

After flagging the rogue antennas we process the calibrator and the scan data through our RFI flagging step (\cref{fig:flow}). The RFI flagging is preformed using the \texttt{TRICOLOR} package \citep{Hugo2022} which implements the `sumthreshold' algorithm \citep{Offringa2010} optimised for MeerKAT. \cref{fig:ff} shows an example of the flag fraction in the MeerKLASS observation from data for an epoch. The left and right panel shows the flag fraction from UHF and L-band respectively. Considering the UHF-band, we see that frequency window between 580 - 880 MHz is relatively clean with the flag fraction mostly below 0.15. The frequency window between 880-1088 MHz appears to be severally contaminated by the GSM links and airport transponders and some of the frequency windows are completely flagged. This affects the observation in terms of achieving the expected sensitivity. The situation is much worrying in L-band (in right panel), where more than half of the frequency channels are flagged due to RFI contamination. Frequency range below 1150 MHz is primarily dominated by the RFI contamination from the GSM bands and aircraft transponders. However the higher frequencies are contaminated by the RFI originating from satellites (\eg GPS, Galileo etc.). We found the pipeline is able to remove all the major RFI contamination successfully, however, the RFI environment somewhat changes in time and as a result the flag fraction can change between one epoch to another for the same observation box.  

\subsection{Calibration}\label{subsec:cal}
The current M-OTF calibration strategy relies upon a primary, a secondary and a polarisation calibrator. It is not efficient to stop and steer the telescope for calibration observation once the constant elevation scanning starts. Due to this limitation, we observe the calibrators at the beginning and toward the end of an epoch for a given observation box. 
We observe a bright and unresolved source that can be used as a primary and band-pass calibrator, whereas another bright  compact and non-variable in phase source is observed to act as a secondary calibrator. Finally we also observe a known polarisation calibrator available in MeerKAT calibrator list \footnote{\href{https://skaafrica.atlassian.net/wiki/spaces/ESDKB/pages/1465548801/Polarisation+calibrators}{MeerKAT Polarisation calibrators.}}. All the calibrators are observed for approximately 2 mins duration. The used calibrator can vary depending upon the time of the observations, however all the calibrator are used from standard MeerKAT calibrator lists\footnote{\href{https://skaafrica.atlassian.net/wiki/spaces/ESDKB/pages/1479802903/UHF+gain+calibrators}{MeerKAT calibrators list.}}. Due to observational constraints some of the earlier M-OTF observation lack an unresolved secondary calibrator (\eg 2021 L-band data), in such a scenario, we use the primary calibrator as the secondary too and perform gain, phase and band-pass calibrations.

We use \texttt{CARACal}\footnote{\href{https://caracal.readthedocs.io/en/latest/index.html}{CARACal }.} \citep{caracal} to produce calibration tables from the primary, secondary and polarisation calibrators. We follow a ‘standard’ calibration strategy to solve the delays, band-passes and gains, and calibrate for the absolute flux using the primary calibrator source. However, considering that no secondary calibrator was observed in L-band, we introduce two additional iteration in solving for the gains (phase and amplitude \& phase) with 2-second time resolution. We solve for the gains using 3.3 MHz (16 channels) resolution. These calibration tables are then applied to the whole M-OTF observation. Most of the MeerKLASS observations are performed at night and significantly distant from the solar activity. Hence we expect the ionospheric effects to be subdominant and has been ignored throughout this paper.

\subsection{Phase centre correction}

As discussed earlier, the delays applied in the correlator are calculated for a single az-el point corresponding to the centre of the scan line. Therefore the delay centre remains fixed at an az-el for the whole fast scanning duration. It is therefore necessary to apply a time dependent phase rotation to the visibilities in order that the phase centre, which is the intrinsic centre of the interferometric map, matches the pointing centre of the antennas. The ``phase centre correction'' step in our pipeline splits the whole measurement set (MS) file of an observation epoch into multiple 2-second snapshot MS files with single time-stamped visibilities. Further it applies a phase-rotation to the visibilities from the observed delay centre to the pointing centre for every visibility integration time ($\delta t$). We use the \texttt{CHGcentre} task in \texttt{WSClean}\footnote{\href{https://wsclean.readthedocs.io/en/latest/index.html}{WSClean }.} to perform the phase rotation \citep{Offringa2014}. The pointing centre is determined from the pointing of a (reference) antenna using the auto-correlation data by matching the auto(antenna pointing) and cross-correlation (visibility) data based on the time stamp.

Finally, we use ``on-the-fly" calibration option in \texttt{CARACal} to apply the calibration tables obtained previously (\cref{subsec:cal}) on each 2-second phase rotated snapshot measurement sets.  At the end of this step the calibrated snapshot visibilities are stored on disc and further processed for imaging. 

\section{Imaging} \label{sec:img}
Calibration and imaging techniques for radio interferometer have developed significantly over the past few decades. There are primarily three stages of calibration that are commonly used in the current imaging techniques. First generation calibration (1GC)  where an amplitude and or phase calibrator is used to estimate the calibrations solution for the entire target field. The 1GC steps include absolute flux calibration, delay, band-pass and gain calibrations. The second generation calibration (2GC) also known as self-calibration is where antenna based complex gains are estimated by comparing the visibility data itself with a model sky derived from the previous iteration of imaging \citep{Pearson_1984}.  This is also known as direction independent (DI) calibration. To estimate and compensate for direction dependent effects, a third generation calibration (3GC) is usually employed.  At present we do not find a need to incorporate 3GC into our M-OTF imaging but we can revisit this in future should it become necessary.

\begin{table}
\begin{tabular}{lllll}
\hline 
\textbf{Parameters} & \textbf{L-band} & \textbf{UHF-band} &  &  \\
\hline 
Pixel size          & $1.5\arcsec \times 1.5\arcsec$     & $3.0\arcsec \times  3.0\arcsec$       &  &  \\
Image size          & $2.15^{\circ}  \times  2.15^{\circ}$       & $4^{\circ}  \times  4^{\circ}$          &  &  \\
Frequency range     & 856-1712 MHz  & 544 - 1088 MHz    &  &  \\
Central Frequency           & 1284 MHz        & 816 MHz           &  &  \\
N Facet             & 10              & 36                &  &  \\
N freq. band        & 7               & 9                 &  & 
\end{tabular}
\caption{This shows the imaging parameters.}
\label{tab:impar}
\end{table}
\subsection{Imaging and Smearing Correction}
\begin{figure}
    \includegraphics[width=0.99\linewidth,]{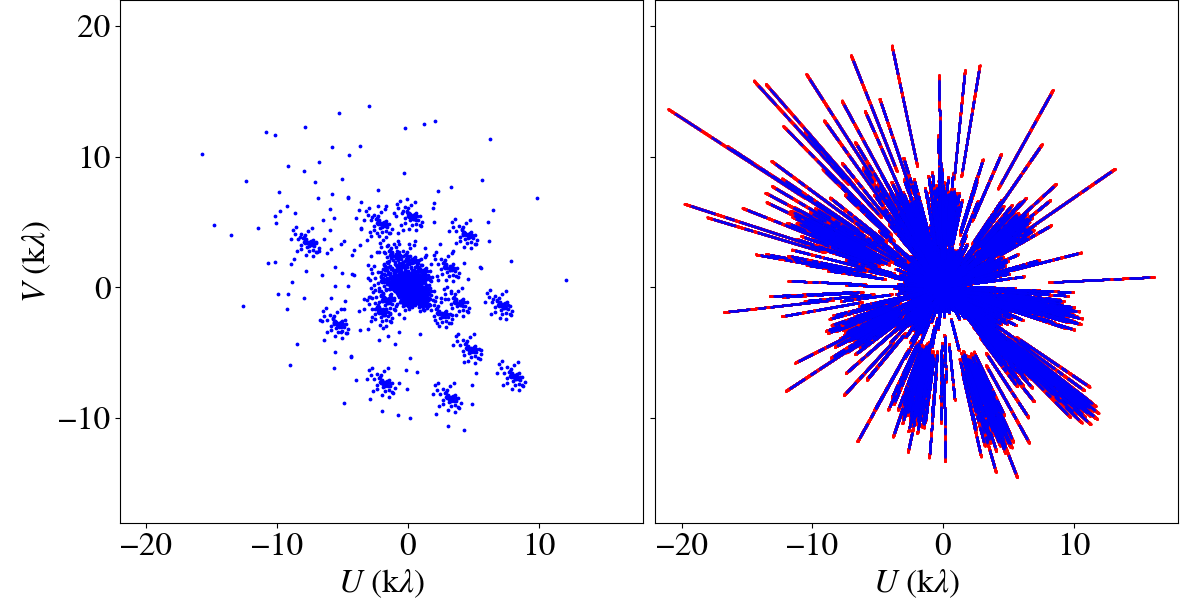}
    \caption{The left panel shows instantaneous $u$-$v$ distribution for a single frequency channel whereas the right panel show the same for the whole UHF-band observed at an elevation of $\sim 35.6^{\circ}$. Blue points show unflagged data, whereas red points mark flagged visibilities.}
    \label{fig:uvdist}
\end{figure}

In this section we introduce the imaging pipeline for M-OTF and discuss the mitigation of the smearing. Each M-OTF 2-second snapshot MSs stored are complete by themselves and standard imaging techniques are applicable on these visibilities to construct an interferometric continuum image. MeerKAT's 64 antennas are distributed over an area approximately 8 km in diameter which is the largest baseline available. 48 of these 64 antennas form a dense core which is approximately 1 km in diameter and the reminder of the antennas are distributed further away, extending the $u$-$v$ coverage. An example of $u$-$v$ coverage for a  snapshot is shown in \cref{fig:uvdist}. The excellent instantaneous $u$-$v$ coverage of the MeerKAT is advantageous for both calibration and imaging. The left panel shows the monochromatic $u$-$v$ coverage whereas in the right panel we show the multi-frequency $u$-$v$ coverage for the 2-second snapshot using the whole UHF-band.

We use \texttt{DDFacet}\footnote{\href{https://github.com/saopicc/DDFacet}{DDFacet}} \citep{Tasse2018} to perform smearing correction and imaging the M-OTF visibilities. The purpose of using an algorithm that relies on faceting is to approximate a wide field-of-view (FoV) with many narrow field images. Considering MeerKAT, although the FoV is not significantly large, we aim to construct images that are larger than the FoV (and far away from the delay centre). A facet based imaging can help us to mitigate different issues such as 
\begin{enumerate}
 
   \item  Considering wide-field imaging the source away from the phase centre are prone to phase error, and this can be accounted for by using non-coplanar facets,  

   \item  For wide-field M-OTF images, we expect the point spread function (PSF) to vary across the image. This can be accounted for, by constructing PSF at each facet centres. 

   \item  Finally \texttt{DDFacet} has been designed for dealing with wide-band spectral deconvolution which is necessary for M-OTF observations given the large bandwidth of MeerKAT.
\end{enumerate}
In this \texttt{DDFacet} framework the dirty images and the PSF are constructed from a set of direction, time and frequency dependent Jones matrices.  The detailed description of this dirty image construction can be found in Section 3 of \citet{Tasse2018}. Even in traditional tracking observations, there will be time and frequency smearing for point sources far away from the delay centre (see \cref{smear_track}). The standard \texttt{DDFacet} version can account for this by forward modelling the effect in each of the facets.

However, as already explained, in current M-OTF setup, the correlator does not track the rotating sky but a constant delay that corresponds to a fixed az-el value during the 2-second integration. This introduces a decorrelation in the visibilities that are much larger than the traditional smearing and these errors dominate our observations (See \cref{sec:smearing}). In other words the smearing causes the effective PSF to be direction dependent. 
A major advantage of \texttt{DDFacet} based imaging and deconvolution is that, we can take advantage of its existing framework to incorporate M-OTF smearing by computing the PSF at each facet centre. This required some modifications to the code. The smearing can be estimated by assuming a point source at the centre of each facet and multiplying the visibilities by the factor $\sinc(\delta \Psi) \sinc(\delta \Phi)$ defined in \cref{eq:vis_def5b}. This effectively gives us a smeared PSF per facet that is used to create an accurate sky model from the M-OTF observations. \cref{tab:impar} shows the parameters used to construct the PSF and later for imaging. The choice of numbers of facets used tries to balance the computational requirements and error introduced due to the facet size. The minimum facet number is chosen such that at least five facets are required to cover the FWHM of the MeerKAT primary beam at the nominal frequency. 

\begin{figure}
    \includegraphics[width=0.49\linewidth,trim={0.5cm 0.1cm 0.5cm 0},clip]{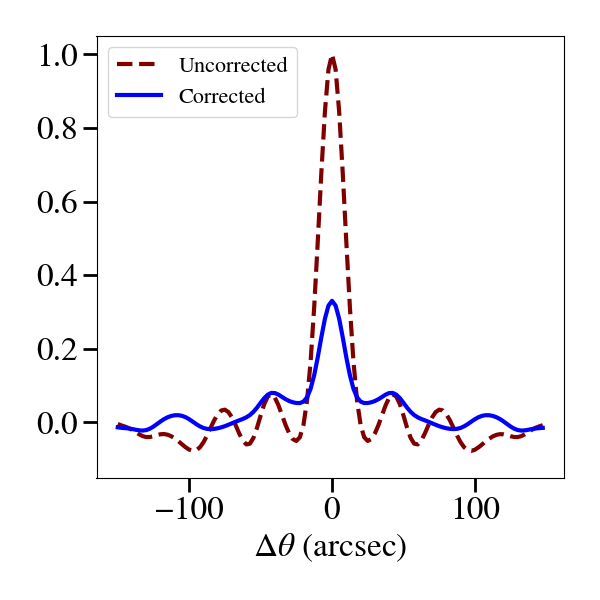}
    \includegraphics[width=0.49\linewidth,trim={0.5cm 0.1cm 0.5cm 0},clip]{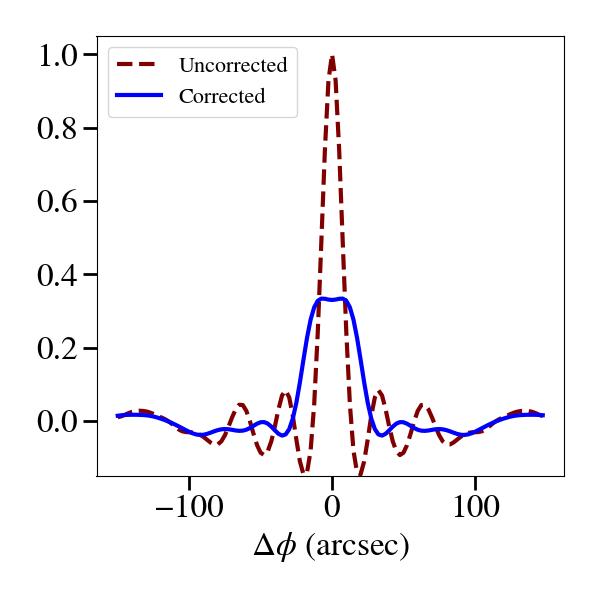}
    \caption{This figure shows a comparison of an effective PSF with a PSF where OTF-smearing effects are accounted. Left and right panels show the cross-section of PSF along RA and DEC directions respectively. Dashed lines show an effective PSF estimated using a 2-second snapshot visibility and Briggs weighting scheme with robustness zero. The solid lines show the PSF when the smearing effects are considered.}
    \label{fig:psf1}
\end{figure}
\cref{fig:psf1} shows a comparison of PSF from our imaging pipeline. The left and right panels show the PSF cross-section along DEC and RA respectively. The solid lines show the PSF with the smearing correction whereas the dashed lines show the PSF without smearing correction. Here we have used 2-second snapshot MS with rotated delay centre coincident with the phase centre of the image. This indicates that the major contribution to the smearing comes from the M-OTF specific issue of delay not tracking the Earth's rotation. Robustness during the PSF construction are set to zero. We find that overall the smeared PSF have a peak value of $\sim 0.5$, which shows the severity of the de-correlation. Further we find that the smearing makes the PSF elongated along the RA ($\Delta \phi$) direction, due to the effect of Earth's rotation. However, the width of the smeared PSF remain similar to the native PSF along DEC ($\Delta \theta$) direction. Considering a single 2-second snapshot PSF, we find that the FWHM of the native PSF is $20.5'' \times 15.2''$ with a position angle (PA) of $40.7^{\circ}$. The smearing corrected PSF has a FWHM of  $42.0'' \times 28.5''$ with ${\rm PA}=73.7^{\circ}$, which implies that the smearing rotates the major axis of the PSF towards the RA direction and the PSF asymmetry increases. This is an extreme example of smearing in M-OTF as this observation is performed close to equator (see \cref{sec:snap}). 

\subsection{Deconvolution}
To proceed with the deconvolution and model building, we use different sets of parameters for UHF-band and L-band observations. The choice of different parameters are given in \cref{tab:impar}. These choices primarily help us to optimize the processing times and the available resources. 

As a first step, we perform a shallow deconvolution using the \texttt{DDFacet} with a limit of two major cycles. In particular we use \texttt{DDFacet} implementation of Subspace Deconvolution algorithm (SSD; \citealt{Tasse2018}) with spectral fitting of n-order Taylor terms (\texttt{SSD2}). For this work we have used a second order polynomial to fit Taylor terms (\texttt{--SSD2-PolyFreqOrder = 2}). To construct the fit we have used ten image bands for degridding (\texttt{--Freq-NDegridBand=10}). One can improve the estimates of the spectral behavior of the model by increasing \texttt{--Freq-NDegridBand} parameter and perform better deconvolution however this dramatically increases computational requirements and is beyond our imaging requirements.

For the imaging we have used a Briggs' weighting scheme \citep{Briggs1995} with robustness value set to zero. This step helps to build model for the bright sources in the field. We use \texttt{MakeMask} tool to construct a mask with the high SNR  sources ($7 -10 \sigma$, depending on the field) from this initial image for a subsequent round of constrained deconvolution. In this second round we proceed with deeper cleaning, the sky model obtained is expected to be reasonably good. We use this second round of deconvolution as the basis for the single round of direction independent (DI) phase and delay self-calibration using \texttt{KillMS} \footnote{\href{https://github.com/saopicc/killMS}{killMS}} \citep{Tasse_2014, Smirnov_2015}. For the current stage of the pipeline we do not consider any direction dependent calibration. These will be addressed in future work.

The dynamic range of the images is limited by the presence of direction dependent effects. These are principally caused by time, frequency and direction-dependent variations of the antenna primary beam pattern, coupled with pointing errors. For the primary beam model to correct the images, we use holographic measurement of MeerKAT beam in UHF band \citep{deVilliers2023}. The main lobe of the MeerKAT primary beam is not circularly symmetric. Considering the facet based imaging and deconvolution framework presented here, we compute per facet PSF on a discontinuous manner in the image domain. Within these direction-dependent facets the primary beam and the decorrelation are assumed to be constant whereas in reality these quantities continuously vary. As an example closer to the half power point the primary beam varies rapidly. This is is partially taken care of by applying a smooth term in the image and one can correct the fluxes by applying these smooth beam correction in the image domain.

\subsection{Snapshot Image Quality} \label{sec:snap}
The immediate and most intuitive product for M-OTF are the 2-second total intensity (Stokes I) snapshot images. The 64 dishes of the MeerKAT in conjunction with cryogenically cooled receiver system provides excellent snapshot $u$-$v$ coverage and low system temperature ($T_{\rm sys}$) which are essential for imaging the 2-second snapshots. An example of  $u$-$v$ coverage for a 816-MHz snapshot is shown in \cref{fig:uvdist}. The left panel shows the monochromatic $u$-$v$ coverage whereas in the right panel we show the multi-frequency $u$-$v$ coverage for the 2-second snapshot in the whole UHF-band.

\begin{figure*}
\hspace{-1cm}
  \begin{subfigure}{.48\textwidth}
    \includegraphics[width=.99\linewidth,trim={0cm 0.1cm 0.0cm 0},clip]{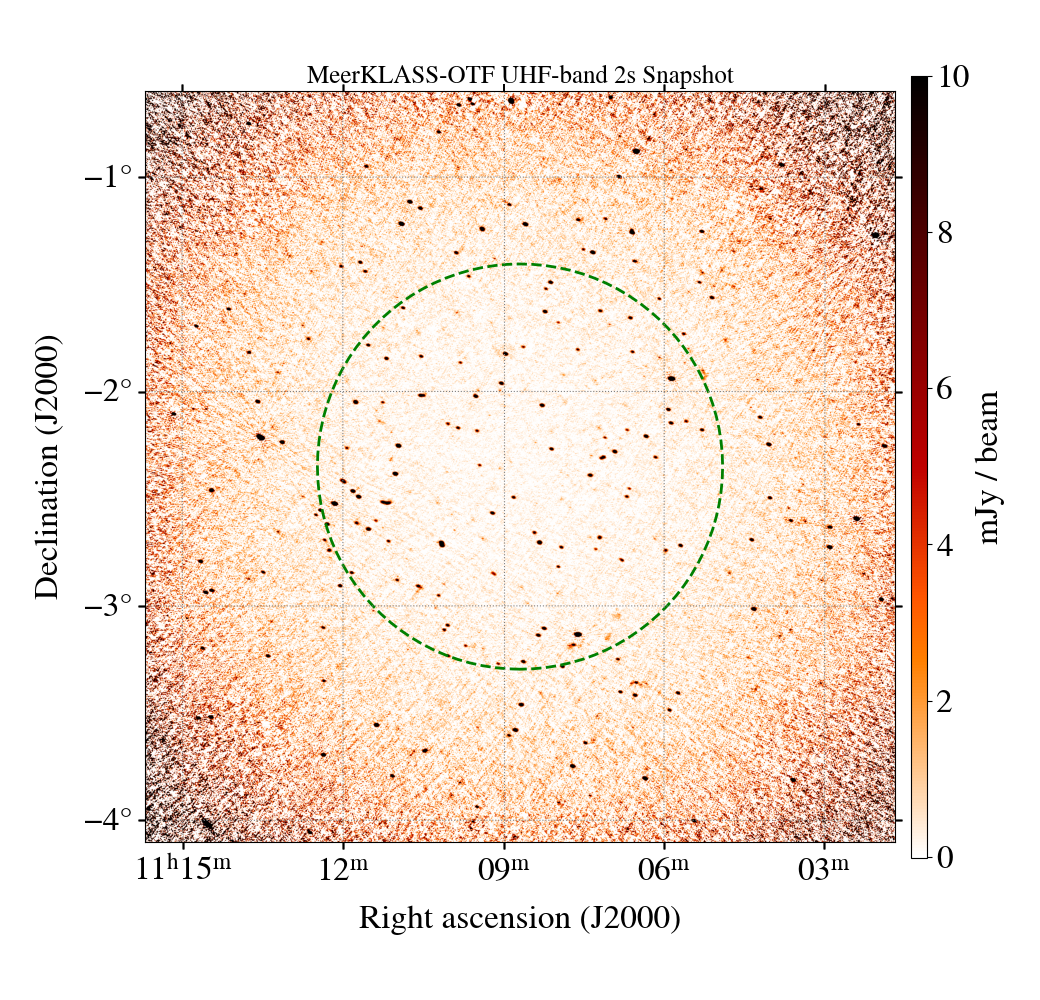}
    \caption{UHF-band MFS snapshot image at 816 MHz}
  \end{subfigure}%
  \begin{subfigure}{.49\textwidth}
    \includegraphics[width=.99\linewidth,trim={0cm 0.1cm 0 0},clip]{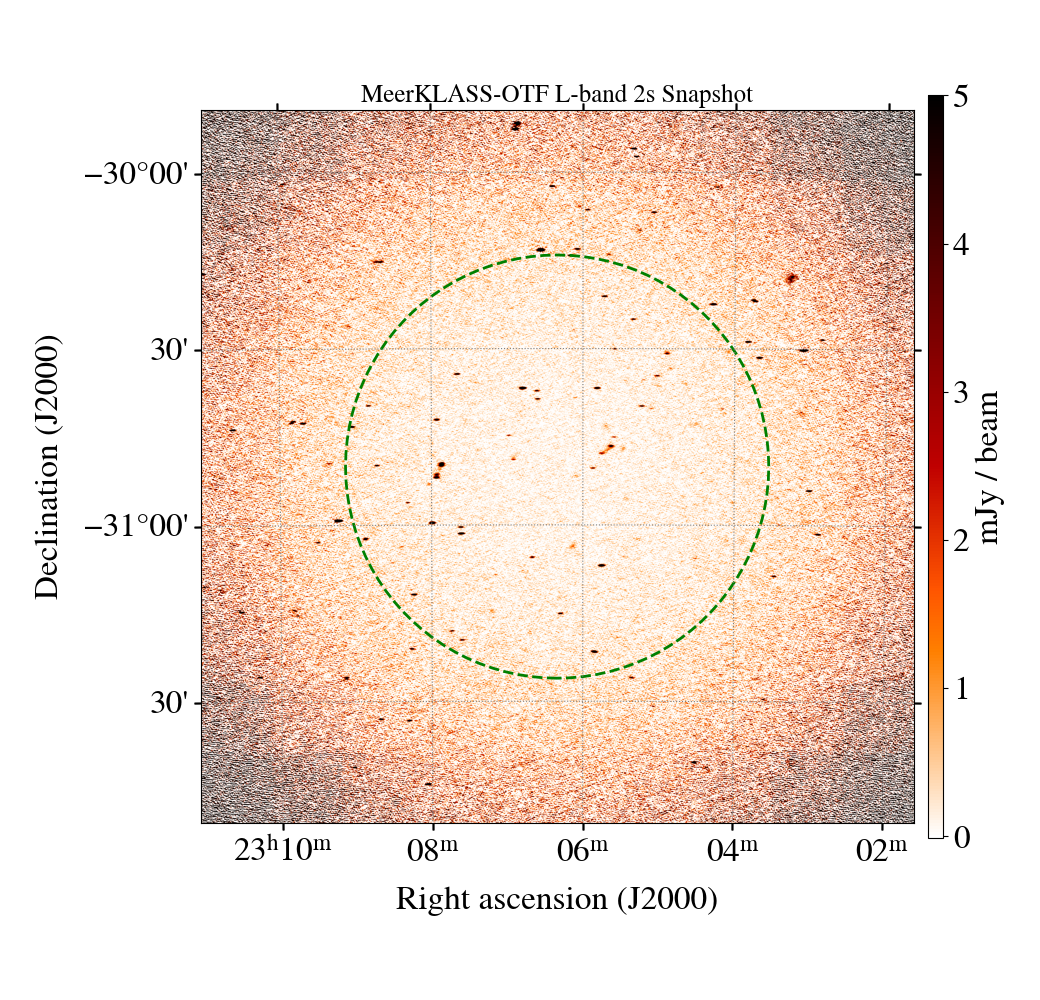}
    \caption{L-band MFS snapshot image at 1284 MHz}
  \end{subfigure}
    \caption{This shows the 2-second snapshot images from UHF-band (left panel) and L-band (right panel). The dashed circles show the FWHM of the primary beam at the nominal frequencies of UHF and L-band. In UHF-band we can achieve r.m.s.  of $387\,\mu{\rm Jy/beam}$ at a resolution of $42.0'' \times 28.0''$, whereas in L-band the r.m.s. is $300\,\mu{\rm Jy/beam}$ at a resolution of $33.7'' \times 10.8''$.}
    \label{fig:snap}
\end{figure*}
\cref{fig:snap} show examples of such 2-second continuum images from M-OTF observation using UHF (left panel) and L-band (right panel). The approximate FWHM of the MeerKAT primary beam are $\Theta_{{\rm PB}, \nu} = 1.8^{\circ}$ at $\nu = 816 {\rm MHz}$ for UHF-band and $\Theta_{{\rm PB}, \nu} = 1.1^{\circ}$ at $\nu = 1284  {\rm MHz}$ for L-band, shown in the figure by dashed circles. The synthesize beam size is $42.0'' \times 28.0''$ in the UHF and $33.8'' \times 10.8''$ in the L-band. The asymmetry in the beam size is primarily caused by two factors, elevation of the observation and  smearing correction. The intrinsic beam \ie without smearing correction are $20.5'' \times 15.2''$ in UHF and $ 18.5'' \times 10.8''$ in L-band. The image is expected to be finer along the major axis if the smearing issue is fixed at the correlator level.

It is worth noting that even with a mere 2-second snapshot images we could identify significant number of sources in continuum. There are no visible artifacts and the bright sources are detectable with high signal-to-noise ratio (SNR). This shows the excellent sensitivity of the MeerKAT as an instrument. The rms values achieved in a source free region are $340 \, \mu{\rm Jy}$ and $260 \, \mu{\rm Jy}$ in UHF and L-band respectively. We measure the thermal noise in these images by taking the rms of the pixel values in {\sc CLEAN}ed, source-free regions away from the main lobe of the primary beam in the images prior to primary beam correction.

\subsection{Visibility plane mosaicking}

The M-OTF 2-second snapshot images are good in quality and can be used for high time cadence science cases, such as search for slow transients. It is possible to produce deep images using the snapshot images corresponding to each 2-second time sample and then mosaic these images together taking into account their changing pointing centres. This in-principle can produce images with lower noise and higher dynamic range. However, as we keep on adding 2-second snapshots to construct deeper images, we are restricted to deconvolution performed at a 2-second timescale. This can restrict the image fidelity, which arises from having a better model for the self-calibration steps due to the better sensitivity of the maps arising from the increased time on sky. To achieve better depth sensitivity and image fidelity we have chosen to perform visibility plane mosaicking (joint deconvolution) for the M-OTF images. We use \texttt{DDFacet} \citep{Tasse2018} to perform smearing correction and imaging the M-OTF visibilities with mosaicking in the visibility plane. The purpose of using an algorithm that relies on faceting is to approximate a wide FoV with many narrow field images.

The key steps for imaging are shown in a flow-chart in the \cref{fig:flow}. To image each block we break down the observed sky region ($\sim 30^{\circ} \times 10^{\circ}$) into multiple smaller rectangular tiles. This is done to reduce processing time and limit the required computational resources. Considering UHF-band observations, these tile sizes are  set to be $4^{\circ} \times 4^{\circ}$.
As a next step of imaging we construct an array of phase centres for imaging and identify all the 2-second snapshot MSs that should contribute to the tile image. 

We identify the boundary of an image tile and any 2-second MS that has a pointing centre (PC) inside a $4^{\circ} \times 4^{\circ}$ region is included for imaging this particular tile, rest of the MSs are not considered for the image tile. Although the tile size is set to $4^{\circ} \times 4^{\circ}$, to de-convolve bright sources that are beyond the tile boundary, we image a larger region in the sky. The size of this larger region is set to $7^{\circ} \times 7^{\circ}$. Then for our final image we make a cutout from this large image to an image of size $4^{\circ} \times 4^{\circ}$.

\begin{figure*}
\hspace{-1cm}
  \begin{subfigure}{.49\textwidth}
    \includegraphics[width=.99\linewidth,trim={2.1cm 0.5cm 0.3cm 0},clip]{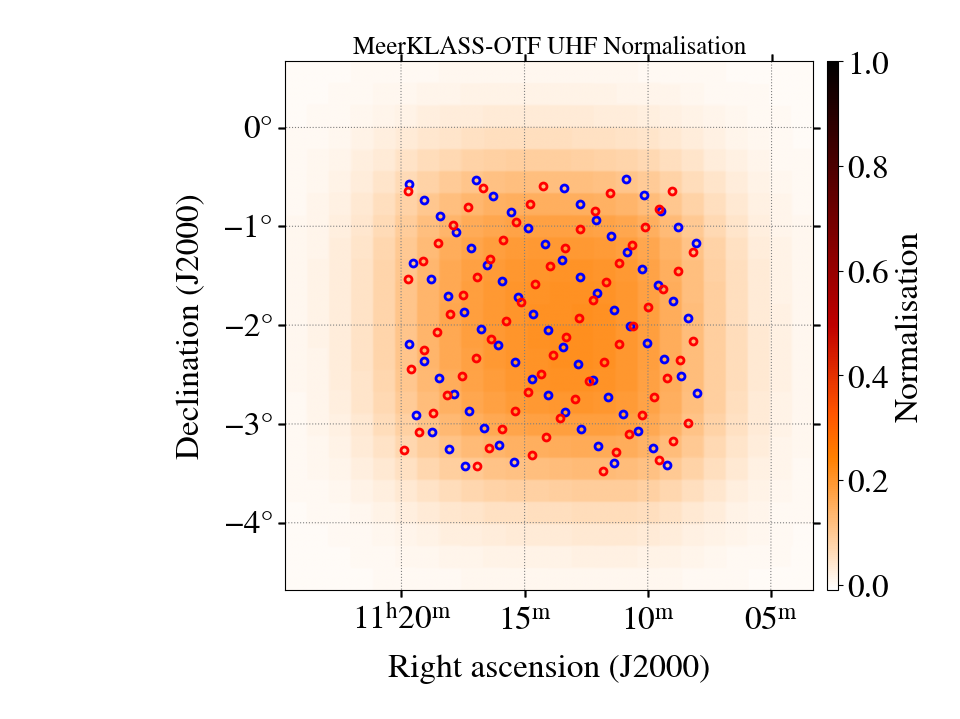}
    \caption{Per facet normalization.}
  \end{subfigure}%
  \begin{subfigure}{.49\textwidth}
    \includegraphics[width=.99\linewidth,trim={2.10cm 0.5cm 0 0},clip]{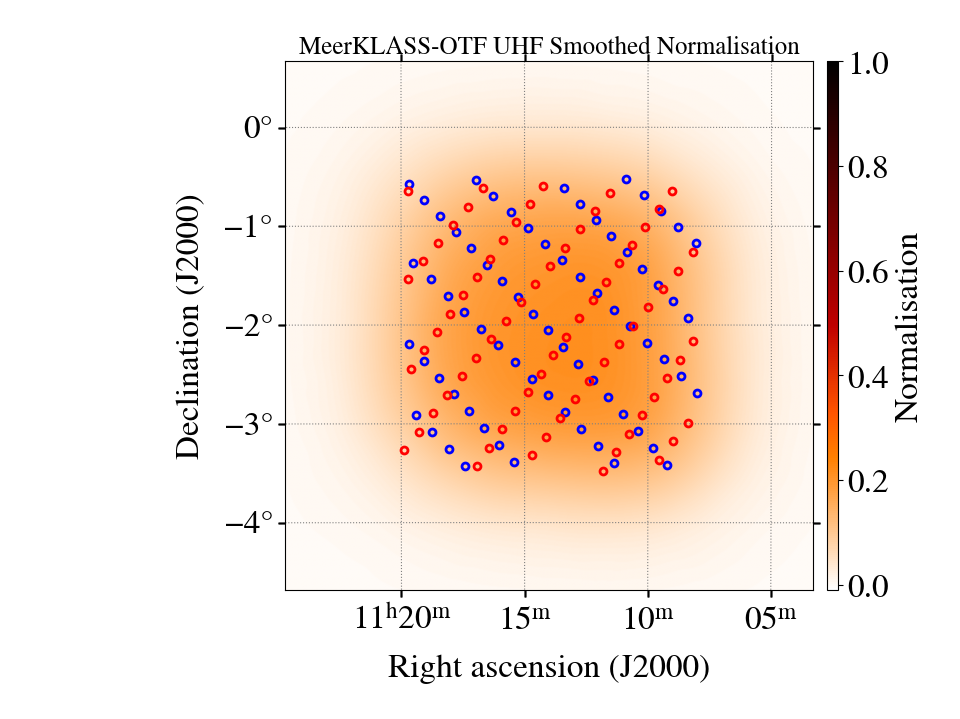}
    \caption{Per facet normalization interpolated to different pixels.}
  \end{subfigure}
    \caption{This shows joint convolution normalisation due to primary beam variation in \texttt{DDFacet}. The points show the pointing direction of the antennas for each 2-second snapshots combined. Blue and red points show the visibilities from rising and setting scans respectively. The left panel show a per-facet normalisation whereas the right panel show the smooth interpolated normalisation. }
    \label{fig:norm}
\end{figure*}
\cref{fig:norm} shows an example of 2-second MS files used when visibilities from two different observations or epochs are combined. Blue circles show the pointing centres from a rising scan where as red circles show the pointing centres from a setting scan. We use the generic hybrid joint deconvolution algorithm, \texttt{SSD2} that uses subspace optimization to perform the deconvolution of MSs with different PCs.  To perform wide-band joint deconvolution for the flux densities as well as the intrinsic spectral properties using \texttt{DDFacet}, one needs to consider direction-time-frequency dependent behavior of the primary beam using Jones matrices. We use UHF-band holographic measurement of the MeerKAT primary beam \citep{deVilliers2023} that provides a model at $60^{\circ}$ elevation over a diameter of $10^{\circ}$ to model the Jones matrices for MeerKAT primary beam. In our case the image is split into $N_f$ rectangular facets (see \cref{tab:impar}) and the primary beam E-Jones is computed at each facet's centre. Left panel of \cref{fig:norm} show primary beam normalization for each facet in the image and the right panel shows the interpolated smooth normalization that is used for the flux correction. Considering the joint deconvolution of the visibilities, that have different pointing direction of the antennas, an effective primary beam model is constructed in the process and we used that for correcting the fluxes. Considering L-band, we use \texttt{EIDOS} software \citep{Asad2021} that provides Zernike-based model over a diameter of $10^{\circ}$ to model the Jones matrices for MeerKAT primary beam.

\section{Mosaicked image quality}\label{sec:result}
The primary data product for the M-OTF are continuum total intensity (stokes I) multi-frequency synthesis (MFS, \citealt{Conway1990}) images and sub-band images.  In this section, we discuss the quality of the M-OTF image noise and sensitivity, astrometry and photometry. We also present some general discussion for the completeness and image fidelity, the degree to which the images are representative of the true sky brightness.

\subsection{Image quality}
\begin{figure*}
\hspace{-1.95cm}
  \begin{subfigure}{.33\textwidth}
    \includegraphics[width=.99\linewidth, trim={0cm 0.1cm 0.0cm 0},clip]{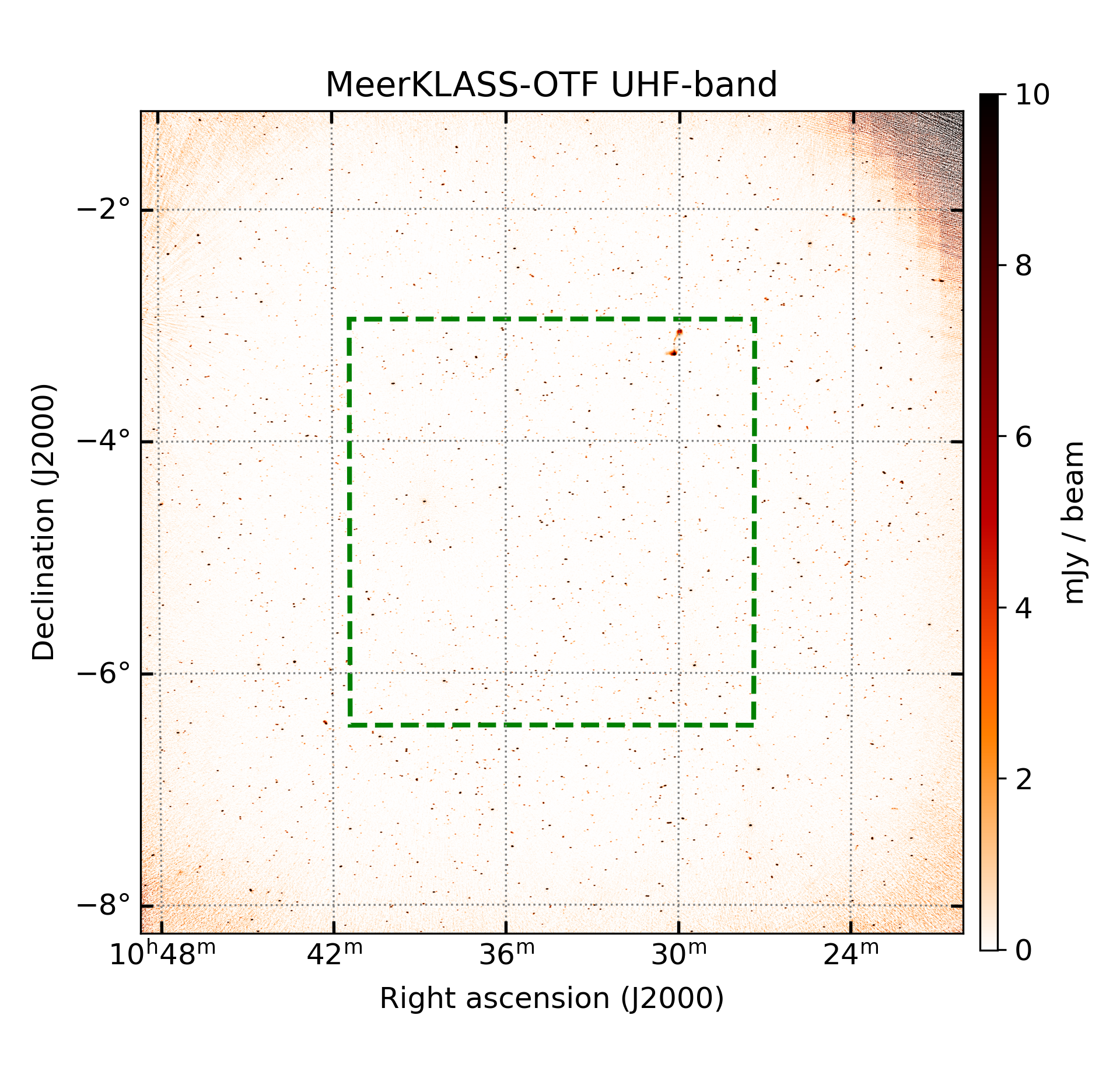}
    \caption{Total intensity large image.}
  \end{subfigure}%
  \begin{subfigure}{.33\textwidth}
    \includegraphics[width=.99\linewidth, trim={0cm 0.1cm 0cm 0},clip]{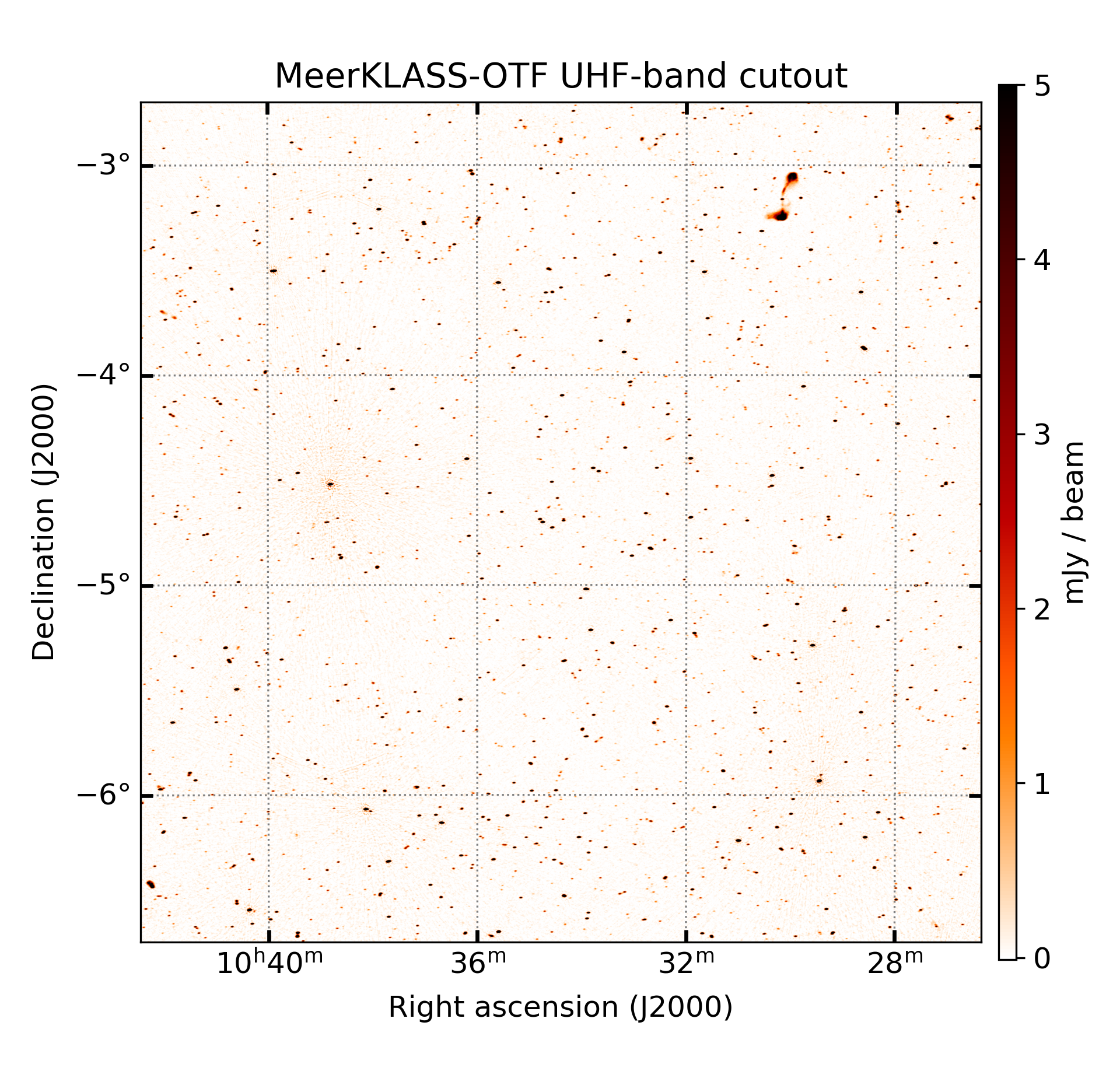}
    \caption{Total intensity cutout image.}
  \end{subfigure}
      \begin{subfigure}{.33\textwidth}
    \includegraphics[width=.99\linewidth, trim={0cm 0.1cm 0cm 0},clip]{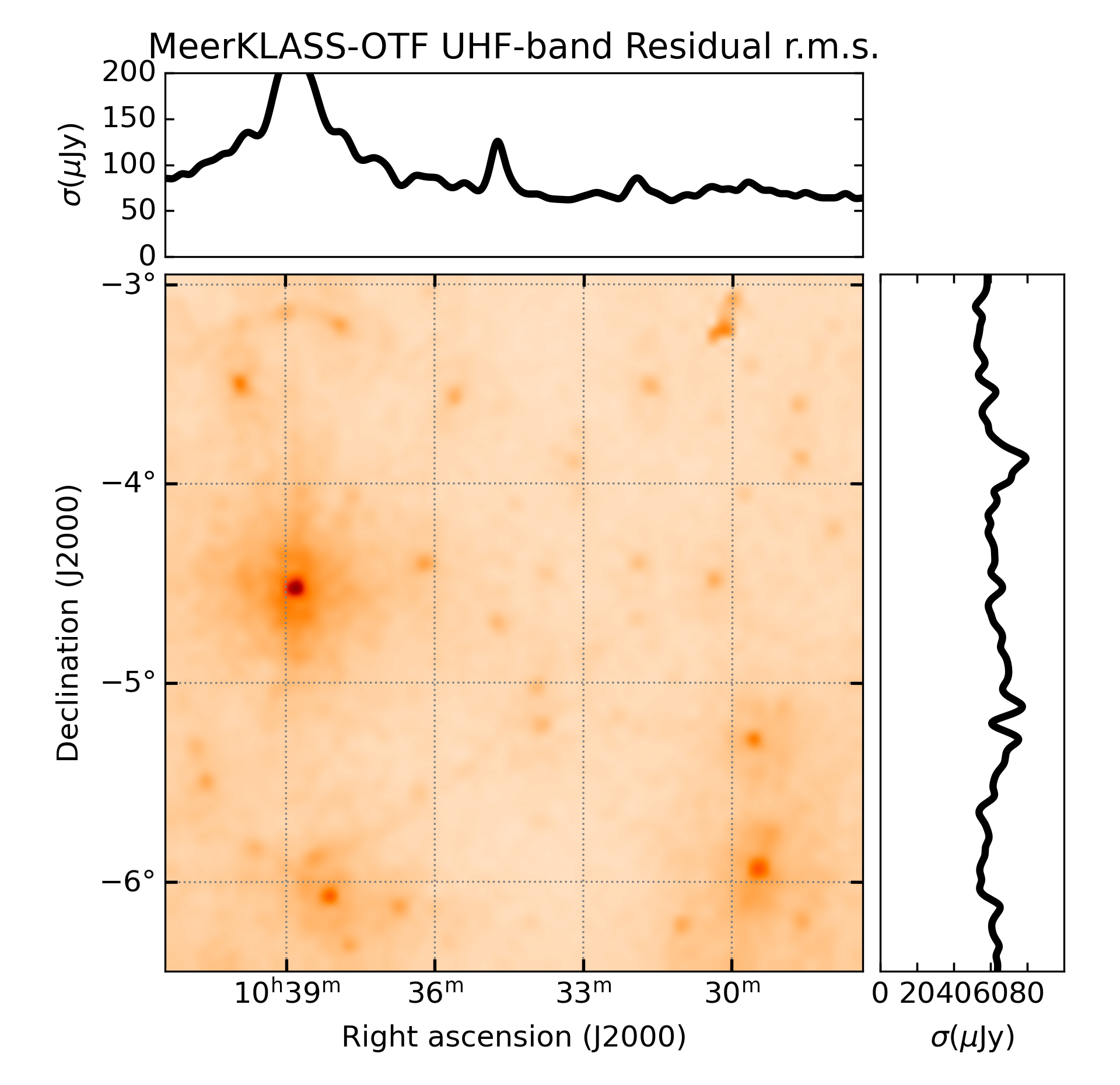}
    \caption{rms of the residual of the cutout image.}
  \end{subfigure}
  \caption{These show the M-OTF final image products in the UHF band. The left panel show $7^{\circ} \times 7^{\circ}$ primary beam corrected total intensity image, the square marked in the image shows the central $4^{\circ} \times 4^{\circ}$ region. We use this marked region to make a cutout that is shown in the middle panel. This $4^{\circ} \times 4^{\circ}$ image is the final product from the pipeline. The right panel show the rms of the residual maps constructed during PyBDSF source finding. We also show a cross-section through the middle of the rms map.}
  \label{fig:UHF1}
\end{figure*}

\cref{fig:UHF1} shows an total intensity mosaic image product for a particular tile (Tile No. 24, \citealt{Paul2025}) in the UHF band DR1. The phase centre of this image is chosen to be $(10h34m24.0s, -04d42m00.0s)$. We have identified all the MS files that lie within a rectangle area of $4^{\circ} \times 4^{\circ}$ around this phase centre for mosaicking. To produce this image we have considered 8 such data blocks that have been observed over 8 different epochs (4 rising and 4 setting). This resulted in $\sim 800$, 2-second snapshot MS files to be combined in the imaging pipeline to produce the image (see \citealt{Paul2025} for more details).

The left panel of the \cref{fig:UHF1} shows the large image produced using the pipeline. The details of the imaging are given in \cref{tab:impar}. We created a large image ($7^{\circ} \times 7^{\circ}$) to deconvolve bright sources that are far away from the phase centre yet can introduce artifacts in the image. Next we produce a $4.0^{\circ} \times 4.0^{\circ}$ cutout from this large image to be used as a final data product for our imaging (shown in the middle panel). This cutout is chosen so that the average rms of a source free region in the image remains constant throughout the image. The right panel show the rms of the residual map of the cutouts and the value of the central pixels are plotted in the edges to show that in a source free region the rms remain constant. Overall this show the image quality achieved using our imaging pipeline is reasonable except the poor resolution due to the smearing.

\begin{figure}
  \hspace{-0.5cm}
    \includegraphics[width=1.1\linewidth, trim={1.5cm 0.5cm 0.cm 0.5cm},clip]{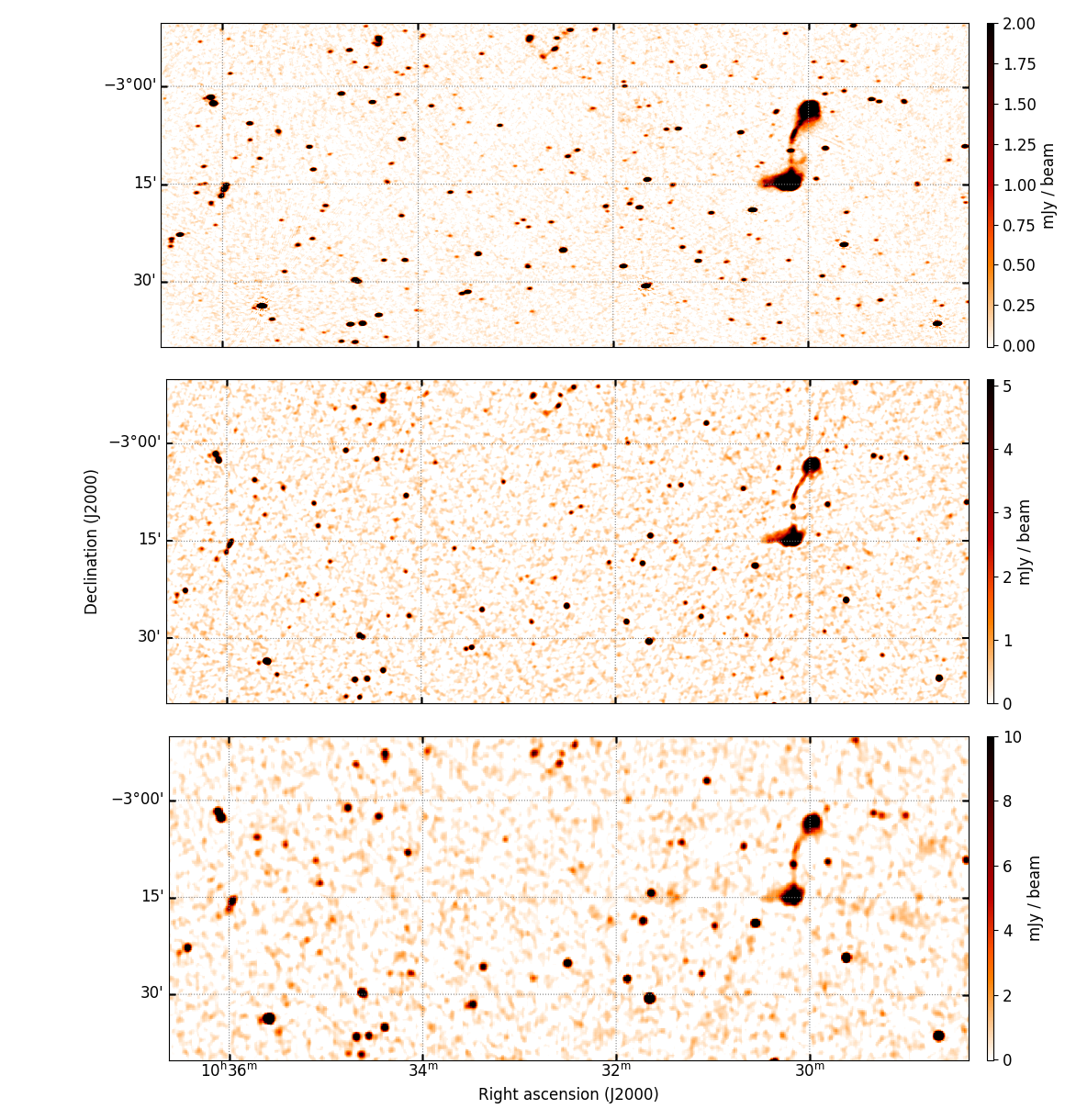}
    \caption{ The top panel shown M-OTF image cutout at a frequency 816 MHz and the synthesized beam size is $30.6 \arcsec \times 16.0 \arcsec$. The middle and the bottom panel shows exactly same region of the sky as shown in top panel for RACS-Low and NVSS respectively. RACS-Low image is at a nominal frequency of 888 MHz with a resolution of $25\arcsec$ whereas NVSS image has a restoring beam of $45\arcsec$ at 1400 MHz.}
    \label{fig:UHF_comp}
\end{figure}

\cref{fig:UHF_comp} shows a comparison of a field imaged by NVSS ($\nu = 1.4 {\rm GHz}$), RACS-Low ($\nu = 888 {\rm MHz}$) and M-OTF ($\nu = 816 {\rm MHz}$). The comparison shows clearly the better sensitivity of the M-OTF observations which led to more sources being detected and to a better definition in the extended objects. It is worth mentioning that the images are made using approximately $30\%$ of the data collected in this field. We also expect to achieve better sensitivity and resolution, once the smearing issue is fixed in the correlator level and all the observed data is combined.

\begin{figure*}
\hspace{-1.9cm}
  \begin{subfigure}{.33\textwidth}
    \includegraphics[width=.99\linewidth, trim={0cm 0.1cm 0.0cm 0},clip]{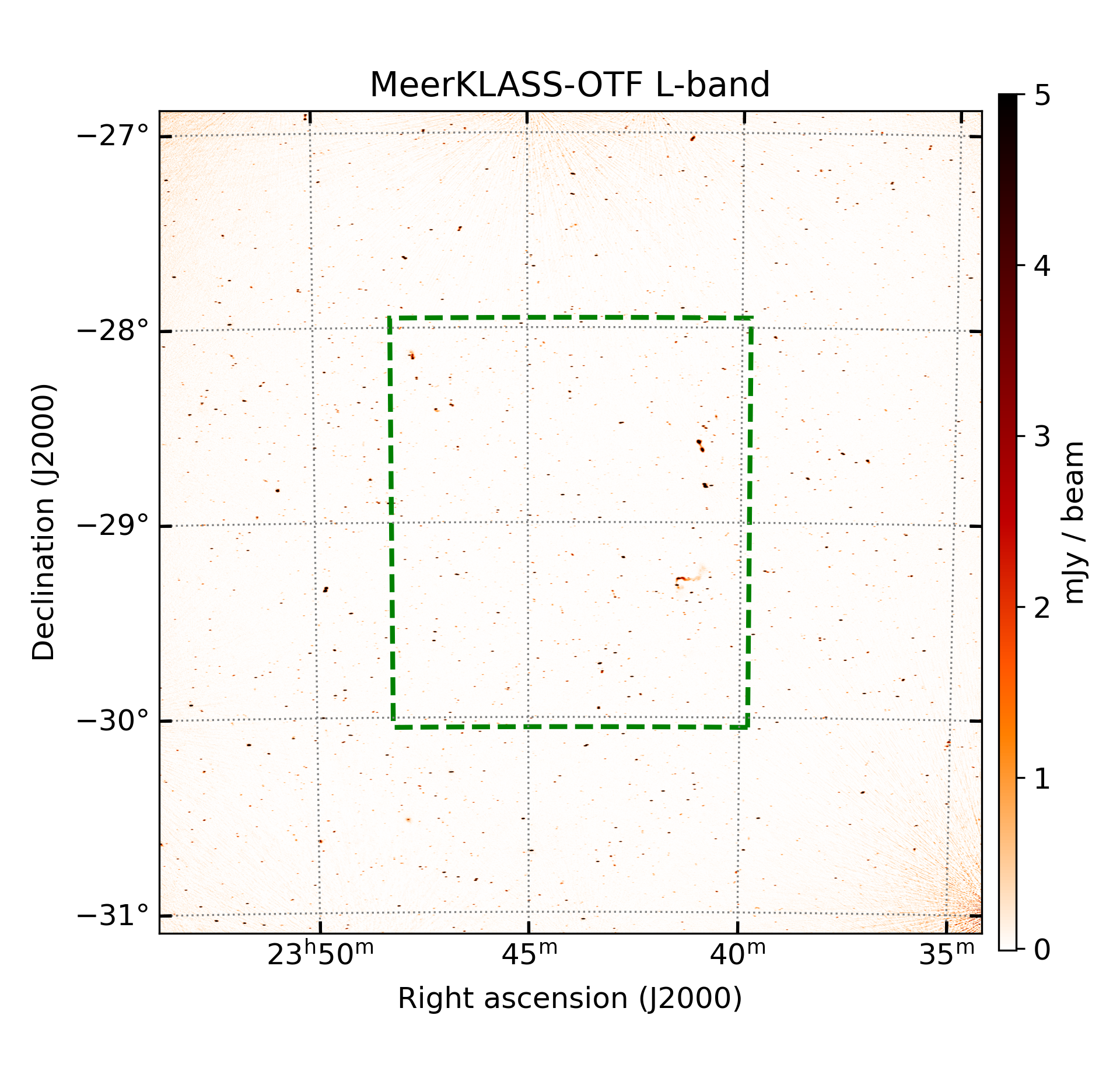}
    \caption{Total intensity large image.}
  \end{subfigure}%
  \begin{subfigure}{.33\textwidth}
  \centering
    \includegraphics[width=.99\linewidth, trim={0cm 0.1cm 0.0cm 0},clip]{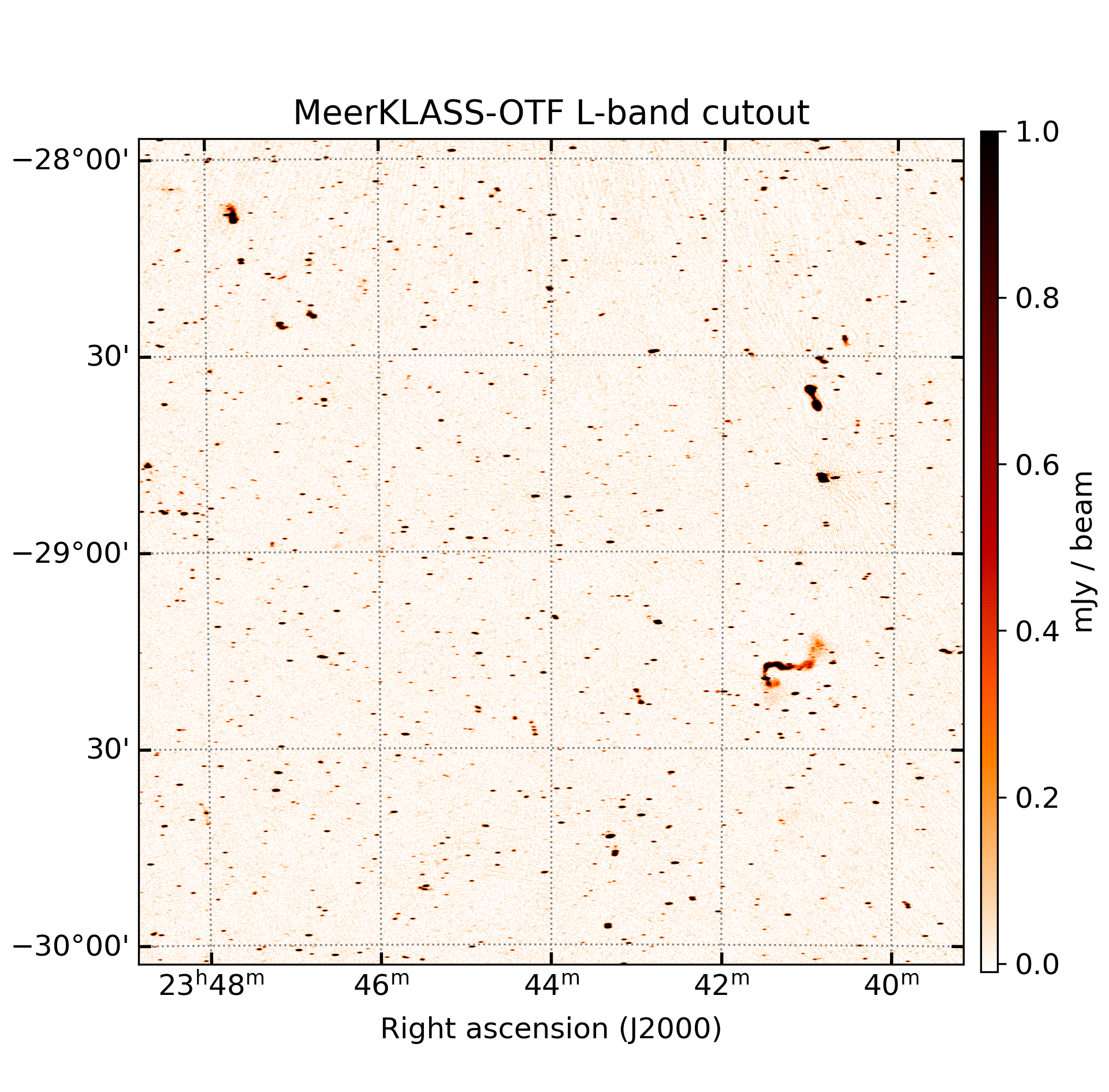}
    \caption{Total intensity cutout image.}
  \end{subfigure}
    \begin{subfigure}{.33\textwidth}
    \includegraphics[width=.99\linewidth, trim={0cm 0.1cm 0.0cm 0},clip]{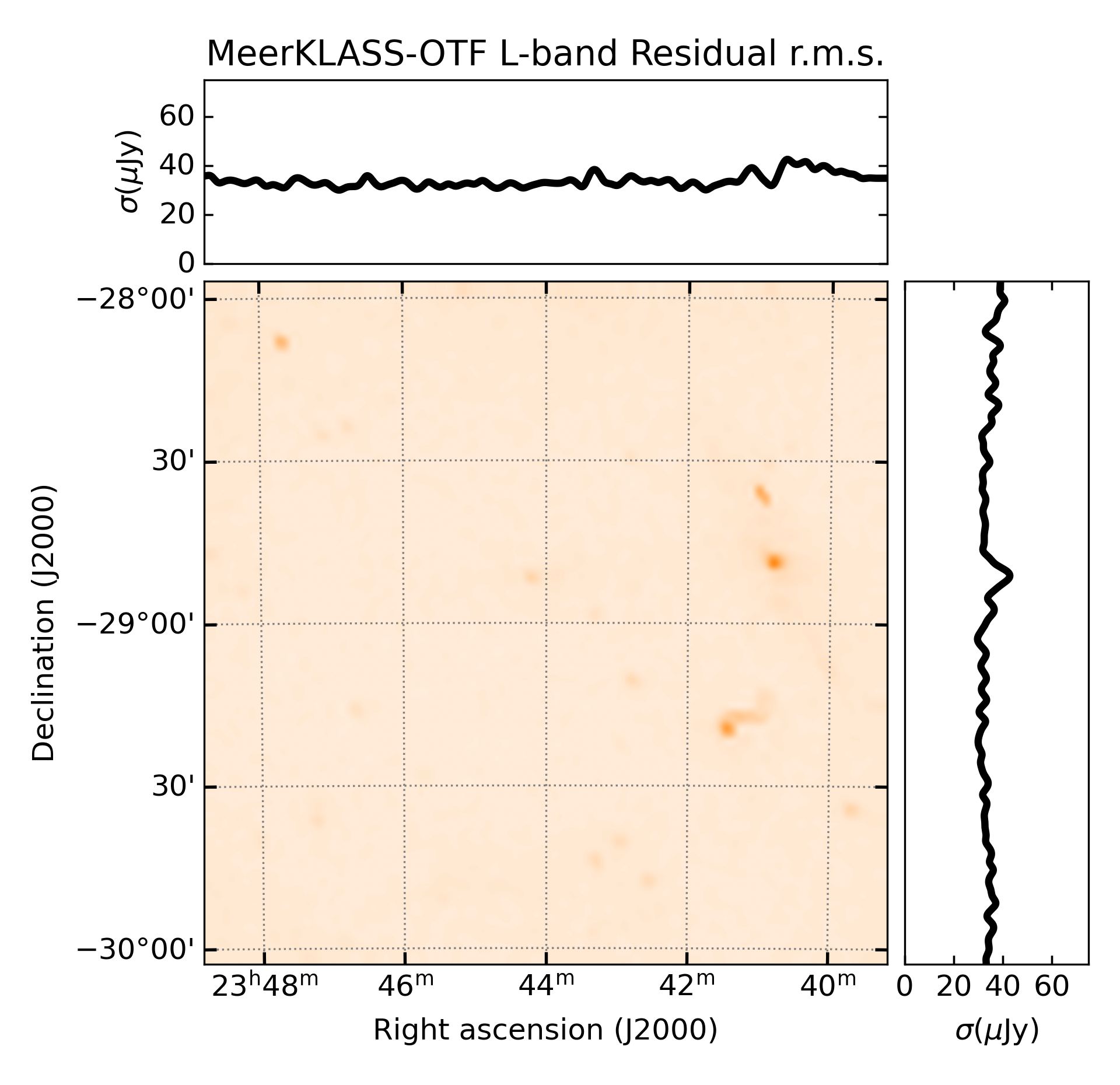}
    \caption{rms of the residual of the cutout image.}
  \end{subfigure}
  \caption{These show the M-OTF final image products in the L band. The left panel show $5.2^{\circ} \times 5.2^{\circ}$ primary beam corrected total intensity image, the square marked in the image shows the central $2.1^{\circ} \times 2.1^{\circ}$ region. We use this marked region to make a cutout that is shown in the middle panel. This $2.1^{\circ} \times 2.1^{\circ}$ image is the final product from the pipeline. The right panel show the rms of the residual maps constructed during PyBDSF source finding. We also show a cross-section through the middle of the rms map.}
  \label{fig:L1}
\end{figure*}

Considering the pilot observation done in 2021 using the L-band of MeerKAT we have performed a similar analysis as in the UHF-band described above and the images are shown in \cref{fig:L1} (see \citealt{Mangla2025} for more details). The left panel of the \cref{fig:L1} shows the large image produced using the pipeline. The details of the imaging are given in \cref{tab:impar}. We created large image ($5.2^{\circ} \times 5.2^{\circ}$) to deconvolve bright sources that are far away from the phase centre yet can introduce artifacts in the images. Next we produce a $2.1^{\circ} \times 2.1^{\circ}$ cutout from this large image to be used as a final data product for our imaging (shown in the middle panel). The rms of the residual map of the cutouts are very encouraging as we find that there is minimal variation in the rms throughout the image.  It is worth mentioning that, during combining different data blocks in the L-band we only have used the rising scans (see \cref{sec:HRD}) and avoided all the setting scans due to absence of any suitable calibrators.

\begin{figure}
  \hspace{-1cm}
    \includegraphics[width=1.15\linewidth, trim={0.0cm 0cm 0.0cm 0cm},clip]{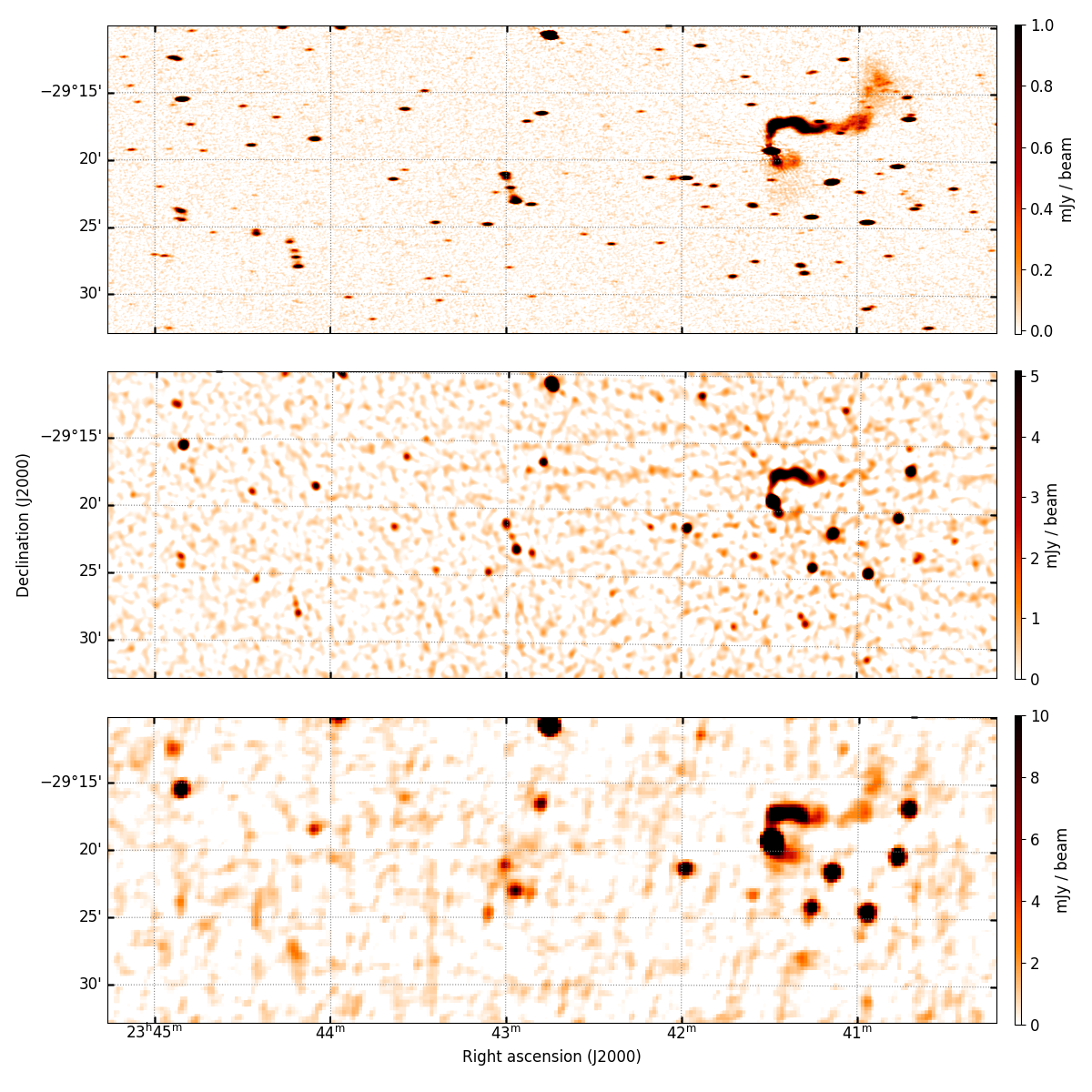}
    \caption{The top panel shown M-OTF image cutout at a frequency 816 MHz and the synthesized beam size is $26.0 \arcsec \times 7.8 \arcsec$. The middle and the bottom panel shows exactly same region of the sky as shown in top panel for RACS-Mid and NVSS respectively. RACS-Low image is at a nominal frequency of 1367.5 MHz with a resolution of $25\arcsec$ whereas NVSS image has a restoring beam of $45\arcsec$ at 1400 MHz.}
    \label{fig:L_comp}
\end{figure}

\cref{fig:L_comp} shows a field imaged by NVSS ($\nu = 1.4 {\rm GHz}$), RACS-Mid ($\nu = 1.3 {\rm GHz}$, resolution = $25\arcsec$) and MeerKLASS ($\nu = 1284 {\rm MHz}$). We see that the M-OTF image quality is significantly better in terms of noise and capturing the extended emission, due to low system temperature of MeerKAT and dense core antenna distribution. We expect the image r.m.s. to improve by a factor of two once we combine all the available observations in this band. Currently, we do not have any plans to conduct further observations in L-band.

\subsection{Astrometry}
To asses the accuracy of the flux scale and source positions in the M-OTF images, we constructed a source catalogue from the M-OTF images. We use the Python Blob Detection and Source Finder (\texttt{PyBDSF} \footnote{\href{https://pybdsf.readthedocs.io/en/latest/index.html}{PyBDSF}.} ;\citealt{Mohan2015}), that constructs the background maps of an image and identifies the islands with emissions greater than the threshold provided  (\texttt{thresh\_isl}). Further these islands are modelled using single or multiple component Gaussian that are used to construct as source model in our catalogue. 

As noted previously, the M-OTF observations suffer from a significant amount of smearing. This leads to an underestimate of the error in the source flux due to differences in the beam area that is convolved with the source model and with the residual  maps. We can calculate the ratio of the beam areas , $B_{\rm R}$, and update the local rms estimate by the factor $\sigma_c  = \sigma/B_{\rm R}$. We choose the threshold for the islands and the pixels to be $\sigma_c = 3$ and 5 respectively, which means we must set the detection thresholds in \texttt{PyBDSF} to $1.5 \times \sigma_c$ = 4.5 and 7.5 respectively. We have used these to run the PyBDSF on the images. The detection parameters for the PyBDSF are:
\begin{verbatim}
bdsf.process_image(image, thresh_isl = 4.5, 
thresh_pix = 7.5, mean_map =`zero', rms_box=(150,30),
rms_map=True, atrous_do=True, 
adaptive_thresh=150, rms_box_bright=(20,7))
\end{verbatim}
where \texttt{image} is the file name of the image, \texttt{thresh\_isl} is the threshold for detecting an island of emission to then be fit with Gaussian components, and \texttt{thresh\_pix} is the criterion which determines which sources are included within the final source catalogue. The \texttt{mean\_map} is the same shape as the input image, reflecting the background emission within the image. We set this to zero at all locations within the image. Finally, \texttt{rms\_box} is a parameter used to quantify the size of the box used to determine the rms in terms of pixels, as well as the step size used to move this sliding box across the image.
There are no surveys with using MeerKAT UHF-band is available that overlaps completely with the M-OTF observation. To quantify the astrometric accuracy of the catalogue produced by the pipeline,we compare them with the RACS-Mid catalogues, that have complete overlap with the MeerKLASS observations. For this purpose we have imposed the following criteria for source selection in both the catalogues; 
\begin{itemize}
    \item We consider sources that have a single Gaussian component in the model (\texttt{Scode=`S'})to minimize any biases introduced by differing angular resolution, or sensitivity to extended emission.
    \item We only consider sources with ${\rm SNR} > 10$.
    \item We impose that the components are isolated \ie there are no identified radio sources within a $60 \arcsec$ radius.
    \item In all cases where a cross-match is performed, we conduct a nearest-neighbour match, and enforce the criteria that the radial separation of the two components must be less than $5 \arcsec$.
\end{itemize}

\begin{figure}
  \begin{subfigure}{.25\textwidth}
  \centering
    \includegraphics[width=.99\linewidth]{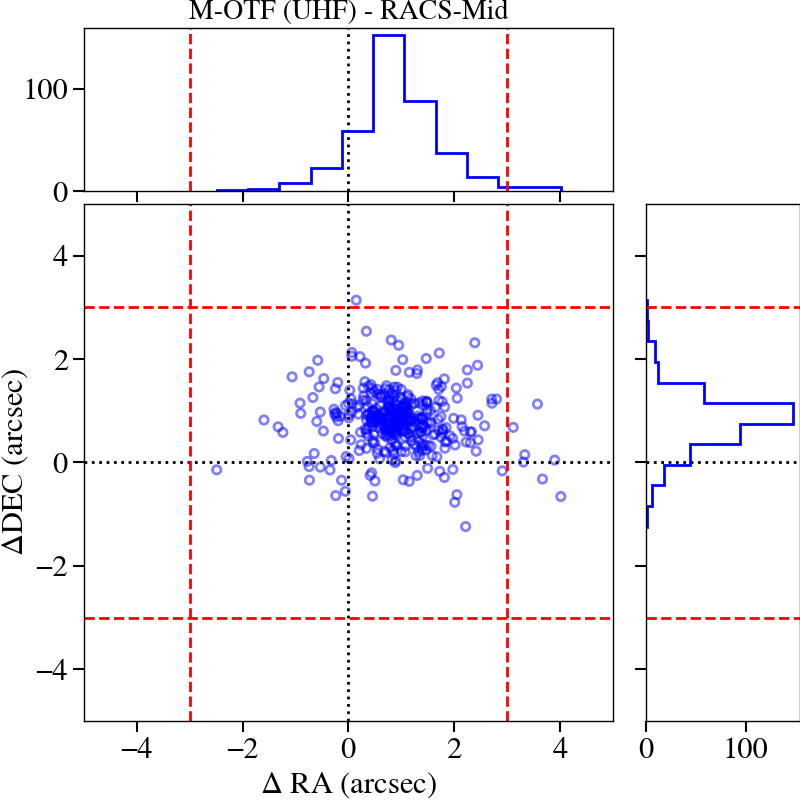}
    \caption{UHF-band}
  \end{subfigure}%
  \begin{subfigure}{.25\textwidth}
  \centering
    \includegraphics[width=.99\linewidth]{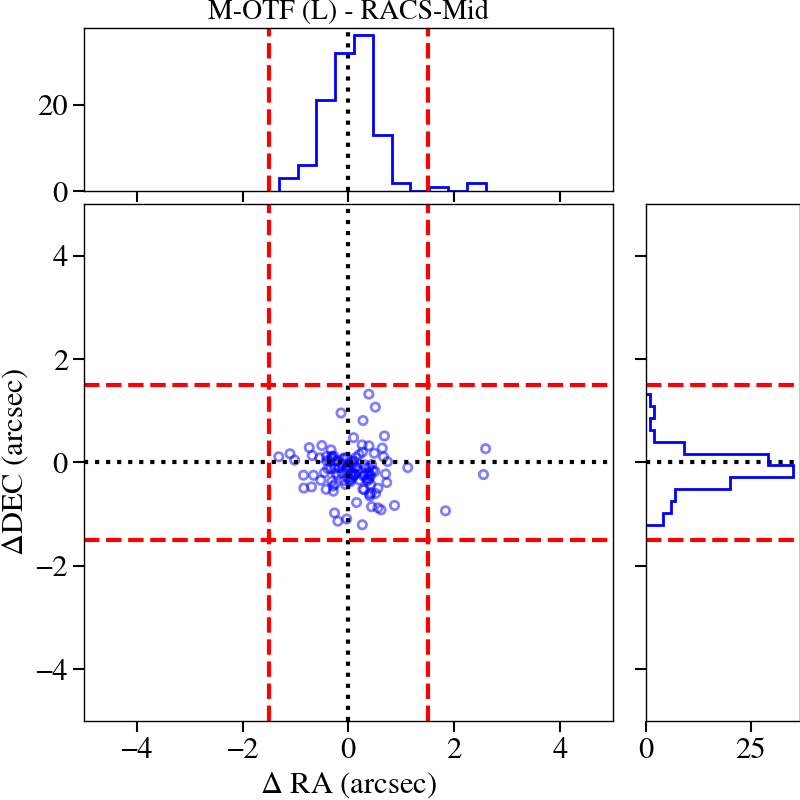}
    \caption{L-band}
  \end{subfigure}
  \caption{Astrometry of M-OTF with RACS-Mid survey. Red dashed lines show the size of one pixel in the images.}
  \label{fig:astro}
\end{figure}
Considering RACS-Mid catalogue, we find that there are 475 sources present with in the tile boundary, which have SNR$>10$ and \texttt{Scode=`S'}.  There are 1193 sources present in the M-OTF catalogue that matches the similar criteria. We found 460 isolated sources in RACS-Mid catalogue. We found 450 of their counterpart in M-OTF catalogue with a separation tolerance of $5\arcsec$. However, among these cross-matches there are only 392 sources are with \texttt{Scode=`S'} and isolated within a radius of $60\arcsec$. \cref{fig:astro} show the results for the astrometric precision between M-OTF and RACS-Mid catalogues, only using these 393 sources. Left panel show the results for the UHF-band. The median deviation along RA and DEC are $0.9 \arcsec$ and $0.5 \arcsec$ respectively. This implies the MeerKLASS catalogue are accurate up to $0.9 \arcsec$, when compared with RACS-Mid. 
We performed a similar comparison with the L-band M-OTF images and the results are shown in the right panel of the \cref{fig:astro}. Here we found that there are 136 number of sources present with in the tile boundary, have SNR$>10$ and \texttt{Scode=`S'} in the RACS-Mid catalogue compared to 730 sources in the MeerKLASS catalogue. We found 113 of them have a match counterpart with the median deviation along RA and DEC are $0.06 \arcsec$ and $-0.53 \arcsec$ respectively.

\begin{figure}
    \centering
    \includegraphics[width=0.49\linewidth]{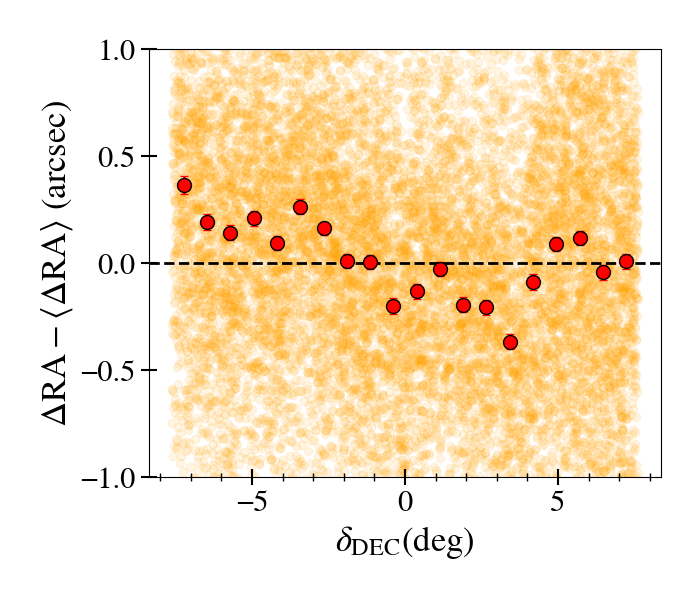}
    \includegraphics[width=0.49\linewidth]{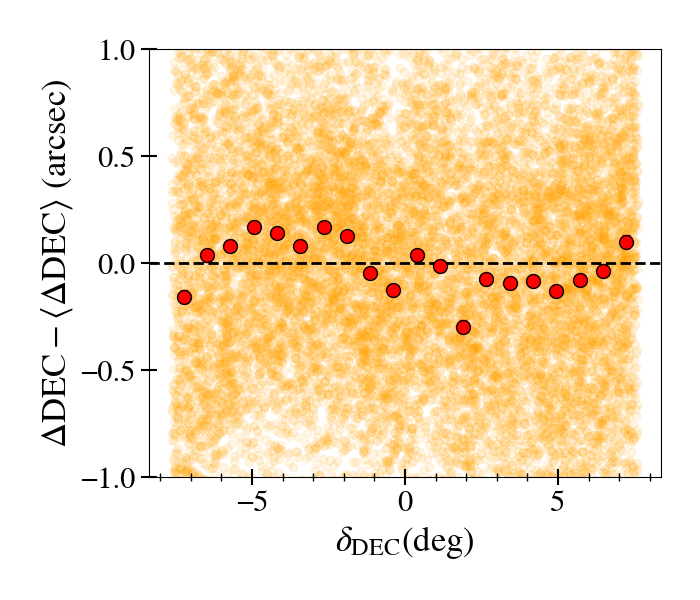}
    \caption{This shows the astrometric variation of M-OTF sources in UHF-band relative to RACS-Mid sources with the separation from the delay centre along declination $\delta_{\rm DEC}$. Left panel shows the variation in RA ($\Delta{\rm RA} - \langle \Delta{\rm RA}\rangle$) and right panel shows variation in DEC ($\Delta{\rm DEC} - \langle \Delta{\rm DEC}\rangle$). The variations are shown in orange points. The red points show the binned median of the orange points and the error bars are the standard deviation of the variation scaled by the square root of the number of sources in each bin.}
    \label{fig:pvr_UHF}
\end{figure}

\begin{figure}
    \centering
    \includegraphics[width=0.49\linewidth]{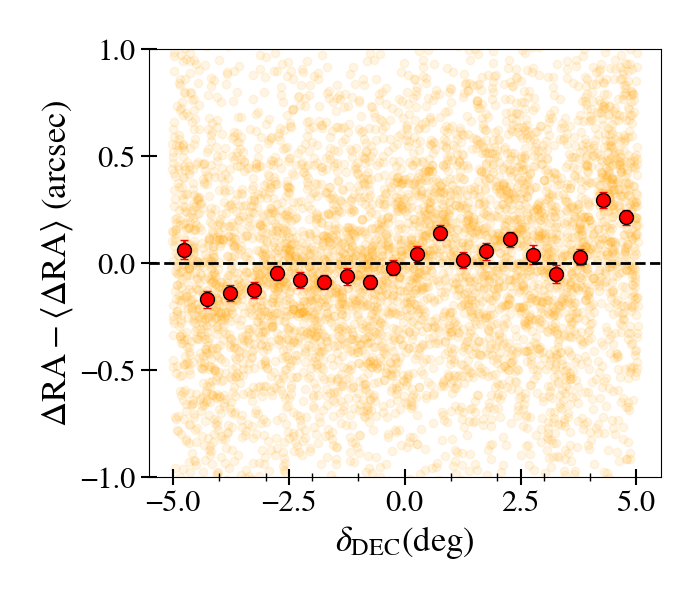}
    \includegraphics[width=0.49\linewidth]{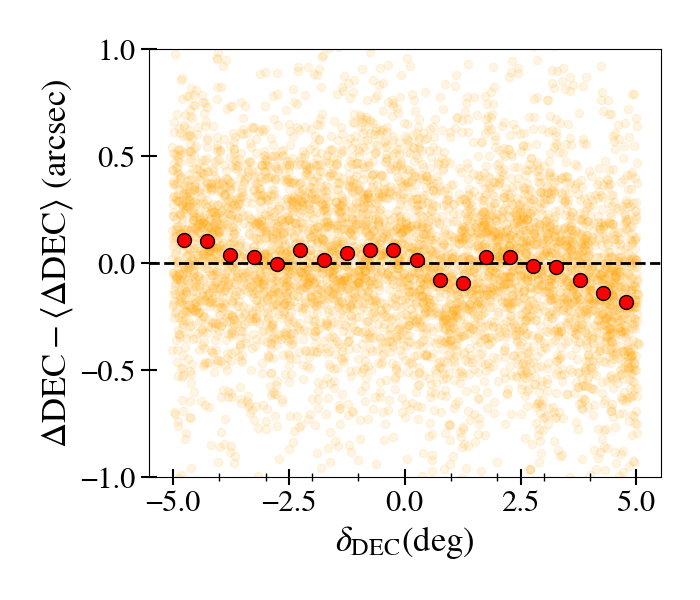}
    \caption{ Same as \cref{fig:pvr_UHF} for L-band images. }
    \label{fig:pvr_L}
\end{figure}
In the current M-OTF setup, the antennas perform a constant elevation scan by rapidly moving back and forth along the azimuth direction (see \cref{fig:scan}). The delay centre for each epoch of observation is fixed at an az-el in the beginning of the constant elevation scan and it traces a line through the centre of the scan region in sky coordinates (\cref{fig:scan}). We phase rotate the visibilities to perform imaging and perform smearing correction to correct for the absence of the sidereal  tracking. We found that given an epoch of observation, the smearing varies with the declination (\cref{fig:sm4}) and it is crucial to investigate if any residual smearing effect left in the M-OTF images. To identify the maximum residual effect, instead of using a single tile, we compare the whole region of M-OTF observation processed till date (see \citealt{Paul2025} and \citealt{Mangla2025}). For this comparison we use the same aforementioned criteria for source cross-matching. The results for UHF-band are shown in \cref{fig:pvr_UHF}. The left panel shows the variation of the difference of the source position in RA direction ($\Delta{\rm RA}$) as a function of distance from the observation's delay centre (at DEC=$-2^{\circ}$) in declination ($\delta_{\rm DEC}$). There is an overall offset between M-OTF and RACS-Mid source positions and we have subtracted the average offset $\langle \Delta{\rm RA}\rangle = 0.82 \arcsec$ to find the residual variation \citep{Paul2025}. The variation $\Delta{\rm RA} - \langle \Delta{\rm RA}\rangle$ are shown in orange point whereas the red points show the median of the $\Delta{\rm RA} - \langle \Delta{\rm RA}\rangle$. To estimate the median we have divided the $\delta_{\rm DEC}$ in 20 linear bins and the error bars are are computed by scaling the standard deviation of the $\Delta{\rm RA} - \langle \Delta{\rm RA}\rangle$ in each bin by square root of the number of the sources with in the bin. The astrometric variation of the sources in declination  $\Delta{\rm DEC} - \langle \Delta{\rm DEC}\rangle$ is shown in the right panel where $\langle \Delta{\rm DEC}\rangle = 0.48 \arcsec$ \citep{Paul2025}. We do not see any distinguishable variation trend with increasing $\delta_{\rm DEC}$ both in RA and DEC. The variation along RA direction is restricted between $\pm 0.4\arcsec$ whereas along declination the variation is even smaller and remains restricted between $-0.3\arcsec$ to $0.2\arcsec$. This indicates that although the observations were perform significantly far away from the delay centre, it does not introduce any significant phase errors ($< \pm 0.5 \arcsec$). We performed similar analysis with the L-band M-OTF images \citep{Mangla2025}  and the results are shown in \cref{fig:pvr_L}. We find that the variation in the positional offset is smaller compared to UHF. However, here we see that the offset has an increasing trend with the $\delta_{\rm DEC}$. This could arise from the fact that there are no secondary calibrator present in the observation which may have resulted in some residual phase error.  It is worth noting the variation ranges between $\pm 0.15 \arcsec$ in DEC and $-0.15 \arcsec$ to $0.3 \arcsec$ in RA direction, which are tiny compared to the average beam size in L-band images (\cref{sec:img}). 

\subsection{Photometry}
As extensively discussed, the nature of the M-OTF observations introduces amplitude smearing and we have introduced a correction for that(see \cref{sec:img}). This makes it crucial to compare the source flux scale with a tracking observation. We do not perform such tracking observation, instead we use other overlapping radio source catalogues to find the accuracy of the flux scale and scatter of measured fluxes. For the comparison of flux density we choose the same set of sources cross-matched with RACS-Mid that are discussed in the last section. RACS-Mid has a nominal frequency of 1367 MHz which is different from the M-OTF images where nominal frequency is 816 MHz. We have scaled the fluxes of the RACS-Mid cross-matched sources to 816 MHz using a fixed spectral index of $\alpha = -0.7$ \citep{Duchesne2023}. 
\begin{figure}
  \begin{subfigure}{.25\textwidth}
  \centering
    \includegraphics[width=.99\linewidth]{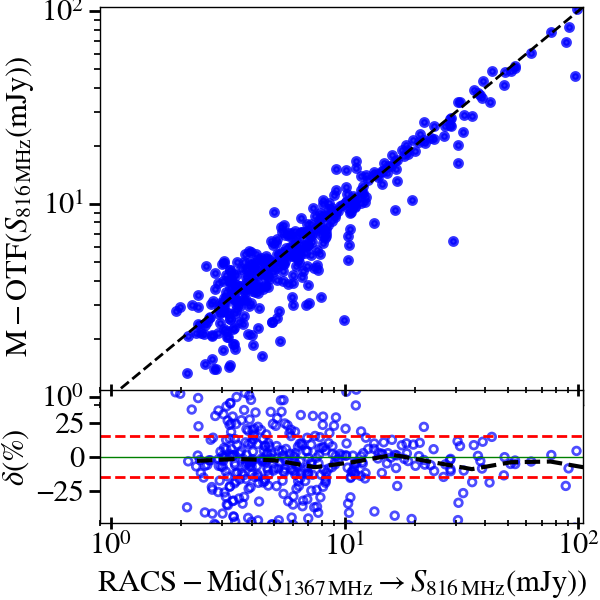}
    \caption{UHF-band}
  \end{subfigure}%
  \begin{subfigure}{.25\textwidth}
  \centering
    \includegraphics[width=.99\linewidth]{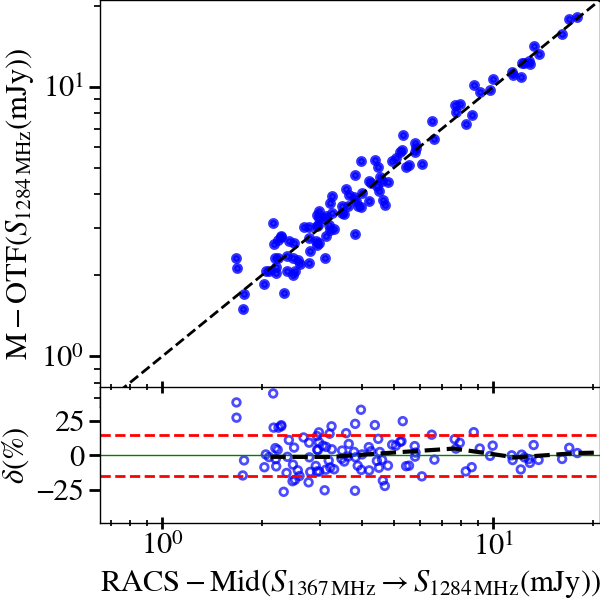}
    \caption{L-band}
  \end{subfigure}
  \caption{Photometry of MeerKLASS OTF with RACS-Mid survey. The dashed line in the top panels show 1:1 flux values, whereas in the bottom panels horizontal dashed lines mark a deviation of 15 percent.}
  \label{fig:phto}
\end{figure}
\cref{fig:phto} summarise the photometry over the MeerKLASS survey images. The comparison with the M-OTF source catalogue from a single tile image in UHF-band and RACS-Mid catalogue are summarised in left, whereas the comparison with M-OTF L-band ($\nu = 1284 {\rm MHz}$) is shown in the right panel. The top panels show the flux comparison between the surveys and the bottom panels show the relative flux difference in percentage ($\Delta$). The flux comparison between M-OTF and RACS-Mid follow an approximate one-to-one relation (shown by dashed line), with larger scatter at the lower flux values. The increasing scatter with decreasing integrated flux density due to the increasing fractional noise contribution is evident. We also see that the median deviation (dashed line in the bottom panel) remains close to zero in both UHF and L-band fluxes that indicates the flux scale are in reasonable agreement.

\begin{figure}
    \centering
    \includegraphics[width=0.49\linewidth]{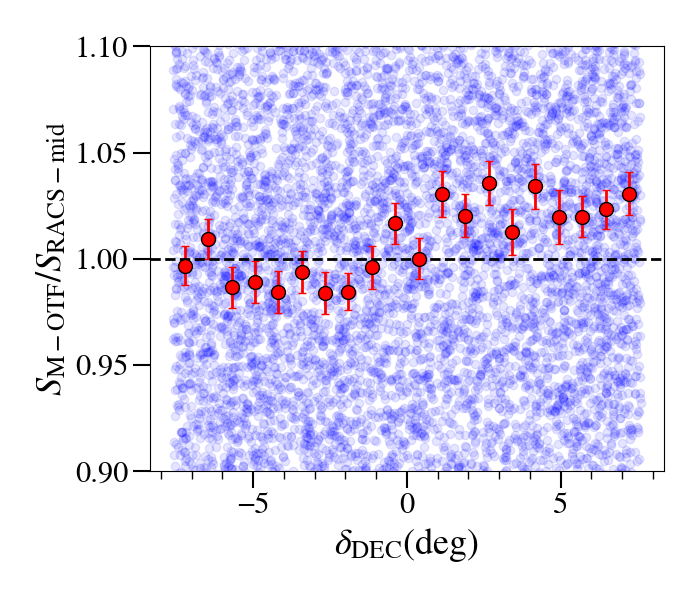}
    \includegraphics[width=0.49\linewidth]{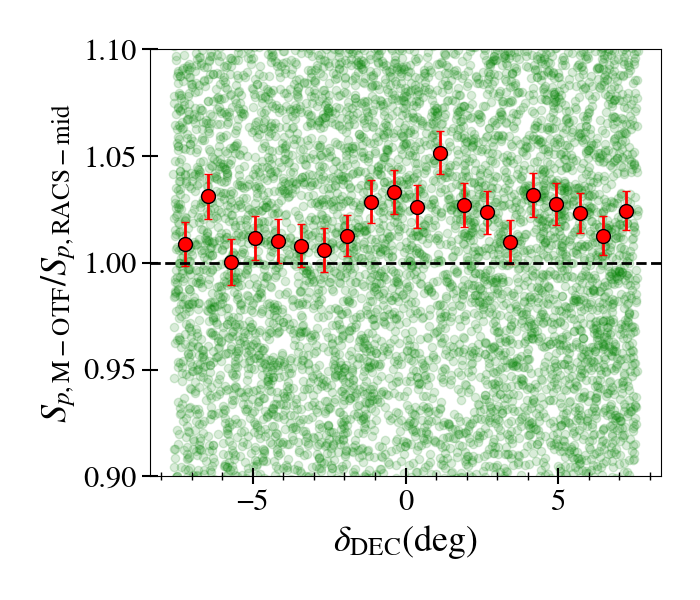}
    \caption{The left panel of the figure shows the ratio of total fluxes of the cross-matched sources between M-OTF and RACS-Mid catalogue as a function of distance from delay centre in DEC ($\delta_{\rm DEC}$). The right panel shows the ratio of the peak fluxes with $\delta_{\rm DEC}$. Each matched sources are shown in blue and green points for total and peak flux ratios respectively. Whereas the median of the distribution is shown by red points and the standard deviation scaled by the square root of the number of the sources are shown by the error-bars. }
    \label{fig:fvr_UHF}
\end{figure}

\begin{figure}
    \centering
    \includegraphics[width=0.49\linewidth]{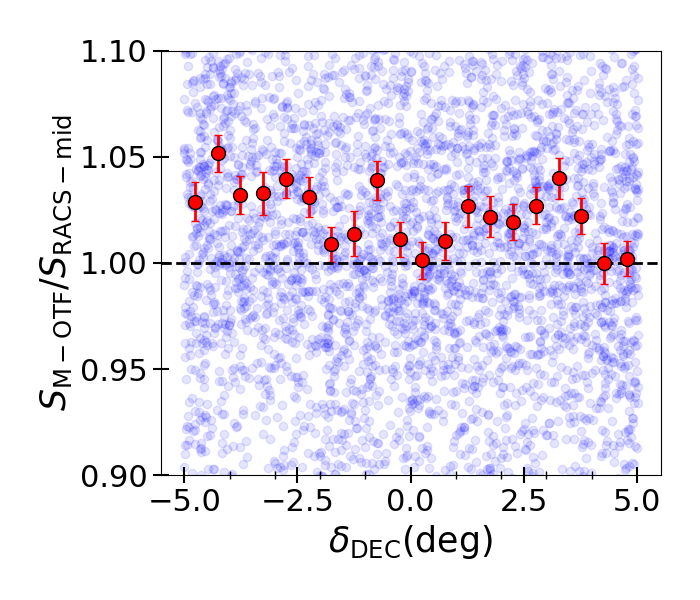}
    \includegraphics[width=0.49\linewidth]{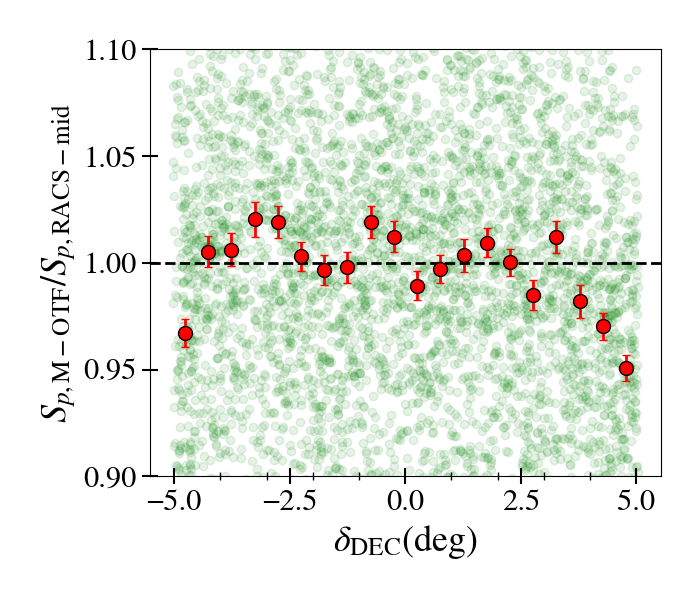}
    \caption{Same as \cref{fig:fvr_UHF} for L-band }
    \label{fig:fvr_L}
\end{figure}

To investigate further, we compare the ratios of flux densities of the cross-matched sources from the whole region of M-OTF observation processed (see \citealt{Paul2025} and \citealt{Mangla2025}) as a function of DEC separation from the delay centre of the observation \citep{Mooley2019}. The left and the right panels of \cref{fig:fvr_UHF} show the total flux and the peak flux ratio respectively. The individual measurements are shown in blue and green points for total and peak flux ratios. We bin the $\delta_{\rm DEC}$ in 20 equispaced linear bins. The median of the flux ratios are plotted in red points and the error-bars show the standard deviation of the flux ratio  scaled by square root of the number of the sources with in each bin. The total flux density ratio varying with $\delta_{\rm DEC}$ show a slight systemic effect ($<4\%$). The variation increases with $|\delta_{\rm DEC}|$ which indicates that the flux density is overestimated by $4\%$ at higher DEC, whereas underestimated by $2\%$ at lower DEC. In the right panel, peak flux ratio shows a similar behaviour with an overall offset of $\sim 2\%$ which indicate the peak flux values in M-OTF images are overestimated by $2\%$  when compared with RACS-Mid. Efforts are going on to understand the cause of the variation and it will be address in subsequent work. We also perform similar analysis for the L-band images and the results are shown in \cref{fig:fvr_L}. Here we find that the ratio of the total flux densities have an offset of $\sim 2\%$ and varies $<5\%$ over the $\delta_{\rm DEC}$ range. Whereas, the peak flux ratio does not show any offset with a similar $5\%$ variation at high DEC. Considering the fact that, here the comparison is performed between two different surveys using two different instruments which can invoke different systematic errors in the measurement, it is difficult to identify any conclusive reasoning for the small variation with $\delta_{\rm DEC}$. In future we plan to compare the flux variation with tracking observations performed using MeerKAT to understand the cause of the variation if any.

\subsection{Flux error estimate}\label{sec:ferr}

\begin{figure}
\hspace{-0.5cm}
  \begin{subfigure}{.25\textwidth}
  \centering
    \includegraphics[width=1.05\linewidth, trim={0cm 0.5cm 0.5cm 0},clip]{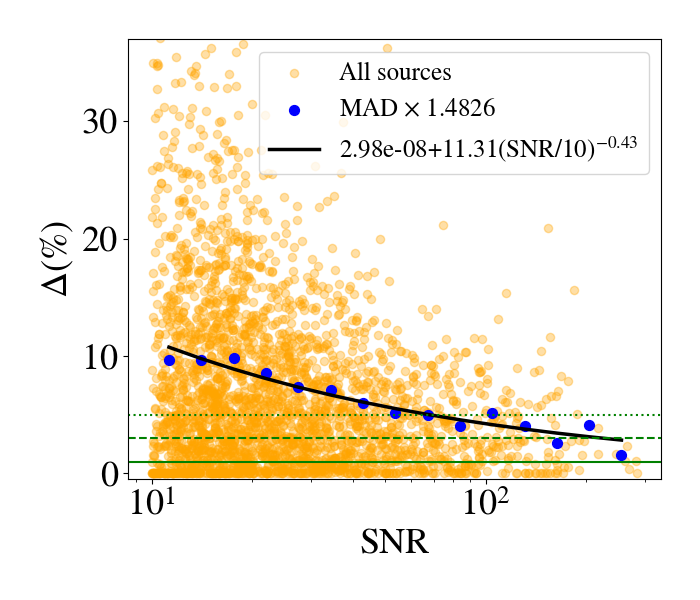}
    \caption{UHF-band}
  \end{subfigure}%
  \begin{subfigure}{.25\textwidth}
  \centering
    \includegraphics[width=.93\linewidth, trim={2.0cm 0.5cm 0.5cm 0},clip]{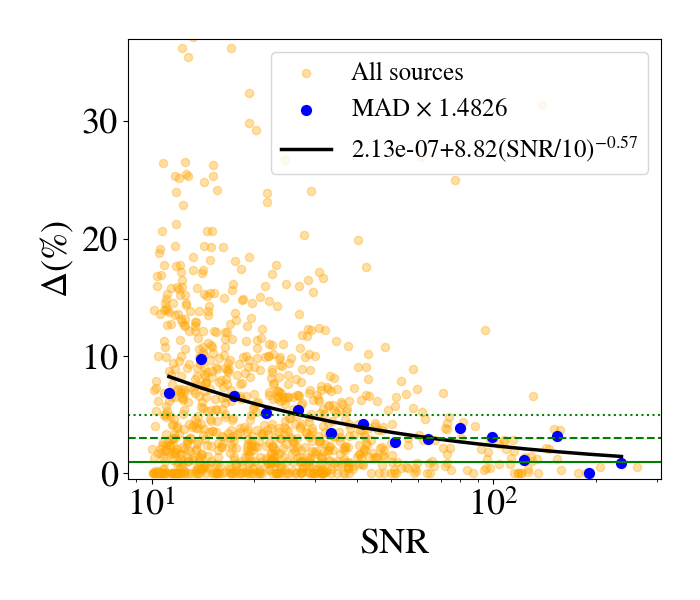}
    \caption{L-band}
  \end{subfigure}
  \caption{Here we show the absolute percentage difference (shown in yellow points) as a function of SNR. We have binned the SNR range into 15  bins in logarithmic bins and estimated the $\sigma_k$ (shown in blue points).Finally we fit the scatter with a simple power law and the black solid line shows the fit. The horizontal solid, dashed and dotted lines correspond to 1, 3, 5 percent respectively.}
  \label{fig:ferr}
\end{figure}
Considering unconventional approach of the M-OTF observations, it is crucial to quantify the uncertainty of the source flux density $\Delta S$ in the M-OTF images. These uncertainties can arise from thermal noise, source variability, imperfect deconvolution, error in absolute flux scale of the calibrator and variation from voltage to flux conversion in the correlator. Situation is further complicated by the fact that the calibrators are observed only before and after constant elevation scans at different elevations, which might lead to different air mass and ionospheric effect. By combining different rising and setting scans we use different combination of weighted primary beam, that is expected to be slightly distorted due to OTF nature of the observations (see \cref{appen1}). However it is difficult to quantify the flux error arising for the motion of antennas. \texttt{PyBDSF} provides us with errors on flux measurement ($\Delta S_i^{\rm fit}$,\texttt{e\_Total\_flux}) which is estimated from the Gaussian fit to the sources. This fitting error can quantify the uncertainty associated with the thermal noise and the deconvolution errors up-to some extent; however it is likely to underestimate the total uncertainty. Particularly the large scatter arising in flux density comparison at lower SNR sources (\cref{fig:phto}) could be due to the aforementioned possible reasoning. To quantify the uncertainties more reliably, we use the comparison between multiply observed sources in M-OTF images. To quantify the scatter, we use repeated observations within one image tile for 8 epochs considered here. We identify SNR$>10$ sources, that have \texttt{S\_code}=`S' and isolated (no other sources within $60\arcsec$). The total flux of $i^{\rm th}$ source at $j^{\rm th}$ epoch is denoted by $S_i^j$ and $\bar{S}_i$ is the median of the total flux estimate over all the epochs where the source satisfies aforementioned criteria. Then the percentage variation in the source flux is defined as $\Delta_i^j = (S_i^j/\bar{S}_i - 1) \times 100 \%$. The SNR for the $i^{\rm th}$ source is ${\rm SNR}_i^j = S_{{\rm P} \, i}^j/(1.5 \times \sigma_{{\rm loc}\, i}^j)$ where $S_{\rm P}$ is the peak flux of the sources, $\sigma_{{\rm loc}\, i}^j$ is the local r.m.s. estimated by \texttt{PyBDSF} and the 1.5 factor compensates the beam variation between rms and the restored sources. We binned the SNR range into 15 logarithmic bins and computed the median,
\begin{equation}
    \Delta_{\rm MAD}^k = {\rm median}(|\Delta_i^j|) \, \, {\rm SNR}_i^j \in \left[{\rm SNR}_{k-1}, {\rm SNR}_k \right).
    \label{eq:ferr}
\end{equation}
Then the standard deviation of the flux is approximated by $\sigma_k = 1.4826 \times \Delta_{\rm MAD}^k$. We use a simple power-law model to estimate the flux uncertainty $\Delta = A \times ({\rm SNR}/10)^B + C$ and the results are shown in \cref{fig:ferr}. The left and the right panel shows the uncertainties for UHF and L-band respectively. The blue points show the $\sigma_k$ for the background yellow points. The black solid lines show the best fit model where $A, B, C$ are $11.3, -0.43, 2.98\times 10^{-8}$ for UHF band and $8.82, -0.57, 2.13\times 10^{-7}$ for L-band. We find that the parameter $B$ is close to $-0.5$ in both the bands which is somewhat expected and the absolute offset is close to zero which indicates a negligible uncertainty for high SNR sources. We use the power-law fits to quantify the systematic uncertainty $\Delta S_i^{\rm sys} = S_i [A({\rm SNR}_i/10)^B + C]/100$. Then the total uncertainty on a source flux is 
\begin{equation}
    \Delta S_i = \sqrt{(\Delta S_i^{\rm fit})^2 + (\Delta S_i^{\rm sys})^2}.
    \label{eq:ferr2}
\end{equation}
It is worth noting that for now we use the same relation for all the M-OTF sources in the catalogue, however this relation may break down for very extended sources. We plan to investigate this in a future study. Further the uncertainty introduced by the flux scale of the calibrator can introduce $3-5\%$ uncertainty in the absolute flux values \citep{Perley_2017} and for now we do not include this in our estimates.
\subsection{Sub-band imaging and Source spectra}
MeerKAT have reasonably large usable bandwidth of $\sim 440$MHz from 580 to 1020 MHz. Given this large fractional bandwidth $\dnu/\nu_c \sim 0.54$ one can estimate the spectral behaviour of a source across the band by computing in-band spectral index. DDFacet deconvolution algorithm \texttt{SSD2}, used in this work can be used to fit different orders of Taylor term to estimate the spectral behaviour. Here we have restricted to estimating Taylor term = 1 which is representative of the spectral index. These spectral index maps can be used to estimate the spectral behaviour of the sources particularly diffused emissions. Given the excellent sensitivity of MeerKAT, it is possible to construct sub-band continuum images along with the MFS images. These images are useful for understanding spectral curvature. Considering the MeerKLASS observations in the UHF-band we have subdivided the whole band into nine sub-bands of equal width ($\nu_s=$ 574, 635, 695, 755, 816, 876, 937, 997, and 1058\,MHz). We found that, due to presence of RFI, a large fraction of data is flagged in the ninth sub-band ($\nu_s=1058\,{\rm MHz}$) and we exclude that from our subsequent analysis. Although L-band has a wider bandwidth compared to UHF-band. We found that a large portion of L-band is RFI contaminated. Considering this, we have subdivided L-band onto seven sub-bands of equal width ($\nu_s=$ 917, 1039, 1161, 1283,1406, 1528 and 1651\,MHz). 
\begin{figure*}
      \begin{subfigure}{.33\textwidth}
  \centering
    \includegraphics[width=.99\linewidth]{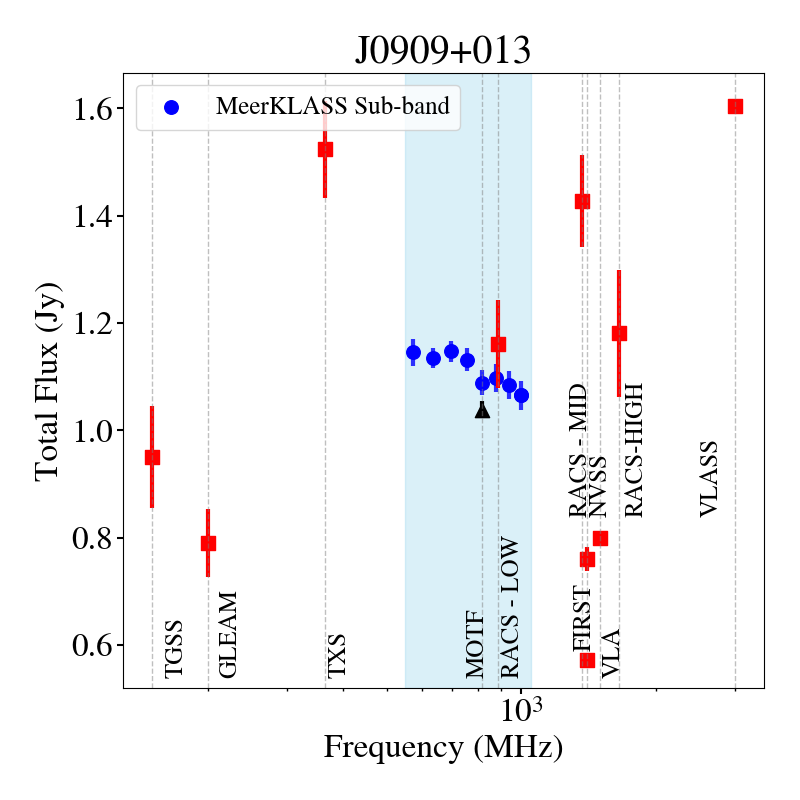}
  \end{subfigure}%
  \begin{subfigure}{.33\textwidth}
  \centering
    \includegraphics[width=.99\linewidth]{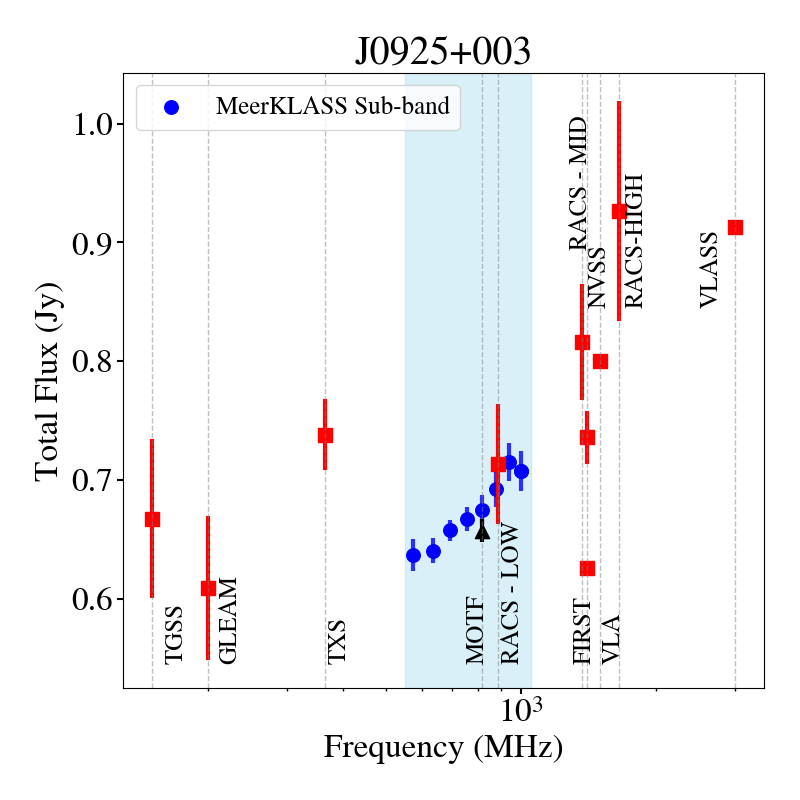}
  \end{subfigure}
    \begin{subfigure}{.33\textwidth}
  \centering
    \includegraphics[width=.99\linewidth]{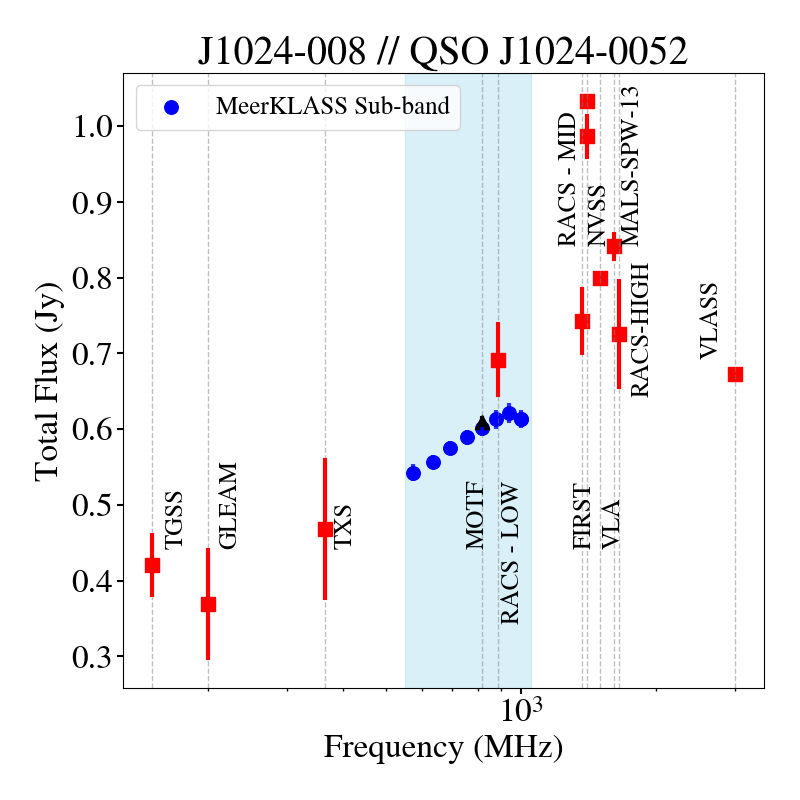}
  \end{subfigure}
  \caption{The figure shows the quality of flux recovered from sub-band images for three sources from VLA calibrator list using the M-OTF UHF-band. Blue points show the \texttt{PyBDSF} recovered total flux using M-OTF sub-band images whereas the red points show measurement from other observations.}
  \label{fig:UHFspec}
\end{figure*}

We could extract in-band source spectra using these images. We have chosen three sources from VLA calibrator list \footnote{\href{https://science.nrao.edu/facilities/vla/observing/callist}{https://science.nrao.edu/facilities/vla/observing/callist}}  to compare the fluxes with other observations that are available. \cref{fig:UHFspec} shows the results for the UHF-band where the blue circles show the M-OTF sub-band flux estimates. The shaded region show the coverage of UHF-band and triangular point show the estimate from the MFS images. Overall we found that it is difficult to match with other observations (red squares) primarily as there is only RACS-Low that overlaps partially with the MeerKLASS sub-bands. However the general trend of the flux evolution matches reasonably with other observations. The error-bars are estimated using the method described in \cref{sec:ferr} and as the these sources are very high SNR sources, the associated errors are small.
\begin{figure*}
      \begin{subfigure}{.33\textwidth}
  \centering
    \includegraphics[width=.99\linewidth]{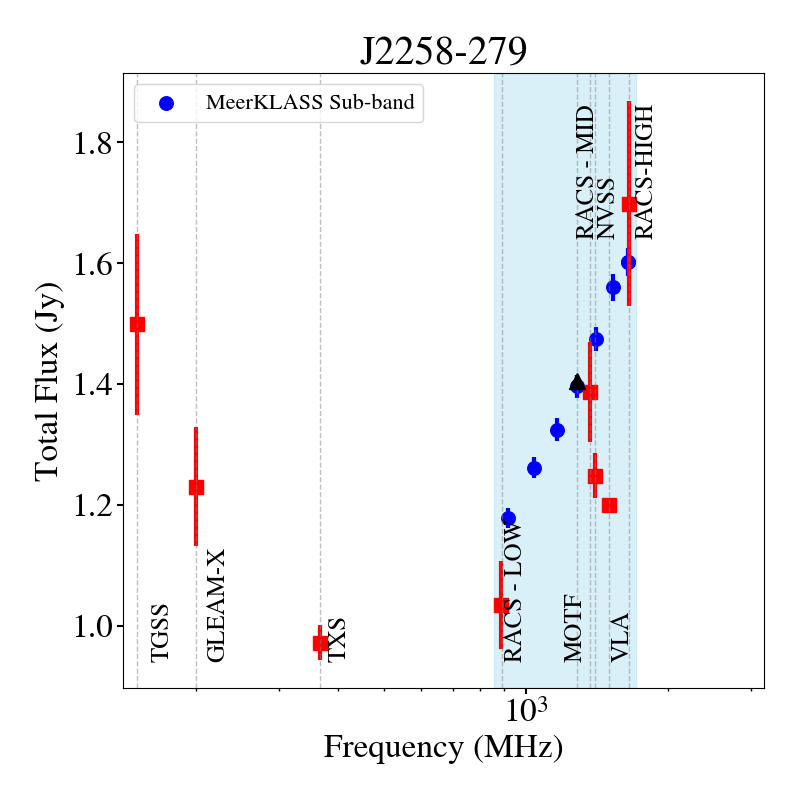}
  \end{subfigure}%
  \begin{subfigure}{.33\textwidth}
  \centering
    \includegraphics[width=.99\linewidth]{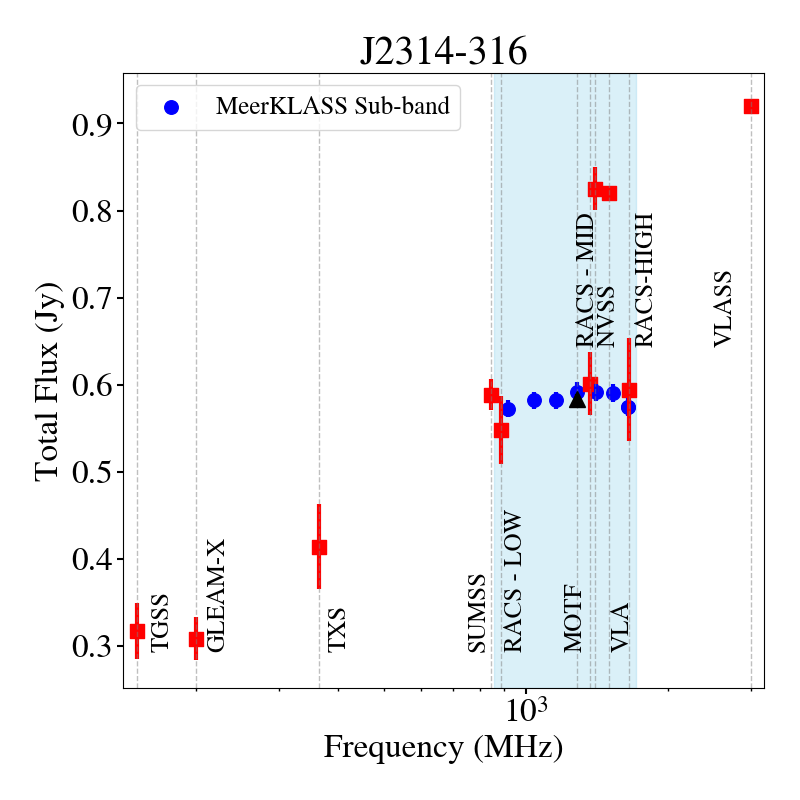}
  \end{subfigure}
    \begin{subfigure}{.33\textwidth}
  \centering
    \includegraphics[width=.99\linewidth]{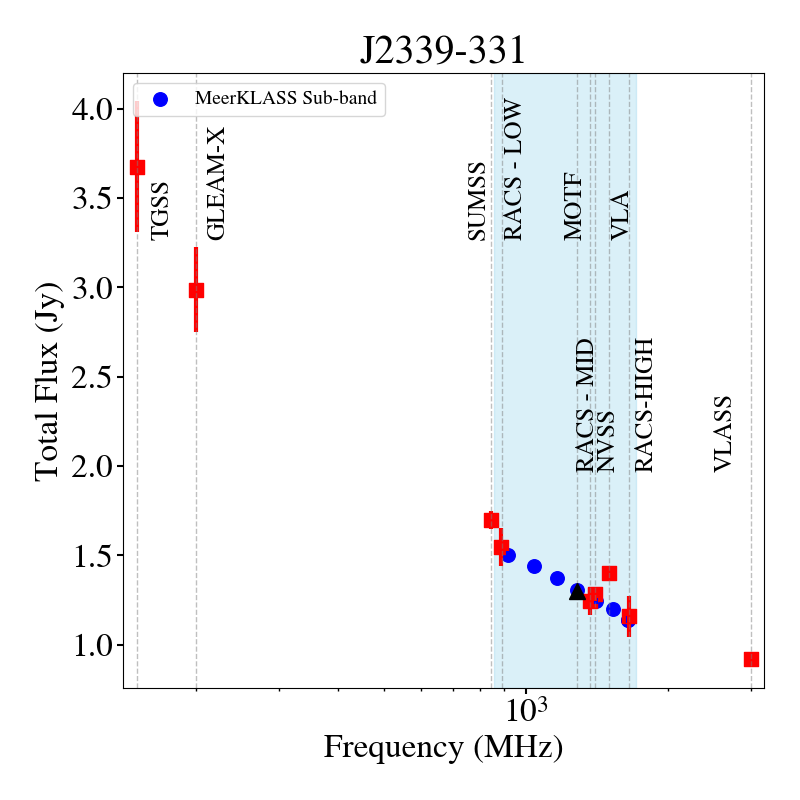}
  \end{subfigure}
  \caption{The figure shows the quality of flux recovered from sub-band images for three sources from VLA calibrator list using the M-OTF L-band. Blue points show the \texttt{PyBDSF} recovered total flux using M-OTF sub-band images whereas the red points show measurement from other observations.}
  \label{fig:Lspec}
\end{figure*}
Unlike UHF-band, sub-band images associated with the L-band are in overlap with other observations such as SUMSS, RACS-LOW, RACS-MID, NVSS, RACS-High, VLA and VLASS, the comparison are shown in the \cref{fig:Lspec}. We find that the sub-band flux estimates are in reasonable agreement with RACS and SUMSS flux values whereas the NVSS and VLA estimates varies significantly. A detailed study on these and comparison with different surveys are addressed in \citealt{Paul2025} and \citealt{Mangla2025}. Overall we see the sub-band images and the fitted spectra can reveal different features in the Southern sky, we plan to explore further in subsequent studies.

\section{MeerKLASS data release plan}\label{sec:DR}
Here we have reported on the first MeerKLASS-OTF images in the UHF and L-band of MeerKAT. The measured visibilities are usually made public with in a year of observations and can be accessed via SARAO archive portal. The results for a fraction of the first stage of the survey dated between 2021-2023 have been processed and going to be released as a part of MeerKLASS Data Release 1 (DR1). The DR1 products will contain:
\begin{enumerate}
    \item Total intensity smearing corrected MFS images at the nominal frequencies of the UHF and L-band.
    \item 9 or 7 Sub-band images for all the total intensity MFS images.
    \item Extracted source model for the MFS images, spectra index maps and the corresponding residuals.
\end{enumerate}

All images will be presented in the FITS format. We will also make the catalogue for the sources available in MeerKLASS DR1 release.

In future we will add the polarised maps for the images in the UHF band. We are working to release the 2-second snapshot images to enable work on transients using MeerKLASS OTF data sets.

\section{Summary}\label{sec:summary}
MeerKLASS is an \HI~ intensity mapping survey that aims to detect 21cm power spectrum from redshift $<1.44$. Alongside the auto-correlation (single dish) visibilities, MeerKLASS also records interferometric visibilities in On-The-Fly (OTF) mode. In this work we introduce the MeerKLASS OTF (M-OTF) survey. We have developed and tested a new calibration and imaging capability for MeerKAT in which antennas are moved back-and-forth at a constant elevation and the visibilities are recorded continuously at 2-second interval. This commensal continuum observation significantly reduces the slew-and-settle overhead compared to a traditional tracking observation. As a consequence we achieve a high survey speed of $\sim 150\,{\rm deg}^2 {\rm hr}^{-1}$ per epoch.

Each MeerKLASS observation lasts approximately 2\,hours and covers $270 - 300$ sq. degrees depending upon declination. We have developed an end-to-end pipeline that flags, calibrates, performs OTF correction and stores the visibilities in 2-second snapshot measurement sets. The relevant steps of this process are described in details in \cref{sec:otf_data}. These 2-second snapshot measurement sets are independent and standard imaging techniques can be used to produce 2-second snapshot images. These are crucial for slow transient search. However, there are limitations to M-OTF observations. In the current format, the delay centre of the observation are fixed during the scanning process, whereas the telescope pointing performs a horizontal raster scan and the earth rotates. This introduces a smearing that affects our images significantly (\cref{sec:smearing}). To overcome these limitations, we have incorporated a smearing correction during the imaging, where the dirty image is deconvolved using a smeared PSF that constructs a correct sky model. Due to this correction, the resolution of the images gets degraded (\cref{sec:img}). MeerKLASS collaboration is acting actively with SARAO to resolve this issue. 

In principle the 2-second images can be combined to achieve a deeper image that is require for producing source catalogue and other scientific objectives. But that limits our ability to deconvolve faint sources. To overcome this we have used \texttt{DDFacet} to perform joint deconvolution (visibility plane mosaicking) in our imaging pipeline. Considering upcoming DR1, we have combined measurements from 8 epochs of observations in the UHF-band. After performing smearing correction, the resulting mosaic image have a resolution of $23.3 \arcsec (\sim \sqrt{32\arcsec \times 17 \arcsec})$ with an rms sensitivity of $35~\mu{\rm Jy}/{\rm beam}$ in the deepest region. In future, with the delay tracking fix implemented, we expect to achieve a resolution of $14\arcsec$ and an average rms of $25~\mu{\rm Jy}/{\rm beam}$ in the UHF-band. However, currently do not have any plans to re-observe the sky region already observed which indicates the continuum maps will have a resolution of $~23\arcsec$ instead of targeted $14\arcsec$ in those regions.We have performed pilot observations in L-band and a subset of the L-band images will also be released in DR1. Here we have combined 8 rising epochs for imaging and after smearing correction, the resulting mosaic image has an rms sensitivity of $33~\mu{\rm Jy}/{\rm beam}$ with an average resolution of $14 \arcsec (\sim \sqrt{25\arcsec \times 8 \arcsec})$.  We expect to reach an rms of $15~\mu{\rm Jy}/{\rm beam}$ once all the data observed in this field are combined. Using this joint deconvolution, we are able to achieve completeness of 90 percent at a flux value of 0.6 mJy at the deepest location of the UHF-band observations analysed in DR1 \citep{Paul2025}. Further we have performed a thorough investigation of the source flux, positional accuracy by comparing with other overlapping available catalogues. We compare the astrometry and flux-density measurements of MeerKLASS catalogue with RACS-Mid. These comparisons show that astrometric offsets are typically sub-arcsec and small compared to the synthesize beam sizes of the M-OTF images (see \cref{fig:astro}, \ref{fig:pvr_UHF} and~\ref{fig:pvr_L}). Comparison of source  flux densities show agreement with RACS-Mid observations at high SNR, whereas the scatter in the flux-density comparison increases at low SNR regime while maintaining the average one-to-one trend (\cref{fig:phto}). We find that the flux ratio between the M-OTF \citep{Paul2025, Mangla2025} and RACS-Mid catalogue has a small variation ($< 5\%$) depending upon the distance from the correlator delay centre (\cref{fig:fvr_UHF} and \ref{fig:fvr_L}). Given that the variations are small, the catalogues are produced from different telescopes and different survey techniques, it is difficult to identify a probable cause. We plan to investigate this using MeerKAT based observations in a subsequent study.

Leveraging high sensitivity and the large bandwidth of the MeerKAT, we produce spectral map and sub-band images alongside the MFS image. The spectral maps are crucial for analysing diffused emissions whereas sub-band images are being used to identify rare sources with anomalous spectral behaviour. MeerKLASS OTF survey is an ideal example of a commensal survey where the telescope time is utilized for multiple science drivers. This potentially be helpful for producing large area sky models and slow transient searches with upcoming SKA-Mid.

\section*{Acknowledgments}

SC thanks Matthias Hoeft for their helpful feedback on the manuscript. SC acknowledges financial support from the South African National Research Foundation (Grant No. 84156) and the Inter-University Institute for Data Intensive Astronomy (IDIA). IDIA is a partnership of the University of Cape Town, the University of Pretoria and the University of the Western Cape. IDIA is registered on the Research Organization Registry with ROR ID 01edhwb26, and on Open Funder Registry with funder ID 100031500.
SM and JM acknowledge the support provided by the German Federal Ministry of Education and Research (BMBF) through the BMBF D-MeerKAT III award (number 05A23WM2). SP acknowledges support from the Science and Technology Facilities Council (STFC) through the Consolidated Grant ST/X001229/1 at the Jodrell Bank Centre for Astrophysics, University of Manchester.
OMS's research is supported by the South African Research Chairs Initiative of the Department of Science and Technology and National Research Foundation (grant No. 81737). 
The authors acknowledge the use of the ilifu cloud computing facility – www.ilifu.ac.za, a partnership between the University of Cape Town, the University of the Western Cape, Stellenbosch University, Sol Plaatje University and the Cape Peninsula University of Technology. The ilifu facility is supported by contributions from IDIA, the Computational Biology division at UCT and the Data Intensive Research Initiative of South Africa (DIRISA).
The MeerKAT telescope is operated by the South African Radio Astronomy Observatory, which is a facility of the National Research Foundation, an agency of the Department of Science and Innovation. We thank the developers of open-source Python libraries \textsc{NumPy} \citep{Numpy}, \textsc{SciPy} \citep{Scipy}, \textsc{Matplotlib} \citep{Matplotlib}, and \textsc{Astropy} \citep{astropy:2022}.

\section*{Data availability}
All the radio observation data used in this study are available in the SARAO Online Archive \href{https://archive.sarao.ac.za}{(https://archive.sarao.ac.za)} with proposal ID SCI-20210212-MS-01 and SCI-20220822-MS-01. 
\par All data products described in \cref{sec:DR} will be made available on the MeerKLASS survey webpage: \url{https://meerklass.org/}

\bibliographystyle{mnras}
\bibliography{mylist}

@misc{ACT2025,
    author = "Aguena, M. and others",
    collaboration = "ACT, DES, HSC",
    title = "{The Atacama Cosmology Telescope: DR6 Sunyaev-Zel'dovich Selected Galaxy Clusters Catalog}",
    eprint = "2507.21459",
    archivePrefix = "arXiv",
    primaryClass = "astro-ph.CO",
    reportNumber = "FERMILAB-PUB-25-0543",
    month = "7",
    year = "2025"
}

@article{Alonso2021,
    author = {Alonso, David and Bellini, Emilio and Hale, Catherine and Jarvis, Matt J and Schwarz, Dominik J},
    title = {Cross-correlating radio continuum surveys and CMB lensing: constraining redshift distributions, galaxy bias, and cosmology},
    journal = {Monthly Notices of the Royal Astronomical Society},
    volume = {502},
    number = {1},
    pages = {876-887},
    year = {2021},
    month = {01},
    issn = {0035-8711},
    doi = {10.1093/mnras/stab046},
    url = {https://doi.org/10.1093/mnras/stab046},
    eprint = {https://academic.oup.com/mnras/article-pdf/502/1/876/38869980/stab046.pdf},
}

@ARTICLE{Andersson2023,
       author = {{Andersson}, Alex and {Lintott}, Chris and {Fender}, Rob and {Bright}, Joe and {Carotenuto}, Francesco and {Driessen}, Laura and {Espinasse}, Mathilde and {Gasealahwe}, Kelebogile and {Heywood}, Ian and {van der Horst}, Alexander J. and {Motta}, Sara and {Rhodes}, Lauren and {Tremou}, Evangelia and {Williams}, David R.~A. and {Woudt}, Patrick and {Zhang}, Xian and {Bloemen}, Steven and {Groot}, Paul and {Vreeswijk}, Paul and {Giarratana}, Stefano and {Saikia}, Payaswini and {Andersson}, Jonas and {Ruiz Arroyo}, Lizzeth and {Baert}, Lo{\"\i}c and {Baumann}, Matthew and {Domainko}, Wilfried and {Eschweiler}, Thorsten and {Forsythe}, Tim and {Gaudenzi}, Sauro and {Ann Grenier}, Rachel and {Iannone}, Davide and {Lahoz}, Karla and {Melville}, Kyle J. and {De Sousa Nascimento}, Marianne and {Navarro}, Leticia and {Parthasarathi}, Sai and {Piilonen} and {Rahman}, Najma and {Smith}, Jeffrey and {Stewart}, B. and {Temoke}, Newton and {Tworek}, Chloe and {Whittle}, Isabelle},
        title = "{Bursts from Space: MeerKAT - the first citizen science project dedicated to commensal radio transients}",
      journal = {\mnras},
         year = 2023,
        month = aug,
       volume = {523},
       number = {2},
        pages = {2219-2235},
          doi = {10.1093/mnras/stad1298},
archivePrefix = {arXiv},
       eprint = {2304.14157},
 primaryClass = {astro-ph.HE},
       adsurl = {https://ui.adsabs.harvard.edu/abs/2023MNRAS.523.2219A}
}

@article{Asad2021,
    author = {Asad, K M B and Girard, J N and de Villiers, M and Ansah-Narh, T and Iheanetu, K and Smirnov, O and Santos, M G and Lehmensiek, R and Jonas, J and de Villiers, D I L and Thorat, K and Hugo, B and Makhathini, S and Jozsa, G I G and Sirothia, S K},
    title = {Primary beam effects of radio astronomy antennas – II. Modelling MeerKAT L-band beams},
    journal = {Monthly Notices of the Royal Astronomical Society},
    volume = {502},
    number = {2},
    pages = {2970-2983},
    year = {2021},
    month = {01},
    issn = {0035-8711},
    doi = {10.1093/mnras/stab104},
    url = {https://doi.org/10.1093/mnras/stab104},
    eprint = {https://academic.oup.com/mnras/article-pdf/502/2/2970/36276175/stab104.pdf},
}

@article{Baan2010,
  author = "Baan, Willem",
  title = "{RFI Mitigation in Radio Astronomy (invited)}",
  doi = "10.22323/1.107.0001",
  journal = "PoS",
  year = 2010,
  volume = "RFI2010",
  pages = "001"
}

@article{Battye2004,
    author = {Battye, Richard A. and Davies, Rod D. and Weller, Jochen},
    title = {Neutral hydrogen surveys for high-redshift galaxy clusters and protoclusters},
    journal = {Monthly Notices of the Royal Astronomical Society},
    volume = {355},
    number = {4},
    pages = {1339-1347},
    year = {2004},
    month = {12},
    issn = {0035-8711},
    doi = {10.1111/j.1365-2966.2004.08416.x},
    url = {https://doi.org/10.1111/j.1365-2966.2004.08416.x},
    eprint = {https://academic.oup.com/mnras/article-pdf/355/4/1339/6275970/355-4-1339.pdf},
}

@ARTICLE{Bharadwaj2001b,
   author = {{Bharadwaj}, S. and {Nath}, B.~B. and {Sethi}, S.~K.},
    title = "{Using HI to Probe Large Scale Structures at z \~{} 3}",
  journal = {J. Astrophys. Astron.},
   eprint = {astro-ph/0003200},
     year = 2001,
    month = mar,
   volume = 22,
    pages = {21},
      doi = {10.1007/BF02933588},
   adsurl = {http://adsabs.harvard.edu/abs/2001JApA...22...21B}
}

@article{Bock1999,
doi = {10.1086/300786},
url = {https://dx.doi.org/10.1086/300786},
year = {1999},
month = {mar},
publisher = {},
volume = {117},
number = {3},
pages = {1578},
author = {Bock, D. C.-J. and Large, M. I. and Sadler, Elaine M.},
title = {SUMSS: A Wide-Field Radio Imaging Survey of the Southern Sky.
I. Science Goals, Survey Design, and Instrumentation},
journal = {The Astronomical Journal}
}

@misc{Braun2019,
      title={Anticipated Performance of the Square Kilometre Array -- Phase 1 (SKA1)}, 
      author={Robert Braun and Anna Bonaldi and Tyler Bourke and Evan Keane and Jeff Wagg},
      year={2019},
      eprint={1912.12699},
      archivePrefix={arXiv},
      primaryClass={astro-ph.IM},
      url={https://arxiv.org/abs/1912.12699}, 
}

@PHDTHESIS{Briggs1995,
       author = {{Briggs}, Daniel Shenon},
        title = "{High fidelity deconvolution of moderately resolved sources}",
       school = {New Mexico Institute of Mining and Technology},
         year = 1995,
        month = jan,
       adsurl = {https://ui.adsabs.harvard.edu/abs/1995PhDT.......238B}
}

@ARTICLE{Caleb2022,
       author = {{Caleb}, Manisha and {Heywood}, Ian and {Rajwade}, Kaustubh and {Malenta}, Mateusz and {Stappers}, Benjamin Willem and {Barr}, Ewan and {Chen}, Weiwei and {Morello}, Vincent and {Sanidas}, Sotiris and {van den Eijnden}, Jakob and {Kramer}, Michael and {Buckley}, David and {Brink}, Jaco and {Motta}, Sara Elisa and {Woudt}, Patrick and {Weltevrede}, Patrick and {Jankowski}, Fabian and {Surnis}, Mayuresh and {Buchner}, Sarah and {Bezuidenhout}, Mechiel Christiaan and {Driessen}, Laura Nicole and {Fender}, Rob},
        title = "{Discovery of a radio-emitting neutron star with an ultra-long spin period of 76 s}",
      journal = {Nature Astronomy},
         year = 2022,
        month = may,
       volume = {6},
        pages = {828-836},
          doi = {10.1038/s41550-022-01688-x},
archivePrefix = {arXiv},
       eprint = {2206.01346},
 primaryClass = {astro-ph.HE},
       adsurl = {https://ui.adsabs.harvard.edu/abs/2022NatAs...6..828C}
}

@INPROCEEDINGS{caracal,
       author = {{J{\'o}zsa}, G.~I.~G. and {White}, S.~V. and {Thorat}, K. and {Smirnov}, O.~M. and {Serra}, P. and {Ramatsoku}, M. and {Ramaila}, A.~J.~T. and {Perkins}, S.~J. and {Maccagni}, F.~M. and {Makhathini}, S. and {Moln{\'a}r}, D.~C. and {Kamphuis}, P. and {Kleiner}, D. and {Hugo}, B.~V. and {de Blok}, W.~J.~G. and {Andati}, L.~A.~L.},
        title = "{MeerKATHI - an End-to-End Data Reduction Pipeline for MeerKAT and Other Radio Telescopes}",
    booktitle = {Astronomical Data Analysis Software and Systems XXIX},
         year = 2020,
       editor = {{Pizzo}, R. and {Deul}, E.~R. and {Mol}, J.~D. and {de Plaa}, J. and {Verkouter}, H.},
       series = {Astronomical Society of the Pacific Conference Series},
       volume = {527},
        month = jan,
        pages = {635},
          doi = {10.48550/arXiv.2006.02955},
archivePrefix = {arXiv},
       eprint = {2006.02955},
 primaryClass = {astro-ph.IM},
       adsurl = {https://ui.adsabs.harvard.edu/abs/2020ASPC..527..635J}
}

@ARTICLE{Chang2008,
   author = {{Chang}, T.-C. and {Pen}, U.-L. and {Peterson}, J.~B. and {McDonald}, P.
	},
    title = "{Baryon Acoustic Oscillation Intensity Mapping of Dark Energy}",
  journal = {Physical Review Letters},
archivePrefix = "arXiv",
   eprint = {0709.3672},
     year = 2008,
    month = mar,
   volume = 100,
   number = 9,
      eid = {091303},
    pages = {091303},
      doi = {10.1103/PhysRevLett.100.091303},
   adsurl = {http://ads.nao.ac.jp/abs/2008PhRvL.100i1303C}
}

@article{Paul2025,
  author = {Paul, Sourabh and MeerKLASS Team},
  title = {The MeerKLASS UHF Continuum Survey – Data Release I},
  journal = {to be submitted {\mnras}},
  year = {2025}
}

@article{Mangla2025,
  author = {Mangla, Sarvesh and MeerKLASS Team},
  title = {The MeerKLASS L-band OTF Continuum Survey: Stokes I Images and Catalogue for the Initial Data Release},
  journal = {to be submitted {\mnras}},
  year = {2025}
}

@ARTICLE{Condon1998,
       author = {{Condon}, J.~J. and {Cotton}, W.~D. and {Greisen}, E.~W. and {Yin}, Q.~F. and {Perley}, R.~A. and {Taylor}, G.~B. and {Broderick}, J.~J.},
        title = "{The NRAO VLA Sky Survey}",
      journal = {\aj},
         year = 1998,
        month = may,
       volume = {115},
       number = {5},
        pages = {1693-1716},
          doi = {10.1086/300337},
       adsurl = {https://ui.adsabs.harvard.edu/abs/1998AJ....115.1693C}
}

@ARTICLE{Conway1990,
       author = {{Conway}, J.~E. and {Cornwell}, T.~J. and {Wilkinson}, P.~N.},
        title = "{Multi-frequency synthesis : a new technique in radio interferometrie imaging.}",
      journal = {\mnras},
         year = 1990,
        month = oct,
       volume = {246},
        pages = {490},
       adsurl = {https://ui.adsabs.harvard.edu/abs/1990MNRAS.246..490C}
}

@article{Cunnington2022,
    author = {Cunnington, Steven and Li, Yichao and Santos, Mario G and Wang, Jingying and Carucci, Isabella P and Irfan, Melis O and Pourtsidou, Alkistis and Spinelli, Marta and Wolz, Laura and Soares, Paula S and Blake, Chris and Bull, Philip and Engelbrecht, Brandon and Fonseca, José and Grainge, Keith and Ma, Yin-Zhe},
    title = "{H i intensity mapping with MeerKAT: power spectrum detection in cross-correlation with WiggleZ galaxies}",
    journal = {\mnras},
    volume = {518},
    number = {4},
    pages = {6262-6272},
    year = {2022},
    month = {10},
    issn = {0035-8711},
    doi = {10.1093/mnras/stac3060},
    url = {https://doi.org/10.1093/mnras/stac3060},
    eprint = {https://academic.oup.com/mnras/article-pdf/518/4/6262/48302259/stac3060.pdf},
}

@article{Cunnington2025,
    author = {MeerKLASS Collaboration  and Bernal, José L and Bull, Philip and Camera, Stefano and Carucci, Isabella P and Chen, Zhaoting and Cunnington, Steven and Engelbrecht, Brandon N and Fonseca, José and Grainge, Keith and Irfan, Melis O and Li, Yichao and Mazumder, Aishrila and Paul, Sourabh and Pourtsidou, Alkistis and Santos, Mario G and Spinelli, Marta and Wang, Jingying and Witzemann, Amadeus and Wolz, Laura},
    title = {MeerKLASS L-band deep-field intensity maps: entering the H i dominated regime},
    journal = {Monthly Notices of the Royal Astronomical Society},
    volume = {537},
    number = {4},
    pages = {3632-3661},
    year = {2025},
    month = {02},
    issn = {0035-8711},
    doi = {10.1093/mnras/staf195},
    url = {https://doi.org/10.1093/mnras/staf195},
    eprint = {https://academic.oup.com/mnras/article-pdf/537/4/3632/61743817/staf195.pdf},
}

@misc{Cunnington2025b,
      title={Revealing cosmological fluctuations in 21cm intensity maps with MeerKLASS: from maps to power spectra}, 
      author={Steven Cunnington and Matilde Barberi-Squarotti and José Luis Bernal and Stefano Camera and Isabella P. Carucci and Zhaoting Chen and José Fonseca and Mario Santos and Marta Spinelli and Jingying Wang and Laura Wolz},
      year={2025},
      eprint={2510.27549},
      archivePrefix={arXiv},
      primaryClass={astro-ph.CO},
      url={https://arxiv.org/abs/2510.27549}, 
}

@article{Deka2024,
doi = {10.3847/1538-4365/acf7b9},
url = {https://dx.doi.org/10.3847/1538-4365/acf7b9},
year = {2024},
month = {feb},
publisher = {The American Astronomical Society},
volume = {270},
number = {2},
pages = {33},
author = {Deka, P. P. and Gupta, N. and Jagannathan, P. and Sekhar, S. and Momjian, E. and Bhatnagar, S. and Wagenveld, J. and Klöckner, H.-R. and Jose, J. and Balashev, S. A. and Combes, F. and Hilton, M. and Borgaonkar, D. and Chatterjee, A. and Emig, K. L. and Gaunekar, A. N. and Józsa, G. I. G. and Klutse, D. Y. and Knowles, K. and Krogager, J.-K. and Mohapatra, A. and Moodley, K. and Muller, Sébastien and Noterdaeme, P. and Petitjean, P. and Salas, P. and Sikhosana, S.},
title = {The MeerKAT Absorption Line Survey (MALS) Data Release. I. Stokes I Image Catalogs at 1–1.4 GHz},
journal = {The Astrophysical Journal Supplement Series}
}

@ARTICLE{DESI_Y1-2025,
       author = {{DESI Collaboration} and {Abdul-Karim}, M. and {Adame}, A.~G. and {Aguado}, D. and {Aguilar}, J. and {Ahlen}, S. and {Alam}, S. and {Aldering}, G. and {Alexander}, D.~M. and {Alfarsy}, R. and {Allen}, L. and {Allende Prieto}, C. and {Alves}, O. and {Anand}, A. and {Andrade}, U. and {Armengaud}, E. and {Avila}, S. and {Aviles}, A. and {Awan}, H. and {Bailey}, S. and {Baleato Lizancos}, A. and {Ballester}, O. and {Bault}, A. and {Bautista}, J. and {BenZvi}, S. and {Beraldo e Silva}, L. and {Bermejo-Climent}, J.~R. and {Beutler}, F. and {Bianchi}, D. and {Blake}, C. and {Blum}, R. and {Bolton}, A.~S. and {Bonici}, M. and {Brieden}, S. and {Brodzeller}, A. and {Brooks}, D. and {Buckley-Geer}, E. and {Burtin}, E. and {Canning}, R. and {Carnero Rosell}, A. and {Carr}, A. and {Carrilho}, P. and {Casas}, L. and {Castander}, F.~J. and {Cereskaite}, R. and {Cervantes-Cota}, J.~L. and {Chaussidon}, E. and {Chaves-Montero}, J. and {Chen}, S. and {Chen}, X. and {Claybaugh}, T. and {Cole}, S. and {Cooper}, A.~P. and {Cousinou}, M. -C. and {Cuceu}, A. and {Davis}, T.~M. and {Dawson}, K.~S. and {de Belsunce}, R. and {de la Cruz}, R. and {de la Macorra}, A. and {de Mattia}, A. and {Deiosso}, N. and {Della Costa}, J. and {Demina}, R. and {Demirbozan}, U. and {DeRose}, J. and {Dey}, A. and {Dey}, B. and {Ding}, J. and {Ding}, Z. and {Doel}, P. and {Douglass}, K. and {Dowicz}, M. and {Ebina}, H. and {Edelstein}, J. and {Eisenstein}, D.~J. and {Elbers}, W. and {Emas}, N. and {Escoffier}, S. and {Fagrelius}, P. and {Fan}, X. and {Fanning}, K. and {Fawcett}, V.~A. and {Fern\textbackslash'andez-Garc\textbackslash'ia}, E. and {Ferraro}, S. and {Findlay}, N. and {Font-Ribera}, A. and {Forero-Romero}, J.~E. and {Forero-S\textbackslash'anchez}, D. and {Frenk}, C.~S. and {G\textbackslash''ansicke}, B.~T. and {Galbany}, L. and {Garc\textbackslash'ia-Bellido}, J. and {Garcia-Quintero}, C. and {Garrison}, L.~H. and {Gazta\textbackslash\raisebox{-0.5ex}\textasciitildenaga}, E. and {Gil-Mar\textbackslash'in}, H. and {Gnedin}, O.~Y. and {Gontcho}, S. Gontcho A and {Gonzalez-Morales}, A.~X. and {Gonzalez-Perez}, V. and {Gordon}, C. and {Graur}, O. and {Green}, D. and {Gruen}, D. and {Gsponer}, R. and {Guandalin}, C. and {Gutierrez}, G. and {Guy}, J. and {Hahn}, C. and {Han}, J.~J. and {Han}, J. and {He}, S. and {Herrera-Alcantar}, H.~K. and {Honscheid}, K. and {Hou}, J. and {Howlett}, C. and {Huterer}, D. and {Ir\textbackslashv\{s\}i\textbackslashv\{c\}}, V. and {Ishak}, M. and {Jacques}, A. and {Jimenez}, J. and {Jing}, Y.~P. and {Joachimi}, B. and {Joudaki}, S. and {Joyce}, R. and {Jullo}, E. and {Juneau}, S. and {Kara\textbackslashc\{c\}ayl\{\textbackslashi\}}, N.~G. and {Karim}, T. and {Kehoe}, R. and {Kent}, S. and {Khederlarian}, A. and {Kirkby}, D. and {Kisner}, T. and {Kitaura}, F. -S. and {Kizhuprakkat}, N. and {Kong}, H. and {Koposov}, S.~E. and {Kremin}, A. and {Krolewski}, A. and {Lahav}, O. and {Lai}, Y. and {Lamman}, C. and {Lan}, T. -W. and {Landriau}, M. and {Lang}, D. and {Lange}, J.~U. and {Lasker}, J. and {Le Goff}, J.~M. and {Le Guillou}, L. and {Leauthaud}, A. and {Levi}, M.~E. and {Li}, S. and {Li}, T.~S. and {Lodha}, K. and {Lokken}, M. and {Luo}, Y. and {Magneville}, C. and {Manera}, M. and {Manser}, C.~J. and {Margala}, D. and {Martini}, P. and {Maus}, M. and {McCullough}, J. and {McDonald}, P. and {Medina}, G.~E. and {Medina-Varela}, L. and {Meisner}, A. and {Mena-Fern\textbackslash'andez}, J. and {Menegas}, A. and {Mezcua}, M. and {Miquel}, R. and {Montero-Camacho}, P. and {Moon}, J. and {Moustakas}, J. and {Mu\textbackslash\raisebox{-0.5ex}\textasciitildenoz-Guti\textbackslash'errez}, A. and {Mu\textbackslash\raisebox{-0.5ex}\textasciitildenoz-Santos}, D. and {Myers}, A.~D. and {Myles}, J. and {Nadathur}, S. and {Najita}, J. and {Napolitano}, L. and {Newman}, J.~A. and {Nikakhtar}, F. and {Nikutta}, R. and {Niz}, G. and {Noriega}, H.~E. and {Padmanabhan}, N. and {Paillas}, E. and {Palanque-Delabrouille}, N. and {Palmese}, A. and {Pan}, J. and {Pan}, Z. and {Parkinson}, D. and {Peacock}, J. and {Percival}, W.~J. and {P\textbackslash'erez-Fern\textbackslash'andez}, A. and {P\textbackslash'erez-R\textbackslash`afols}, I. and {Peterson}, P.},
        title = "{Data Release 1 of the Dark Energy Spectroscopic Instrument}",
      journal = {arXiv e-prints},
         year = 2025,
        month = mar,
          eid = {arXiv:2503.14745},
        pages = {arXiv:2503.14745},
          doi = {10.48550/arXiv.2503.14745},
archivePrefix = {arXiv},
       eprint = {2503.14745},
 primaryClass = {astro-ph.CO},
       adsurl = {https://ui.adsabs.harvard.edu/abs/2025arXiv250314745D}
}

@article{deVilliers2023,
doi = {10.3847/1538-3881/acabc3},
url = {https://dx.doi.org/10.3847/1538-3881/acabc3},
year = {2023},
month = {feb},
publisher = {The American Astronomical Society},
volume = {165},
number = {3},
pages = {78},
author = {de Villiers, Mattieu S.},
title = {MeerKAT Holography Measurements in the UHF, L, and S Bands},
journal = {The Astronomical Journal}
}

@article{Duchesne2023,
   title={The Rapid ASKAP Continuum Survey IV: continuum imaging at 1367.5 MHz and the first data release of RACS-mid},
   volume={40},
   ISSN={1448-6083},
   url={http://dx.doi.org/10.1017/pasa.2023.31},
   DOI={10.1017/pasa.2023.31},
   journal={Publications of the Astronomical Society of Australia},
   publisher={Cambridge University Press (CUP)},
   author={Duchesne, S. W. and Thomson, A. J. M. and Pritchard, J. and Lenc, E. and Moss, V. A. and McConnell, D. and Wieringa, M. H. and Whiting, M. T. and Wang, Z. and Wang, Y. and Rose, K. and Raja, W. and Murphy, Tara and Leung, J. K. and Huynh, M. T. and Hotan, A. W. and Hodgson, T. and Heald, G. H.},
   year={2023} }

@article{duchesne2025,
  title={The Rapid ASKAP Continuum Survey (RACS) VI: The RACS-high 1655.5 MHz images and catalogue},
  author={Duchesne, SW and Ross, K and Thomson, AJM and Lenc, E and Murphy, Tara and Galvin, TJ and Hotan, AW and Moss, VA and Whiting, Matthew T},
  journal={arXiv preprint arXiv:2501.04978},
  year={2025}}

@article{Engelbrecht2024,
    author = {Engelbrecht, Brandon N and Santos, Mario G and Fonseca, José and Li, Yichao and Wang, Jingying and Irfan, Melis O and Harper, Stuart E and Grainge, Keith and Bull, Philip and Carucci, Isabella P and Cunnington, Steven and Pourtsidou, Alkistis and Spinelli, Marta and Wolz, Laura},
    title = {Radio frequency interference from radio navigation satellite systems: simulations and comparison to MeerKAT single-dish data},
    journal = {Monthly Notices of the Royal Astronomical Society},
    volume = {536},
    number = {1},
    pages = {1035-1055},
    year = {2024},
    month = {11},
    issn = {0035-8711},
    doi = {10.1093/mnras/stae2649},
    url = {https://doi.org/10.1093/mnras/stae2649},
    eprint = {https://academic.oup.com/mnras/article-pdf/536/1/1035/61049770/stae2649.pdf},
}

@INPROCEEDINGS{Gupta2016,
       author = {{Gupta}, N. and {Srianand}, R. and {Baan}, W. and {Baker}, A.~J. and {Beswick}, R.~J. and {Bhatnagar}, S. and {Bhattacharya}, D. and {Bosma}, A. and {Carilli}, C. and {Cluver}, M. and {Combes}, F. and {Cress}, C. and {Dutta}, R. and {Fynbo}, J. and {Heald}, G. and {Hilton}, M. and {Hussain}, T. and {Jarvis}, M. and {Jozsa}, G. and {Kamphuis}, P. and {Kembhavi}, A. and {Kerp}, J. and {Kloeckner}, H.~R. and {Krogager}, J. and {Kulkarni}, V.~P. and {Ledoux}, C. and {Mahabal}, A. and {Mauch}, T. and {Moodley}, K. and {Momjian}, E. and {Morganti}, R. and {Noterdaeme}, P. and {Oosterloo}, T. and {Petitjean}, P. and {Schroeder}, A. and {Serra}, P. and {Sievers}, J. and {Spekkens}, K. and {Vaisanen}, P. and {van der Hulst}, T. and {Vivek}, M. and {Wang}, J. and {Wong}, O.~I. and {Zungu}, A.~R.},
        title = "{The MeerKAT Absorption Line Survey (MALS)}",
    booktitle = {MeerKAT Science: On the Pathway to the SKA},
         year = 2016,
        month = jan,
          eid = {14},
        pages = {14},
          doi = {10.22323/1.277.0014},
archivePrefix = {arXiv},
       eprint = {1708.07371},
 primaryClass = {astro-ph.GA},
       adsurl = {https://ui.adsabs.harvard.edu/abs/2016mks..confE..14G}
}

@article{Hale2021,
   title={The Rapid ASKAP Continuum Survey Paper II: First Stokes I Source Catalogue Data Release},
   volume={38},
   ISSN={1448-6083},
   url={http://dx.doi.org/10.1017/pasa.2021.47},
   DOI={10.1017/pasa.2021.47},
   journal={Publications of the Astronomical Society of Australia},
   publisher={Cambridge University Press (CUP)},
   author={Hale, Catherine L. and McConnell, D. and Thomson, A. J. M. and Lenc, E. and Heald, G. H. and Hotan, A. W. and Leung, J. K. and Moss, V. A. and Murphy, T. and Pritchard, J. and Sadler, E. M. and Stewart, A. J. and Whiting, M. T.},
   year={2021} }

@misc{Hale2024,
      title={MIGHTEE: The Continuum Survey Data Release 1}, 
      author={C. L. Hale and I. Heywood and M. J. Jarvis and I. H. Whittam and P. N. Best and Fangxia An and R. A. A. Bowler and I. Harrison and A. Matthews and D. J. B. Smith and A. R. Taylor and M. Vaccari},
      year={2024},
      eprint={2411.04958},
      archivePrefix={arXiv},
      primaryClass={astro-ph.GA},
      url={https://arxiv.org/abs/2411.04958}, 
}

@article{Harper2018,
    author = {Harper, Stuart E and Dickinson, Clive},
    title = {Potential impact of global navigation satellite services on total power H i intensity mapping surveys},
    journal = {Monthly Notices of the Royal Astronomical Society},
    volume = {479},
    number = {2},
    pages = {2024-2036},
    year = {2018},
    month = {06},
    issn = {0035-8711},
    doi = {10.1093/mnras/sty1495},
    url = {https://doi.org/10.1093/mnras/sty1495},
    eprint = {https://academic.oup.com/mnras/article-pdf/479/2/2024/25142146/sty1495.pdf},
}

@article{Heckman2014,
   author = "Heckman, Timothy M. and Best, Philip N.",
   title = "The Coevolution of Galaxies and Supermassive Black Holes: Insights from Surveys of the Contemporary Universe", 
   journal= "Annual Review of Astronomy and Astrophysics",
   year = "2014",
   volume = "52",
   number = "Volume 52, 2014",
   pages = "589-660",
   doi = "https://doi.org/10.1146/annurev-astro-081913-035722",
   url = "https://www.annualreviews.org/content/journals/10.1146/annurev-astro-081913-035722",
   publisher = "Annual Reviews",
   issn = "1545-4282",
   type = "Journal Article",
  }

@article{Heywood2021,
   title={MIGHTEE: total intensity radio continuum imaging and the COSMOS/XMM-LSS Early Science fields},
   volume={509},
   ISSN={1365-2966},
   url={http://dx.doi.org/10.1093/mnras/stab3021},
   DOI={10.1093/mnras/stab3021},
   number={2},
   journal={Monthly Notices of the Royal Astronomical Society},
   publisher={Oxford University Press (OUP)},
   author={Heywood, I and Jarvis, M J and Hale, C L and Whittam, I H and Bester, H L and Hugo, B and Kenyon, J S and Prescott, M and Smirnov, O M and Tasse, C and Afonso, J M and Best, P N and Collier, J D and Deane, R P and Frank, B S and Hardcastle, M J and Knowles, K and Maddox, N and Murphy, E J and Prandoni, I and Randriamampandry, S M and Santos, M G and Sekhar, S and Tabatabaei, F and Taylor, A R and Thorat, K},
   year={2021},
   month=oct, pages={2150–2168} }

@ARTICLE{Hilton2021,
       author = {{Hilton}, M. and {Sif{\'o}n}, C. and {Naess}, S. and {Madhavacheril}, M. and {Oguri}, M. and {Rozo}, E. and {Rykoff}, E. and {Abbott}, T.~M.~C. and {Adhikari}, S. and {Aguena}, M. and {Aiola}, S. and {Allam}, S. and {Amodeo}, S. and {Amon}, A. and {Annis}, J. and {Ansarinejad}, B. and {Aros-Bunster}, C. and {Austermann}, J.~E. and {Avila}, S. and {Bacon}, D. and {Battaglia}, N. and {Beall}, J.~A. and {Becker}, D.~T. and {Bernstein}, G.~M. and {Bertin}, E. and {Bhandarkar}, T. and {Bhargava}, S. and {Bond}, J.~R. and {Brooks}, D. and {Burke}, D.~L. and {Calabrese}, E. and {Carrasco Kind}, M. and {Carretero}, J. and {Choi}, S.~K. and {Choi}, A. and {Conselice}, C. and {da Costa}, L.~N. and {Costanzi}, M. and {Crichton}, D. and {Crowley}, K.~T. and {D{\"u}nner}, R. and {Denison}, E.~V. and {Devlin}, M.~J. and {Dicker}, S.~R. and {Diehl}, H.~T. and {Dietrich}, J.~P. and {Doel}, P. and {Duff}, S.~M. and {Duivenvoorden}, A.~J. and {Dunkley}, J. and {Everett}, S. and {Ferraro}, S. and {Ferrero}, I. and {Fert{\'e}}, A. and {Flaugher}, B. and {Frieman}, J. and {Gallardo}, P.~A. and {Garc{\'\i}a-Bellido}, J. and {Gaztanaga}, E. and {Gerdes}, D.~W. and {Giles}, P. and {Golec}, J.~E. and {Gralla}, M.~B. and {Grandis}, S. and {Gruen}, D. and {Gruendl}, R.~A. and {Gschwend}, J. and {Gutierrez}, G. and {Han}, D. and {Hartley}, W.~G. and {Hasselfield}, M. and {Hill}, J.~C. and {Hilton}, G.~C. and {Hincks}, A.~D. and {Hinton}, S.~R. and {Ho}, S. -P.~P. and {Honscheid}, K. and {Hoyle}, B. and {Hubmayr}, J. and {Huffenberger}, K.~M. and {Hughes}, J.~P. and {Jaelani}, A.~T. and {Jain}, B. and {James}, D.~J. and {Jeltema}, T. and {Kent}, S. and {Knowles}, K. and {Koopman}, B.~J. and {Kuehn}, K. and {Lahav}, O. and {Lima}, M. and {Lin}, Y. -T. and {Lokken}, M. and {Loubser}, S.~I. and {MacCrann}, N. and {Maia}, M.~A.~G. and {Marriage}, T.~A. and {Martin}, J. and {McMahon}, J. and {Melchior}, P. and {Menanteau}, F. and {Miquel}, R. and {Miyatake}, H. and {Moodley}, K. and {Morgan}, R. and {Mroczkowski}, T. and {Nati}, F. and {Newburgh}, L.~B. and {Niemack}, M.~D. and {Nishizawa}, A.~J. and {Ogando}, R.~L.~C. and {Orlowski-Scherer}, J. and {Page}, L.~A. and {Palmese}, A. and {Partridge}, B. and {Paz-Chinch{\'o}n}, F. and {Phakathi}, P. and {Plazas}, A.~A. and {Robertson}, N.~C. and {Romer}, A.~K. and {Carnero Rosell}, A. and {Salatino}, M. and {Sanchez}, E. and {Schaan}, E. and {Schillaci}, A. and {Sehgal}, N. and {Serrano}, S. and {Shin}, T. and {Simon}, S.~M. and {Smith}, M. and {Soares-Santos}, M. and {Spergel}, D.~N. and {Staggs}, S.~T. and {Storer}, E.~R. and {Suchyta}, E. and {Swanson}, M.~E.~C. and {Tarle}, G. and {Thomas}, D. and {To}, C. and {Trac}, H. and {Ullom}, J.~N. and {Vale}, L.~R. and {Van Lanen}, J. and {Vavagiakis}, E.~M. and {De Vicente}, J. and {Wilkinson}, R.~D. and {Wollack}, E.~J. and {Xu}, Z. and {Zhang}, Y.},
        title = "{The Atacama Cosmology Telescope: A Catalog of >4000 Sunyaev-Zel{\textquoteright}dovich Galaxy Clusters}",
      journal = {\apjs},
         year = 2021,
        month = mar,
       volume = {253},
       number = {1},
          eid = {3},
        pages = {3},
          doi = {10.3847/1538-4365/abd023},
archivePrefix = {arXiv},
       eprint = {2009.11043},
 primaryClass = {astro-ph.CO},
       adsurl = {https://ui.adsabs.harvard.edu/abs/2021ApJS..253....3H}
}

@article{Hopkins_2025, 
title={The Evolutionary Map of the Universe: A new radio atlas for the southern hemisphere sky}, 
volume={42}, 
DOI={10.1017/pasa.2025.10042}, 
journal={Publications of the Astronomical Society of Australia}, 
author={Hopkins, Andrew and Kapinska, Anna and Marvil, Joshua and Vernstrom, Tessa and Collier, Jordan and Norris, Ray and Gordon, Yjan and Duchesne, Stefan and Rudnick, Lawrence and Gupta, Nikhel and et al.}, 
year={2025}, 
pages={e071}
}

@INPROCEEDINGS{Hugo2022,
       author = {{Hugo}, Benjamin V. and {Perkins}, S. and {Merry}, B. and {Mauch}, T. and {Smirnov}, O.~M.},
        title = "{Tricolour: An Optimized SumThreshold Flagger for MeerKAT}",
    booktitle = {Astronomical Data Analysis Software and Systems XXX},
         year = 2022,
       editor = {{Ruiz}, Jose Enrique and {Pierfedereci}, Francesco and {Teuben}, Peter},
       series = {Astronomical Society of the Pacific Conference Series},
       volume = {532},
        month = jul,
        pages = {541},
          doi = {10.48550/arXiv.2206.09179},
archivePrefix = {arXiv},
       eprint = {2206.09179},
 primaryClass = {astro-ph.IM},
       adsurl = {https://ui.adsabs.harvard.edu/abs/2022ASPC..532..541H}
}

@article{Hurley-Walker2022,
   title={GaLactic and Extragalactic All-sky Murchison Widefield Array survey eXtended (GLEAM-X) I: Survey description and initial data release},
   volume={39},
   ISSN={1448-6083},
   url={http://dx.doi.org/10.1017/pasa.2022.17},
   DOI={10.1017/pasa.2022.17},
   journal={Publications of the Astronomical Society of Australia},
   publisher={Cambridge University Press (CUP)},
   author={Hurley-Walker, N. and Galvin, T. J. and Duchesne, S. W. and Zhang, X. and Morgan, J. and Hancock, P. J. and An, T. and Franzen, T. M. O. and Heald, G. and Ross, K. and Vernstrom, T. and Anderson, G. E. and Gaensler, B. M. and Johnston-Hollitt, M. and Kaplan, D. L. and Riseley, C. J. and Tingay, S. J. and Walker, M.},
   year={2022} }

@article{Intema2016,
	author = {Intema, H. T. and Jagannathan, P. and Mooley, K. P. and Frail, D. A.},
	title = {The GMRT 150 MHz all-sky radio survey - First alternative data release TGSS ADR1},
	DOI= "10.1051/0004-6361/201628536",
	url= "https://doi.org/10.1051/0004-6361/201628536",
	journal = {A\&A},
	year = 2017,
	volume = 598,
	pages = "A78",
}

@ARTICLE{Jonas2009,
  author={Jonas, Justin L.},
  journal={Proceedings of the IEEE}, 
  title={MeerKAT—The South African Array With Composite Dishes and Wide-Band Single Pixel Feeds}, 
  year={2009},
  volume={97},
  number={8},
  pages={1522-1530},
  doi={10.1109/JPROC.2009.2020713}}

@INPROCEEDINGS{Jonas2016,
       author = {{Jonas}, J. and {MeerKAT Team}},
        title = "{The MeerKAT Radio Telescope}",
    booktitle = {MeerKAT Science: On the Pathway to the SKA},
         year = 2016,
        month = jan,
          eid = {1},
        pages = {1},
          doi = {10.22323/1.277.0001},
       adsurl = {https://ui.adsabs.harvard.edu/abs/2016mks..confE...1J}
}

@article{Kondapally2022,
    author = {Kondapally, Rohit and Best, Philip N and Cochrane, Rachel K and Sabater, José and Duncan, Kenneth J and Hardcastle, Martin J and Haskell, Paul and Mingo, Beatriz and Röttgering, Huub J A and Smith, Daniel J B and Williams, Wendy L and Bonato, Matteo and Calistro Rivera, Gabriela and Gao, Fangyou and Hale, Catherine L and Małek, Katarzyna and Miley, George K and Prandoni, Isabella and Wang, Lingyu},
    title = {Cosmic evolution of low-excitation radio galaxies in the LOFAR two-metre sky survey deep fields},
    journal = {Monthly Notices of the Royal Astronomical Society},
    volume = {513},
    number = {3},
    pages = {3742-3767},
    year = {2022},
    month = {04},
    issn = {0035-8711},
    doi = {10.1093/mnras/stac1128},
    url = {https://doi.org/10.1093/mnras/stac1128},
    eprint = {https://academic.oup.com/mnras/article-pdf/513/3/3742/43693720/stac1128.pdf},
}

@article{Lacy2020,
doi = {10.1088/1538-3873/ab63eb},
url = {https://dx.doi.org/10.1088/1538-3873/ab63eb},
year = {2020},
month = {jan},
publisher = {The Astronomical Society of the Pacific},
volume = {132},
number = {1009},
pages = {035001},
author = {Lacy, M. and Baum, S. A. and Chandler, C. J. and Chatterjee, S. and Clarke, T. E. and Deustua, S. and English, J. and Farnes, J. and Gaensler, B. M. and Gugliucci, N. and Hallinan, G. and Kent, B. R. and Kimball, A. and Law, C. J. and Lazio, T. J. W. and Marvil, J. and Mao, S. A. and Medlin, D. and Mooley, K. and Murphy, E. J. and Myers, S. and Osten, R. and Richards, G. T. and Rosolowsky, E. and Rudnick, L. and Schinzel, F. and Sivakoff, G. R. and Sjouwerman, L. O. and Taylor, R. and White, R. L. and Wrobel, J. and Andernach, H. and Beasley, A. J. and Berger, E. and Bhatnager, S. and Birkinshaw, M. and Bower, G. C. and Brandt, W. N. and Brown, S. and Burke-Spolaor, S. and Butler, B. J. and Comerford, J. and Demorest, P. B. and Fu, H. and Giacintucci, S. and Golap, K. and Güth, T. and Hales, C. A. and Hiriart, R. and Hodge, J. and Horesh, A. and Ivezić, Ž. and Jarvis, M. J. and Kamble, A. and Kassim, N. and Liu, X. and Loinard, L. and Lyons, D. K. and Masters, J. and Mezcua, M. and Moellenbrock, G. A. and Mroczkowski, T. and Nyland, K. and O’Dea, C. P. and O’Sullivan, S. P. and Peters, W. M. and Radford, K. and Rao, U. and Robnett, J. and Salcido, J. and Shen, Y. and Sobotka, A. and Witz, S. and Vaccari, M. and Weeren, R. J. van and Vargas, A. and Williams, P. K. G. and Yoon, I.},
title = {The Karl G. Jansky Very Large Array Sky Survey (VLASS). Science Case and Survey Design},
journal = {Publications of the Astronomical Society of the Pacific}
}

@article{Mauch2003,
    author = {Mauch, T. and Murphy, T. and Buttery, H. J. and Curran, J. and Hunstead, R. W. and Piestrzynski, B. and Robertson, J. G. and Sadler, E. M.},
    title = {SUMSS: a wide-field radio imaging survey of the southern sky – II. The source catalogue},
    journal = {Monthly Notices of the Royal Astronomical Society},
    volume = {342},
    number = {4},
    pages = {1117-1130},
    year = {2003},
    month = {07},
    issn = {0035-8711},
    doi = {10.1046/j.1365-8711.2003.06605.x},
    url = {https://doi.org/10.1046/j.1365-8711.2003.06605.x},
    eprint = {https://academic.oup.com/mnras/article-pdf/342/4/1117/2819646/342-4-1117.pdf},
}

@article{Mauch2007,
    author = {Mauch, Tom and Sadler, Elaine M.},
    title = {Radio sources in the 6dFGS: local luminosity functions at 1.4 GHz for star-forming galaxies and radio-loud AGN},
    journal = {Monthly Notices of the Royal Astronomical Society},
    volume = {375},
    number = {3},
    pages = {931-950},
    year = {2007},
    month = {01},
    issn = {0035-8711},
    doi = {10.1111/j.1365-2966.2006.11353.x},
    url = {https://doi.org/10.1111/j.1365-2966.2006.11353.x},
    eprint = {https://academic.oup.com/mnras/article-pdf/375/3/931/40677396/mnras_375_3_931.pdf},
}

@article{McConnell2020,
   title={The Rapid ASKAP Continuum Survey I: Design and first results},
   volume={37},
   ISSN={1448-6083},
   url={http://dx.doi.org/10.1017/pasa.2020.41},
   DOI={10.1017/pasa.2020.41},
   journal={Publications of the Astronomical Society of Australia},
   publisher={Cambridge University Press (CUP)},
   author={McConnell, D. and Hale, C. L. and Lenc, E. and Banfield, J. K. and Heald, George and Hotan, A. W. and Leung, James K. and Moss, Vanessa A. and Murphy, Tara and O’Brien, Andrew and Pritchard, Joshua and Raja, Wasim and Sadler, Elaine M. and Stewart, Adam and Thomson, Alec J. M. and Whiting, M. and Allison, James R. and Amy, S. W. and Anderson, C. and Ball, Lewis and Bannister, Keith W. and Bell, Martin and Bock, Douglas C.-J. and Bolton, Russ and Bunton, J. D. and Chippendale, A. P. and Collier, J. D. and Cooray, F. R. and Cornwell, T. J. and Diamond, P. J. and Edwards, P. G. and Gupta, N. and Hayman, Douglas B. and Heywood, Ian and Jackson, C. A. and Koribalski, Bärbel S. and Lee-Waddell, Karen and McClure-Griffiths, N. M. and Ng, Alan and Norris, Ray P. and Phillips, Chris and Reynolds, John E. and Roxby, Daniel N. and Schinckel, Antony E. T. and Shields, Matt and Tremblay, Chenoa and Tzioumis, A. and Voronkov, M. A. and Westmeier, Tobias},
   year={2020} }

@MISC{Mohan2015,
       author = {{Mohan}, Niruj and {Rafferty}, David},
        title = "{PyBDSF: Python Blob Detection and Source Finder}",
 howpublished = {Astrophysics Source Code Library, record ascl:1502.007},
         year = 2015,
        month = feb,
          eid = {ascl:1502.007},
        pages = {ascl:1502.007},
archivePrefix = {ascl},
       eprint = {1502.007},
       adsurl = {https://ui.adsabs.harvard.edu/abs/2015ascl.soft02007M}
}

@article{Mooley2019,
doi = {10.3847/1538-4357/aaef7c},
url = {https://dx.doi.org/10.3847/1538-4357/aaef7c},
year = {2018},
month = {dec},
publisher = {The American Astronomical Society},
volume = {870},
number = {1},
pages = {25},
author = {Mooley, K. P. and Myers, S. T. and Frail, D. A. and Hallinan, G. and Butler, B. and Kimball, A. and Golap, K.},
title = {The Caltech-NRAO Stripe 82 Survey (CNSS). II. On-the-fly Mosaicking Methodology},
journal = {The Astrophysical Journal}
}

@article{Norris2011, 
title={EMU: Evolutionary Map of the Universe}, 
volume={28}, 
DOI={10.1071/AS11021}, 
number={3}, 
journal={Publications of the Astronomical Society of Australia}, 
author={Norris, Ray P. and Hopkins, A. M. and Afonso, J. and Brown, S. and Condon, J. J. and Dunne, L. and Feain, I. and Hollow, R. and Jarvis, M. and Johnston-Hollitt, M. and et al.}, 
year={2011}, 
pages={215–248}
}

@article{Norris2021, 
title={The Evolutionary Map of the Universe pilot survey}, volume={38}, DOI={10.1017/pasa.2021.42}, journal={Publications of the Astronomical Society of Australia}, author={Norris, Ray P. and Marvil, Joshua and Collier, J. D. and Kapińska, Anna D. and O’Brien, Andrew N. and Rudnick, L. and Andernach, Heinz and Asorey, Jacobo and Brown, Michael J. I. and Brüggen, Marcus and et al.}, year={2021}, pages={e046}}

@ARTICLE{Offringa2010,
       author = {{Offringa}, A.~R. and {de Bruyn}, A.~G. and {Biehl}, M. and {Zaroubi}, S. and {Bernardi}, G. and {Pandey}, V.~N.},
        title = "{Post-correlation radio frequency interference classification methods}",
      journal = {\mnras},
         year = 2010,
        month = jun,
       volume = {405},
       number = {1},
        pages = {155-167},
          doi = {10.1111/j.1365-2966.2010.16471.x},
archivePrefix = {arXiv},
       eprint = {1002.1957},
 primaryClass = {astro-ph.IM},
       adsurl = {https://ui.adsabs.harvard.edu/abs/2010MNRAS.405..155O}
}

@article{Offringa2014,
    author = {Offringa, A. R. and McKinley, B. and Hurley-Walker, N. and Briggs, F. H. and Wayth, R. B. and Kaplan, D. L. and Bell, M. E. and Feng, L. and Neben, A. R. and Hughes, J. D. and Rhee, J. and Murphy, T. and Bhat, N. D. R. and Bernardi, G. and Bowman, J. D. and Cappallo, R. J. and Corey, B. E. and Deshpande, A. A. and Emrich, D. and Ewall-Wice, A. and Gaensler, B. M. and Goeke, R. and Greenhill, L. J. and Hazelton, B. J. and Hindson, L. and Johnston-Hollitt, M. and Jacobs, D. C. and Kasper, J. C. and Kratzenberg, E. and Lenc, E. and Lonsdale, C. J. and Lynch, M. J. and McWhirter, S. R. and Mitchell, D. A. and Morales, M. F. and Morgan, E. and Kudryavtseva, N. and Oberoi, D. and Ord, S. M. and Pindor, B. and Procopio, P. and Prabu, T. and Riding, J. and Roshi, D. A. and Shankar, N. Udaya and Srivani, K. S. and Subrahmanyan, R. and Tingay, S. J. and Waterson, M. and Webster, R. L. and Whitney, A. R. and Williams, A. and Williams, C. L.},
    title = "{wsclean: an implementation of a fast, generic wide-field imager for radio astronomy}",
    journal = {Monthly Notices of the Royal Astronomical Society},
    volume = {444},
    number = {1},
    pages = {606-619},
    year = {2014},
    month = {08},
    issn = {0035-8711},
    doi = {10.1093/mnras/stu1368},
    url = {https://doi.org/10.1093/mnras/stu1368},
    eprint = {https://academic.oup.com/mnras/article-pdf/444/1/606/18501772/stu1368.pdf},
}

@article{Pearson_1984,
   author = "Pearson, T. J. and Readhead, A. C. S.",
   title = "Image Formation by Self-Calibration in Radio Astronomy", 
   journal= "Annual Review of Astronomy and Astrophysics",
   year = "1984",
   volume = "22",
   number = "Volume 22, 1984",
   pages = "97-130",
   doi = "https://doi.org/10.1146/annurev.aa.22.090184.000525",
   url = "https://www.annualreviews.org/content/journals/10.1146/annurev.aa.22.090184.000525",
   publisher = "Annual Reviews",
   issn = "1545-4282",
   type = "Journal Article",
  }

@article{Perley_2017,
doi = {10.3847/1538-4365/aa6df9},
url = {https://doi.org/10.3847/1538-4365/aa6df9},
year = {2017},
month = {may},
publisher = {The American Astronomical Society},
volume = {230},
number = {1},
pages = {7},
author = {Perley, R. A. and Butler, B. J.},
title = {An Accurate Flux Density Scale from 50 MHz to 50 GHz},
journal = {The Astrophysical Journal Supplement Series}
}

@inproceedings{Santos2017,
      author         = "Santos, Mario G. and others",
      title          = "{MeerKLASS: MeerKAT Large Area Synoptic Survey}",
      booktitle      = "{Proceedings, MeerKAT Science: On the Pathway to the SKA
                        (MeerKAT2016): Stellenbosch, South Africa, May 25-27,
                        2016}",
      collaboration  = "MeerKLASS",
      year           = "2017",
      eprint         = "1709.06099",
      archivePrefix  = "arXiv",
      primaryClass   = "astro-ph.CO",
      SLACcitation   = "%%CITATION = ARXIV:1709.06099;%%"
}

@article{SKA2020,
    author = "Bacon, David J. and others",
    collaboration = "SKA",
    title = "{Cosmology with Phase 1 of the Square Kilometre Array: Red Book 2018: Technical specifications and performance forecasts}",
    eprint = "1811.02743",
    archivePrefix = "arXiv",
    primaryClass = "astro-ph.CO",
    doi = "10.1017/pasa.2019.51",
    journal = "Publ. Astron. Soc. Austral.",
    volume = "37",
    pages = "e007",
    year = "2020"
}

@ARTICLE{Smirnov2011,
   author = {{Smirnov}, O.~M.},
    title = "{Revisiting the radio interferometer measurement equation. II. Calibration and direction-dependent effects}",
  journal = {\aap},
archivePrefix = "arXiv",
   eprint = {1101.1765},
 primaryClass = "astro-ph.IM",
     year = 2011,
    month = mar,
   volume = 527,
      eid = {A107},
    pages = {A107},
      doi = {10.1051/0004-6361/201116434},
   adsurl = {http://adsabs.harvard.edu/abs/2011A%26A...527A.107S}
}

@article{Smirnov_2015,
    author = {Smirnov, O. M. and Tasse, C.},
    title = {Radio interferometric gain calibration as a complex optimization problem},
    journal = {Monthly Notices of the Royal Astronomical Society},
    volume = {449},
    number = {3},
    pages = {2668-2684},
    year = {2015},
    month = {04},
    issn = {0035-8711},
    doi = {10.1093/mnras/stv418},
    url = {https://doi.org/10.1093/mnras/stv418},
    eprint = {https://academic.oup.com/mnras/article-pdf/449/3/2668/9381057/stv418.pdf},
}

@ARTICLE{Smirnov2024,
       author = {{Smirnov}, O.~M. and {Stappers}, B.~W. and {Tasse}, C. and {Bester}, H.~L. and {Bignall}, H. and {Walker}, M.~A. and {Caleb}, M. and {Rajwade}, K.~M. and {Buchner}, S. and {Woudt}, P. and {Ivchenko}, M. and {Roth}, L. and {Noordam}, J.~E. and {Camilo}, F.},
        title = "{The RATT PARROT: serendipitous discovery of a peculiarly scintillating pulsar in MeerKAT imaging observations of the Great Saturn - Jupiter Conjunction of 2020. I. Dynamic imaging and data analysis}",
      journal = {\mnras},
         year = 2024,
        month = mar,
       volume = {528},
       number = {4},
        pages = {6517-6537},
          doi = {10.1093/mnras/stae303},
       adsurl = {https://ui.adsabs.harvard.edu/abs/2024MNRAS.528.6517S}
}

@article{Tasse_2014,
	author = {{Tasse, C.}},
	title = {Nonlinear Kalman filters for calibration in radio interferometry},
	DOI= "10.1051/0004-6361/201423503",
	url= "https://doi.org/10.1051/0004-6361/201423503",
	journal = {A&A},
	year = 2014,
	volume = 566,
	pages = "A127",
	month = "",
}

@article{Tasse2018,
   title={Faceting for direction-dependent spectral deconvolution},
   volume={611},
   ISSN={1432-0746},
   url={http://dx.doi.org/10.1051/0004-6361/201731474},
   DOI={10.1051/0004-6361/201731474},
   journal={Astronomy &amp; Astrophysics},
   publisher={EDP Sciences},
   author={Tasse, C. and Hugo, B. and Mirmont, M. and Smirnov, O. and Atemkeng, M. and Bester, L. and Hardcastle, M. J. and Lakhoo, R. and Perkins, S. and Shimwell, T.},
   year={2018},
   month=mar, pages={A87} }

@article{Wang2021,
    author = {Wang, Jingying and Santos, Mario G and Bull, Philip and Grainge, Keith and Cunnington, Steven and Fonseca, José and Irfan, Melis O and Li, Yichao and Pourtsidou, Alkistis and Soares, Paula S and Spinelli, Marta and Bernardi, Gianni and Engelbrecht, Brandon},
    title = {H i intensity mapping with MeerKAT: calibration pipeline for multidish autocorrelation observations},
    journal = {Monthly Notices of the Royal Astronomical Society},
    volume = {505},
    number = {3},
    pages = {3698-3721},
    year = {2021},
    month = {05},
    issn = {0035-8711},
    doi = {10.1093/mnras/stab1365},
    url = {https://doi.org/10.1093/mnras/stab1365},
    eprint = {https://academic.oup.com/mnras/article-pdf/505/3/3698/38711683/stab1365.pdf},
}

@article{Whittam2023,
    author = {Whittam, I H and Prescott, M and Hale, C L and Jarvis, M J and Heywood, I and An, Fangxia and Glowacki, M and Maddox, N and Marchetti, L and Morabito, L K and Adams, N J and Bowler, R A A and Hatfield, P W and Varadaraj, R G and Collier, J and Frank, B and Taylor, A R and Santos, M G and Vaccari, M and Afonso, J and Ao, Y and Delhaize, J and Knowles, K and Kolwa, S and Randriamampandry, S M and Randriamanakoto, Z and Smirnov, O and Smith, D J B and White, S V},
    title = {MIGHTEE: Multi-wavelength counterparts in the COSMOS field},
    journal = {Monthly Notices of the Royal Astronomical Society},
    volume = {527},
    number = {2},
    pages = {3231-3245},
    year = {2023},
    month = {10},
    issn = {0035-8711},
    doi = {10.1093/mnras/stad3307},
    url = {https://doi.org/10.1093/mnras/stad3307},
    eprint = {https://academic.oup.com/mnras/article-pdf/527/2/3231/56343751/stad3307.pdf},
}

@ARTICLE{Wyithe2008,
   author = {{Wyithe}, J.~S.~B. and {Loeb}, A. and {Geil}, P.~M.},
    title = "{Baryonic acoustic oscillations in 21-cm emission: a probe of dark energy out to high redshifts}",
  journal = {\mnras},
archivePrefix = "arXiv",
   eprint = {0709.2955},
     year = 2008,
    month = jan,
   volume = 383,
    pages = {1195-1209},
      doi = {10.1111/j.1365-2966.2007.12631.x},
   adsurl = {http://adsabs.harvard.edu/abs/2008MNRAS.383.1195W}
}

@article{Wayth2015, 
title={GLEAM: The GaLactic and Extragalactic All-Sky MWA Survey}, 
volume={32}, 
DOI={10.1017/pasa.2015.26}, 
journal={Publications of the Astronomical Society of Australia}, 
author={Wayth, R. B. and Lenc, E. and Bell, M. E. and Callingham, J. R. and Dwarakanath, K. S. and Franzen, T. M. O. and For, B.-Q. and Gaensler, B. and Hancock, P. and Hindson, L. and et al.}, 
year={2015}, 
pages={e025}
}

@ARTICLE{Numpy,
       author = {{Harris}, Charles R. and {Millman}, K. Jarrod and {van der Walt}, St{\'e}fan J. and {Gommers}, Ralf and {Virtanen}, Pauli and {Cournapeau}, David and {Wieser}, Eric and {Taylor}, Julian and {Berg}, Sebastian and {Smith}, Nathaniel J. and {Kern}, Robert and {Picus}, Matti and {Hoyer}, Stephan and {van Kerkwijk}, Marten H. and {Brett}, Matthew and {Haldane}, Allan and {del R{\'\i}o}, Jaime Fern{\'a}ndez and {Wiebe}, Mark and {Peterson}, Pearu and {G{\'e}rard-Marchant}, Pierre and {Sheppard}, Kevin and {Reddy}, Tyler and {Weckesser}, Warren and {Abbasi}, Hameer and {Gohlke}, Christoph and {Oliphant}, Travis E.},
        title = "{Array programming with NumPy}",
      journal = {\nat},
         year = 2020,
        month = sep,
       volume = {585},
       number = {7825},
        pages = {357-362},
          doi = {10.1038/s41586-020-2649-2},
archivePrefix = {arXiv},
       eprint = {2006.10256},
 primaryClass = {cs.MS},
       adsurl = {https://ui.adsabs.harvard.edu/abs/2020Natur.585..357H}
}

@ARTICLE{Scipy,
       author = {{Virtanen}, Pauli and {Gommers}, Ralf and {Oliphant}, Travis E. and {Haberland}, Matt and {Reddy}, Tyler and {Cournapeau}, David and {Burovski}, Evgeni and {Peterson}, Pearu and {Weckesser}, Warren and {Bright}, Jonathan and {van der Walt}, St{\'e}fan J. and {Brett}, Matthew and {Wilson}, Joshua and {Millman}, K. Jarrod and {Mayorov}, Nikolay and {Nelson}, Andrew R.~J. and {Jones}, Eric and {Kern}, Robert and {Larson}, Eric and {Carey}, C.~J. and {Polat}, {\.I}lhan and {Feng}, Yu and {Moore}, Eric W. and {VanderPlas}, Jake and {Laxalde}, Denis and {Perktold}, Josef and {Cimrman}, Robert and {Henriksen}, Ian and {Quintero}, E.~A. and {Harris}, Charles R. and {Archibald}, Anne M. and {Ribeiro}, Ant{\^o}nio H. and {Pedregosa}, Fabian and {van Mulbregt}, Paul and {SciPy 1. 0 Contributors}},
        title = "{SciPy 1.0: fundamental algorithms for scientific computing in Python}",
      journal = {Nature Methods},
         year = 2020,
        month = feb,
       volume = {17},
        pages = {261-272},
          doi = {10.1038/s41592-019-0686-2},
archivePrefix = {arXiv},
       eprint = {1907.10121},
 primaryClass = {cs.MS},
       adsurl = {https://ui.adsabs.harvard.edu/abs/2020NatMe..17..261V}
}

@ARTICLE{Matplotlib,
       author = {{Hunter}, John D.},
        title = "{Matplotlib: A 2D Graphics Environment}",
      journal = {Computing in Science and Engineering},
         year = 2007,
        month = may,
       volume = {9},
       number = {3},
        pages = {90-95},
          doi = {10.1109/MCSE.2007.55},
       adsurl = {https://ui.adsabs.harvard.edu/abs/2007CSE.....9...90H}
}

@ARTICLE{astropy:2022,
       author = {{Astropy Collaboration} and {Price-Whelan}, Adrian M. and {Lim}, Pey Lian and {Earl}, Nicholas and {Starkman}, Nathaniel and {Bradley}, Larry and {Shupe}, David L. and {Patil}, Aarya A. and {Corrales}, Lia and {Brasseur}, C.~E. and {N{"o}the}, Maximilian and {Donath}, Axel and {Tollerud}, Erik and {Morris}, Brett M. and {Ginsburg}, Adam and {Vaher}, Eero and {Weaver}, Benjamin A. and {Tocknell}, James and {Jamieson}, William and {van Kerkwijk}, Marten H. and {Robitaille}, Thomas P. and {Merry}, Bruce and {Bachetti}, Matteo and {G{"u}nther}, H. Moritz and {Aldcroft}, Thomas L. and {Alvarado-Montes}, Jaime A. and {Archibald}, Anne M. and {B{'o}di}, Attila and {Bapat}, Shreyas and {Barentsen}, Geert and {Baz{'a}n}, Juanjo and {Biswas}, Manish and {Boquien}, M{'e}d{'e}ric and {Burke}, D.~J. and {Cara}, Daria and {Cara}, Mihai and {Conroy}, Kyle E. and {Conseil}, Simon and {Craig}, Matthew W. and {Cross}, Robert M. and {Cruz}, Kelle L. and {D'Eugenio}, Francesco and {Dencheva}, Nadia and {Devillepoix}, Hadrien A.~R. and {Dietrich}, J{"o}rg P. and {Eigenbrot}, Arthur Davis and {Erben}, Thomas and {Ferreira}, Leonardo and {Foreman-Mackey}, Daniel and {Fox}, Ryan and {Freij}, Nabil and {Garg}, Suyog and {Geda}, Robel and {Glattly}, Lauren and {Gondhalekar}, Yash and {Gordon}, Karl D. and {Grant}, David and {Greenfield}, Perry and {Groener}, Austen M. and {Guest}, Steve and {Gurovich}, Sebastian and {Handberg}, Rasmus and {Hart}, Akeem and {Hatfield-Dodds}, Zac and {Homeier}, Derek and {Hosseinzadeh}, Griffin and {Jenness}, Tim and {Jones}, Craig K. and {Joseph}, Prajwel and {Kalmbach}, J. Bryce and {Karamehmetoglu}, Emir and {Ka{l}uszy{'n}ski}, Miko{l}aj and {Kelley}, Michael S.~P. and {Kern}, Nicholas and {Kerzendorf}, Wolfgang E. and {Koch}, Eric W. and {Kulumani}, Shankar and {Lee}, Antony and {Ly}, Chun and {Ma}, Zhiyuan and {MacBride}, Conor and {Maljaars}, Jakob M. and {Muna}, Demitri and {Murphy}, N.~A. and {Norman}, Henrik and {O'Steen}, Richard and {Oman}, Kyle A. and {Pacifici}, Camilla and {Pascual}, Sergio and {Pascual-Granado}, J. and {Patil}, Rohit R. and {Perren}, Gabriel I. and {Pickering}, Timothy E. and {Rastogi}, Tanuj and {Roulston}, Benjamin R. and {Ryan}, Daniel F. and {Rykoff}, Eli S. and {Sabater}, Jose and {Sakurikar}, Parikshit and {Salgado}, Jes{'u}s and {Sanghi}, Aniket and {Saunders}, Nicholas and {Savchenko}, Volodymyr and {Schwardt}, Ludwig and {Seifert-Eckert}, Michael and {Shih}, Albert Y. and {Jain}, Anany Shrey and {Shukla}, Gyanendra and {Sick}, Jonathan and {Simpson}, Chris and {Singanamalla}, Sudheesh and {Singer}, Leo P. and {Singhal}, Jaladh and {Sinha}, Manodeep and {Sip{H{o}}cz}, Brigitta M. and {Spitler}, Lee R. and {Stansby}, David and {Streicher}, Ole and {{{S}}umak}, Jani and {Swinbank}, John D. and {Taranu}, Dan S. and {Tewary}, Nikita and {Tremblay}, Grant R. and {Val-Borro}, Miguel de and {Van Kooten}, Samuel J. and {Vasovi{'c}}, Zlatan and {Verma}, Shresth and {de Miranda Cardoso}, Jos{'e} Vin{'i}cius and {Williams}, Peter K.~G. and {Wilson}, Tom J. and {Winkel}, Benjamin and {Wood-Vasey}, W.~M. and {Xue}, Rui and {Yoachim}, Peter and {Zhang}, Chen and {Zonca}, Andrea and {Astropy Project Contributors}},
        title = "{The Astropy Project: Sustaining and Growing a Community-oriented Open-source Project and the Latest Major Release (v5.0) of the Core Package}",
      journal = {\apj},
         year = 2022,
        month = aug,
       volume = {935},
       number = {2},
          eid = {167},
        pages = {167},
          doi = {10.3847/1538-4357/ac7c74},
archivePrefix = {arXiv},
       eprint = {2206.14220},
 primaryClass = {astro-ph.IM},
       adsurl = {https://ui.adsabs.harvard.edu/abs/2022ApJ...935..167A}
}

\appendix

\section{Visibility smearing - mathematical formalism}
\label{appen0}
Calculations of the visibility time and frequency smearing can be found in \citet{Smirnov2011} and \citet{Tasse2018}. In this appendix we review the formalism with a focus on the MeerKAT OTF setup. Throughout this discussion, many different pointing centres are used and often confused. Therefore, it is worth giving a brief description of our definitions. First we have the telescope pointing, e.g. where the dishes are pointing. It is assumed that the primary beam will be 1 towards that direction. If the dishes move, the beam will change on a sky based coordinate system. Even in a tracking observation, a non symmetric beam will be time dependent in this coordinate system due to rotation. We then have the delay centre, the direction towards which the delays should add up to zero. And finally, we have the phase centre, which is used to set the coordinate system over which we add up all the measured visibilities in order to make an image. In a single tracking observation, these three pointings are usually the same. But for instance, if we want to combine a set of different tracking observations through mosaicking, then the phase centre will be different from those pointings. 

Let us use the equatorial coordinate system so that the sky is constant in time (except for transients). The $z$ axis points along the North Celestial Pole (NCP), so the angle from the NCP is $\theta = 90^{\circ}-\delta$ where $\delta$ is the standard declination (negative towards the South). The $x$ axis points along $RA=0^{\circ}$ (right ascension) and the $y$ axis along $RA=90^{\circ}$ (East) - right handed system.
Take $\bt$ as the baseline vector coordinates in this system, with $(x,y,z)$ its components along the same axis (note that we are not dividing by the wavelength $\lambda$ here). The baseline coordinates will change with time due to Earth's rotation.
Usually, for this setup, a coordinate system with the $z$ axis along the direction of the telescope pointing is more common. But for now, it is simpler to use this system. They are related through a rotation.

\subsection{Instantaneous Visibility}
In a radio interferometric observation, cross-correlation visibilities are the primary measured quantity. Considering an observation at time $t$ and frequency $\nu$ the instantaneous visibility function for two antennas can be written as,
\begin{equation}
    \visbnut = \int d\Omega \, \Irnut \Brnut e^{- 2 \pi i \nubc \bt \cdot \r},
    \label{eq:vis_def1}
\end{equation}
where $\Irnut$ is the specific intensity of the sky in the direction $\r$ and $\Brnut$ is the primary beam response of the antennas along $\r$. The integral is performed over solid angle $\Omega$. The primary beam is a function of time and frequency and depends on the telescope pointing direction.

For a single point source with flux $S_{\rm p}$ at position $\r_p$, the equation above simplifies to:
\begin{equation}
    \visbnut = S_{\rm p}\, \Brpnut e^{- 2 \pi i \nubc \bt \cdot \r_p}.
    \label{eq:vis_ps1}
\end{equation}
Note that we are not assuming here that the telescope is pointing in the direction of the source, so that $\Brnut$ is not necessarily 1. The phase $\phi=2 \pi \nubc \bt \cdot \r_p$, is only zero if the source direction is perpendicular to the baseline, e.g. at zenith.

\subsection{Averaged Visibility}
In a real setup, the visibilities need to be averaged over a certain time, $\delta t$ (the time dump) and frequency, $\delta\nu$ (the channel width). For our setup, $\delta t$ is 2 seconds and $\delta \nu$ varies between 0.1 and 0.2 MHz. This averaging creates a problem because the phase in \cref{eq:vis_def1} or \cref{eq:vis_ps1} changes with time due to the sky rotation. Even for a point source at zenith, the phase will quickly become non-zero. The effect is more pronounced for longer baselines. As the correlator integrates over time and frequency, the combination of different phases will lead to a de-correlation and reduction of the measured flux. This is commonly called smearing (in time or frequency) or fringe washing. The effect is often described in terms of the geometric delay: $\tau_g = (\mathbfit{b} \cdot \r)/c = b\sin\theta/c$. MeerKAT has a maximum baseline of 8000 m giving a maximum delay of 25.7 $\mu$s.

The way around this is for the correlator to add an instrumental delay, $\tau_i$, that cancels $\tau_g$ before averaging. Since $\tau_g$ changes with time as the sky rotates, $\tau_i$ will also have to be continuously adjusted - this is called delay tracking. Obviously this delay tracking cannot be continuous so the cancellation is not perfect but it is completely negligible for systems like MeerKAT. 

However, one should note that the delay cancellation only happens in a specific direction, usually towards the telescope pointing direction. So point sources away from the pointing centre will still face smearing. It will be useful to consider this more common case before we move to the OTF setup.

\subsection{Smearing in a tracking observation}
\label{smear_track}
Let us assume that the correlator is adding delay tracking as described above towards a sky direction $\r_d$ (and that this is continuous for simplification). When averaging over time and frequency we get:
\begin{multline}
    \visbnutm = \frac{1}{\dt \dnu} \int d\Omega \, \int_{\nu_m - \dnu/2}^{\nu_m + \dnu/2} \, d\nu \int_{t_m - \dt/2}^{t_m + \dt/2} dt\\
    \times \Irnut \Brnut e^{- 2 \pi i \nubc \bt \cdot (\r-\r_d)},
    \label{eq:vis_def2}
\end{multline}
where $t_m$ and $\nu_m$ are the central values of the intervals.

The integrals over time and frequency can be solved as in eq. 23 of \citet{Smirnov2011} or eqs 36-38 in \citet{Tasse2018}. We start by simplifying the setup with the following assumptions:
\begin{enumerate}
    \item We can assume that the specific intensity of the sky $\Irnut$ and the primary beam $\Brnut$ does not change significantly within the small interval of $\dt$ and $\dnu$. Since we are using sky based coordinates, $\Irnut$ should indeed be constant in time and the frequency change should be quite small over the channel width. In a tracking observation, the primary beam is also quite constant in the sky coordinate system. The situation will change with OTF which will be discussed later.
    \item We can further assume that the $\dt$ is small enough that the time dependence of the baselines can be well approximated to first order:
    \begin{equation}
        \bt \approx \btm + (t-t_m) \, \dbdtm
        \label{eq:bl1}
    \end{equation}
\end{enumerate}

\cref{eq:vis_def2} then simplifies to
\begin{multline}
    \visbnutm \approx \frac{1}{\dt \dnu} \int d\Omega \, \Irnutm \Brnutm \\ 
    \times \int_{- \dnu/2}^{+ \dnu/2} \, d\nu \, e^{- 2 \pi i (\frac{\nu + \nu_m}{c}) \btm \cdot (\r-\r_d)}\\
     \times \int_{- \dt/2}^{+ \dt/2} dt \, e^{- 2 \pi i t \numbc \dbdtm \cdot (\r-\r_d)},
    \label{eq:vis_def3}
\end{multline}
where we have also approximated $(\nu + \nu_m)t\approx \nu_m t$, given the very small intervals under integration.

Using:
\begin{equation}
    \frac{1}{\delta x} \int_{-\frac{\delta x}{2}}^{\frac{\delta x}{2}} e^{i\alpha x} dx = \text{sinc}\left(\alpha \frac{\delta x}{2}\right)
    = \text{sinc}\left(- \alpha \frac{\delta x}{2}\right),
\end{equation}
with $\text{sinc}(x)\equiv \text{sin}(x)/x$, we finally have,
\begin{multline}
    \visbnutm = \int d\Omega \, \Irnutm \Brnutm  \\
    \times e^{- 2 \pi i \numbc \btm \cdot [\r - \r_d]} \, \sinc(\delta \Psi) \, \sinc(\delta \Phi),   
    \label{eq:vis_def4}
\end{multline}
where,
\begin{align}
    &\delta \Psi  = \pi \dt \numbc \dbdtm \cdot [\r-\r_d] \label{eq:vis_def5a} \\
    &\delta \Phi = \pi \frac{\dnu}{c} \btm \cdot [\r - \r_d].
    \label{eq:vis_def5b}
\end{align}

We can again look at the simple case of a single point source with flux $S_{\rm p}$ at position $\r_p$. We get:
\begin{multline}
    \visbnutm = S_{\rm p}\, \Brpnut e^{- 2 \pi i \numbc \btm \cdot [\r_p - \r_d]} \\
    \times \sinc(\delta \Psi) \, \sinc(\delta \Phi),
    \label{eq:vis_ps2}
\end{multline}
with
\begin{align}
    &\delta \Psi  = \pi \dt \numbc \dbdtm \cdot [\r_p-\r_d] \\
    &\delta \Phi = \pi \frac{\dnu}{c} \btm \cdot [\r_p - \r_d].
    \label{eq:phases_track}
\end{align}
The exponential outside is just a constant phase which can be removed. For a point source at the delay centre, $\r_p=\r_d$ and we recover the flux exactly. However, for sources away from the delay centre, there is going to be a reduction in flux given by the $\sinc$ functions. It has been shown that for tracking observations, time-smearing is not a significant consideration for MeerKAT.  Details of relevant simulations and analysis can be found in the MeerKAT time smearing simulation report\footnote{https://archive-gw-1.kat.ac.za/public/meerkat/MeerKAT-Time-Smearing-Simulation.pdf}.

\subsection{The case for OTF}
In the current MeerKAT OTF configuration, the observation is performed at a constant elevation ($el$) and the pointing centre of the antennas move back and forth along azimuth ($az$) (see section \ref{sec:HRD} for further details). One may think that the correlator should update the delays so to track the dish movement exactly. However, what matters is the change in the phase during the integration interval and that is due to the change in the baseline vector, $\bt$. Therefore, the requirement is still that the delays track the sky rotation. There is also a change in the primary beam from the dish movements which affects the overall amplitude (not the phase - see Eq.~\ref{eq:vis_def2}), but that should be a small effect as long as the integration interval is small. This can also be corrected using an effective primary beam - see appendix~\ref{appen1} for a discussion. The point is that the delay tracking within the integration time is independent of the scanning speed. However it is true that the update of the delay centre between time dumps during the scan should be tied to that scanning speed.

In the current setup, the integration time $\dt$ is 2 seconds, during which the dishes move about 10 arcmin. Ideally one would want delay tracking to be applied towards the dish direction in the middle of the 2 second interval. The delay centre should then be updated to the middle of the next 2 seconds and so on, with the shift aligned with the time dumps. This is problematic for MeerKAT since it requires some level of synchronization between the dish pointing and the correlator delays. Updating the delay centre every 2 seconds has its own challenges due to data rates. We could in principle calculate all the delay jumps beforehand but again that would require non-trivial changes to the current system. MeerKAT adopts a much simpler approach: it just stops delay tracking when doing scans. There is one constant delay applied corresponding to the initial sky position at the middle of the $az$-$el$ coordinates, but there are no further updates. This might sound quite hopeless for our goals, but the situation is ``saved" up to a certain point by our small time and frequency intervals. We discuss next the mathematical setup for this case.

As mentioned, there is a constant delay applied in the MeerKAT OTF setup, given by: $\tau_i^0 = \mathbfit{b}(t_0) \cdot \r_c(t_0)/c$, where $\mathbfit{b}(t_0)$ is the baseline vector at the start of the scan and $\r_c(t_0)$ is the centre of the $az$-$el$ scan coordinates also at the start of the scan (e.g. it uses the constant elevation and the centre of the azimuth interval). We note that both $\mathbfit{b}$ and $\r_c$ are constant in the $az$-$el$ system but will become time dependent in the equatorial coordinate system we have been considering. They are related by a rotation that does not change the dot product. We can therefore write this delay using baseline vectors at any other time: $\tau_i^0 = \mathbfit{b}(t) \cdot \r_c(t)/c$. The phase corresponding to this delay will still have a frequency dependence and is given by
\begin{equation}
\phi_c(\nu) = 2 \pi \nu \tau_i^0 = 2 \pi \nubc \mathbfit{b}(t_0) \cdot \r_c(t_0) = 2 \pi \nubc \mathbfit{b}(t) \cdot \r_c(t).    
\end{equation}
The derivation then follows closely what we did for the tracking case. We start with:
\begin{multline}
    \visbnutm = \frac{1}{\dt \dnu} \int d\Omega \, \int_{\nu_m - \dnu/2}^{\nu_m + \dnu/2} \, d\nu \, e^{i \phi_c(\nu)} \int_{t_m - \dt/2}^{t_m + \dt/2} dt\\
    \times \Irnut \Brnut e^{- 2 \pi i \nubc \bt \cdot \rt},
\end{multline}
and following through with the same approximations, noting that now we do not have a time dependent instrumental delay, we get
\begin{multline}
    \visbnutm = \int d\Omega \, \Irnutm \Brnutm  \\
    \times e^{- 2 \pi i \numbc \btm \cdot [\r - \rcm]} \, \sinc(\delta \Psi) \, \sinc(\delta \Phi),   
\end{multline}
where,
\begin{align}
    &\delta \Psi  = \pi \dt \numbc \dbdtm \cdot \r \label{eq:vis_def5c} \\
    &\delta \Phi = \pi \frac{\dnu}{c} \btm \cdot [\r - \rcm].
        \label{eq:phases_otf}
\end{align}

The main difference compared to the tracking case is that we no longer have a delay direction $\r_d$ for the time smearing as in \cref{eq:phases_track}. This means that time smearing is only small for directions perpendicular to $\dbdt$. For frequency smearing the situation is less dramatic as $\rcm$ is fairly close to the pointing direction (at most 5 degrees away in the current setup). 

During the writing of this paper, a new fix has been developed that should improve the situation for future data. The new setup will track a fixed sky position during each scan line and then update. The delay centre is then the middle of each scan line. The smearing equations are then the same as for a tracking observation (Eq.~\ref{smear_track}) with the difference that we sometimes have to image farther away from the delay centre than what is usual in standard tracking observations.
This setup will give the same frequency smearing as above but with a much better behaved time smearing corresponding to angular separations under 5 degrees. This will lead to a total smearing of less than 5\%.

\subsection{uvw coordinates}
So far we have been assuming an equatorial coordinate system. Usually the coordinates of $\bf b$ are taken in a rotated coordinate system pointing towards the phase centre (usually also the delay and pointing centre), $\r_{ph}$. They are also divided by the wavelength $\lambda$ and called ${\bf u} = {\bf b}/\lambda = (u,v,w)$, with $w$ ($z$ axis) pointing towards the phase centre, $v$ ($y$ axis), perpendicular to $w$ and along the plane defined by the North Pole and $w$ (so towards the NCP) and $u$ ($x$ axis), perpendicular to $v$ and $w$ and pointing East (in a right handed coordinate system). This rotation is time independent, and since all the expressions above only depend on internal "dot" products, they will stay the same. 

In the new coordinate system, $\r_{ph}=(0,0,1)$. Therefore, towards the phase centre we have ${\bf u}\cdot\r_{ph} = w$. In terms of the time derivative of $\bf b$ or $\bf u$, since its change in time is only due to the Earth rotation, We can write $\dudtm = \utm \times \oc $, where $\oc$ is the angular velocity vector of the Earth’s rotation (along the North pole and with amplitude $\oc = 2\pi/24/3600\ s^{-1}$ or 15 arcsec/s). In the $(uvw)$ coordinate system, we have:
\begin{equation*}
    \oc = \oc (0,  \cos\delta_0, \sin\delta_0),
\end{equation*}
where $\delta_0$ is the declination of the phase centre. 
We then have,
\begin{align}
    &\dudtm = -(v \omega_w - w \omega_v, w \omega_u - u \omega_w, u \omega_v - v \omega_u)\nonumber \\
    &= -\oc (v \sin\delta_0 - w \cos\delta_0, -u \sin\delta_0, u \cos\delta_0),
    \label{eq:dudt}
\end{align}
and $\dudtm\cdot\r_{ph} = \frac{dw}{dt} = - u \oc \cos\delta_0$. This is known as the fringe frequency.

Let us now consider again the simple case of a single point source for the two situations, standard tracking and OTF. Taking the point source at the centre of the beam (along the pointing direction), and ignoring any overall phase, both cases can be written as
\begin{equation}
    \visbnutm = S_{\rm p}\, \sinc(\delta \Psi) \, \sinc(\delta \Phi).
\end{equation}

For the tracking case, let us take the phase centre along the delay centre, $\r_{ph}=\r_d$. Then:
\begin{multline}
    \delta \Psi  = \pi \dt \dudtm \cdot [\r_p-\r_d]  
    = -\pi \dt \oc \big[l(v \sin\delta_0 - w \cos\delta_0)\\ -u m \sin\delta_0 + u (\sqrt{1-l^2-m^2}-1)\cos\delta_0\big],\\
    \delta \Phi = \pi \frac{\dnu}{\nu} \utm \cdot [\r_p - \r_d] \\ 
    = \pi \frac{\dnu}{\nu} [u l + v m + w(\sqrt{1-l^2-m^2}-1)],
\end{multline}
where $(l,m,n)$ are the coordinates of the unit point source vector $\r_p$ (with $n=\sqrt{1-l^2-m^2}$).

For the OTF case, let us take the phase centre along the pointing direction which is also the point source position in this case, $\r_{ph}=\r_p$. Then:
\begin{multline}
    \delta \Psi  = \pi \dt \dudtm \cdot \r_p = -\pi \dt u \oc \cos\delta_0\\
    \delta \Phi = \pi \frac{\dnu}{\nu} \utm \cdot [\r_p - \rcm]\\
    = -\pi \frac{\dnu}{\nu} \big[u l_c + v m_c + w(\sqrt{1-l_c^2-m_c^2}-1)\big],
\end{multline}
where $(l_c,m_c,n_c)$ are the direction cosines of the time dependent delay centre. We can see that time smearing, which is the main source of smearing for the OTF case, is a function of $u$ and declination.

\subsection{Baseline vectors}
In order to define the physical baseline vectors, we use a coordinate system that is bound to the Earth (e.g. does not change with the Earth rotation). We choose the $z$ axis along the NCP, the $x$ axis pointing along $HA=0$ (zero hour-angle, which is observer specific) and the $y$ axis pointing towards $HA=-6\ $ hours (East). This is again a right handed system. This is telescope specific and we can get its values by knowing the latitude, longitude and altitude of the dish positions and by picking an array centre to define $HA=0$. The $(x,y)$ points are on the plane of declination $\delta = 0$. We can related the baseline vector in this coordinate system $(B_x, B_y, B_z)$ with the $(uvw)$ coordinates through:
\begin{align*}
    \begin{bmatrix}
        u\\
        v\\
        w
    \end{bmatrix} = \frac{\nu}{c}
    \begin{bmatrix}
    \sin H_0 & \cos H_0 & 0\\
    -\sin\delta_0\cos H_0 & \sin\delta_0\sin H_0 & \cos \delta_0\\
    \cos\delta_0\cos H_0 & - \cos\delta_0\sin H_0 & \sin \delta_0
    \end{bmatrix}
    \begin{bmatrix}
        B_x\\
        B_y\\
        B_z
    \end{bmatrix},
\end{align*}
where $H_0$ is the local hour angle of the phase centre and $\delta_0$ is its declination.

\section{Smearing of Primary Beam}
\label{appen1}
Here we derive an expression for the fractional change in the flux density of sources if the motion of the primary beam within a snapshot (duration in which the delay centre is constant). Describing the primary beam with a 2D Gaussian function we have
\begin{equation}
    B_{\nu}(x,y) = e^{-x^2/(2 \sigx^2) -(y^2)/(2\sigy^2)}
\end{equation}
where x and y are Cartesian coordinates in the image plane defined with respect to the pointing centre. For data taken over the interval from $t - \dt/2$ to $t + \dt/2$ at OTF scanning rate $\Theta$, a visibility centreed at an offset point $(x_0, y_0)$ in the primary beam will see an effective beam value of:

\begin{align}
    &B_{\nu, {\rm eff}}(x_0, y_0; \Dx, \Dy ) = \frac{1}{\Dx \Dy} \int^{x_0 + \Dx/2}_{x_0 - \Dx/2} dx  \int^{y_0 + \Dy/2}_{y_0 - \Dy/2} dy B_{\nu}(x, y) \\
    &= \frac{\pi \sigx \sigy}{2 \Dx \Dy} \left[ {\rm Erf}\left( \frac{x_0 + \Dx/2}{\sqrt{2}\sigx} \right) - {\rm Erf}\left( \frac{x_0 - \Dx/2}{\sqrt{2}\sigx} \right) \right] \nonumber \\
    &\times \left[ {\rm Erf}\left( \frac{y_0 + \Dy/2}{\sqrt{2}\sigy} \right) - {\rm Erf}\left( \frac{y_0 - \Dy/2}{\sqrt{2}\sigy} \right) \right] 
    \label{eq:beff}
\end{align}

where $\sigx = \sigy = {\rm FWHM}_{{\rm pb}, \nu}/2.355$ is the standard deviation of the primary beam at $\nu$, $\Dx$ and $\Dy$ are the slew of the antennas within time interval $\dt$. The change in the total flux densities due to this beam smearing and the fractional error, $B_{\nu, {\rm eff}}/B_{\nu}$ is shown in \cref{fig:bsm}.

\begin{figure}
    \includegraphics[width=0.95\linewidth]{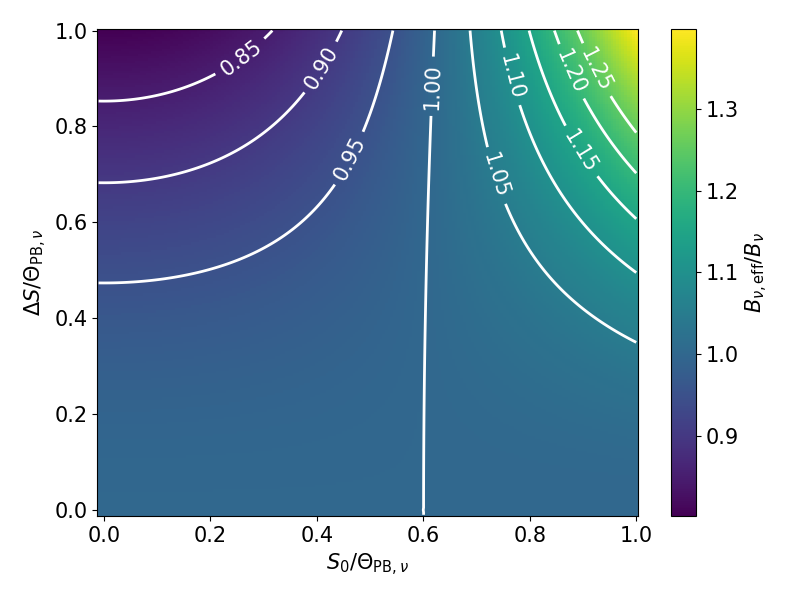}
    \caption{ M-OTF smeared beam at 1.7 GHz. The plot shows the fractional change in the flux density with respect to the true flux density when only a single, time invariant, primary beam correction is applied to each scan.}
    \label{fig:bsm}
\end{figure}

\label{lastpage}
\end{document}